\documentclass[fleqn,usenatbib]{mnras} 

\usepackage{mathptmx}

\usepackage[T1]{fontenc}

\DeclareRobustCommand{\VAN}[3]{#2}
\let\VANthebibliography\thebibliography
\def\thebibliography{\DeclareRobustCommand{\VAN}[3]{##3}\VANthebibliography}

\usepackage{color}
\usepackage{xcolor}
\usepackage{hyperref}
\usepackage{graphicx}	
\usepackage{amsmath}	
\usepackage{amssymb}	

\title[The OIII emission in Seyfert 1 galaxies]{Spectroscopic study of the [OIII]$\lambda$5007 profile in Seyfert 1 galaxies}

\author[Schmidt et al.]{
Eduardo O. Schmidt,$^{1,2,3}$\thanks{E-mail: eduardo.schmidt@unc.edu.ar}
Laura D. Baravalle,$^{1,3}$
Adriana R. Rodr\'iguez-Kamenetzky $^{1,3}$
\\
$^{1}$Instituto de Astronomía Teórica y Experimental, CONICET–UNC, Córdoba, Laprida 854, Córdoba, Argentina\\
$^{2}$Observatorio Astron\'omico de C\'ordoba, Universidad Nacional de C\'ordoba, Laprida 854, X5000BGR, C\'ordoba, Argentina.\\
$^{3}$Consejo de Investigaciones Científicas y Técnicas de la República Argentina, Buenos Aires, Argentina
}

\pubyear{2020}

\begin{document}
\label{firstpage}
\pagerange{\pageref{firstpage}--\pageref{lastpage}}
\maketitle

\begin{abstract}

The spectra of active galactic nuclei usually exhibit wings in some emission lines, such as [OIII]$\lambda\lambda$5007,4959, with these wings generally being blueshifted and related to strong winds and outflows.
 The aim of this work was to analyse the [OIII] emission lines in broad line Seyfert 1 (BLS1) galaxies in order to detect the presence of wings, and to study the [OIII] line properties and their possible connection with the central engine. In addition, we attempted to compare the black hole mass distribution in both BLS1 galaxies with symmetric and blue-asymmetric [OIII] profiles.  
 For this purpose, we carried out a spectroscopic study of a sample of 45 nearby southern BLS1 galaxies from the 6 Degree Field Galaxy survey. 
 The [OIII] emission lines were well fitted using a single Gaussian function in 23 galaxies, while 22 objects presented a wing component and required a double-Gaussian decomposition. 
 By computing the radial velocity difference between the wing and core centroids (i.e. $\Delta$v), we found 18 galaxies exhibiting blueshifted wings, 2 objects presenting red wings and 2 galaxies showing symmetric wings ($\Delta$v$= 0$). Moreover, $\Delta$v was slightly correlated with the black hole mass. In addition, we computed the radial velocity difference of the blue-side full extension of the wing relative to the centroid of the core component through the \emph{blue emission} parameter, which revealed a correlation with black hole mass, in agreement with previous results reported for narrow line galaxies. Finally, in our sample, similar black hole mass distributions were observed in both BLS1 galaxies with symmetric and blueshifted asymmetric [OIII] profiles.

\end{abstract}




\section{Introduction}

Broad line Seyfert 1 (BLS1) galaxies are a class of active galactic nuclei (AGN) which exhibit broad permitted emission lines, and show a wide range of emission at all wavelengths with negligible stellar light contamination \citep[e.g.,][]{Netzer2013}. Some type-1 AGNs present asymmetric features in several emission lines, with this behaviour being observed in a wide range of wavelengths, such as X-rays \citep[e.g.,][]{Chelouche2005}, UV \citep[e.g.,][]{Sulentic2007}, optical \citep[e.g.,][]{VeronCetty2001,Komossa2008}, IR \citep[e.g.,][]{RA2006}, and radio \citep[e.g.,][]{Morganti2005,Gallimore2006}. In particular, Seyfert 1 galaxies usually exhibit asymmetric blueshifted [OIII]$\lambda\lambda$4959,5007 emission lines \citep[e.g.,][]{Komossa2007}. Asymmetric profiles can be mainly described by the combination of a core component and a blue wing \citep[e.g.,][]{ VeronCetty2001,Komossa2007}. This core component is emitted from the narrow line region (NLR) and has a typical full width at half maximum (FWHM) of few hundred km s$^{-1}$, while the blue wing originates closer to the active nucleus and has higher values of FWHM, typically in the range $\sim$500$-$1500 km s$^{-1}$ \citep[e.g.,][]{VeronCetty2001,Komossa2007,Schmidt2018,Oio2019, Cooke2020}. 

Winds and outflows are associated with asymmetric emission lines, which have been observed in massive AGNs at very different redshifts \citep{Morganti2007, Nesvadba2011}. The most invoked mechanisms mentioned to explain their generation are radiation-pressure and magnetic driving \citep[see e.g.,][for a review]{Konigl2006}.
These are known to play a very important role in the evolution of galaxies due to the transport of material (e.g., gas, mass, metals) into the interstellar medium of the host galaxy, and the intergalactic or intracluster medium \citep[e.g.,][]{Colbert1996, Churazov2001, Moll2007}, as well as because of the large amounts of energy they feed back into their environment \citep{DiMatteo2005}.

Astrophysical outflows at different scales are closely linked to matter accretion onto compact objects. 
In the extreme case of AGNs, the intense radiation emitted is due to material accretion towards a supermassive black hole \citep{Lynden1969, Begelman1984}. For this accretion to take place, the removal of angular momentum is required \citep[e.g.,][]{Shlosman1990, Maiolino2000, Schmidt2019b}, which can be achieved not only through gravitational torques, but also via outflows or winds  \citep{Bridle1984}. Thus, AGN winds may in turn regulate the accretion rate onto the black hole and remove gas available for star formation from the bulge \citep{Ciotti2001, Hopkins2005, Kormendy2013, Heckman2014, Fiore2017}.\\
In this context, the study of asymmetric line profiles not only provides tools for understanding the evolution of galaxies, but can also give information on nuclear kinematics. By analysing [Fe] high ionization emission lines, \cite{RA2006} studied outflows in 6 Seyfert galaxies (two of which classified as Seyfert 1), and found that the size of the emitting region depends on the ionization potential: the higher the ionization potential, the more compact the emitting region is. In addition, a study of 39 BLS1 and 55 narrow line Seyfert 1 (NLS1) galaxies of the Sloan Digital Sky Survey \citep{Abazajian2005} revealed a correlation between the line blueshift and the ionization potential \citep{Komossa2007}. Moreover, \cite{Zakamska2016} studied a sample of 4 high redshifted quasars, and observed high blueshift in the [OIII] lines. \cite{Ebrero2016} studied the Seyfert 1 galaxy NGC~985 and reported a fast ($\sim$6000 km s$^{-1}$) outflow, which probably originated very close to the broad line region (BLR). There are also some previous studies on individual objects that have reported outflows in Seyfert 1 galaxies such as Mrk 590 \citep{Gupta2015}, NGC 4051 \citep{King2012}, Mrk 509 \citep{Phillips1983,Cappi2009}, NGC 5548 \citep{Wrobel1994,Crenshaw2009} and NGC 4151 \citep{Crenshaw2007}. 
To date, various theoretical models have been proposed to try to explain the mechanisms responsible for the transfer of material to the black hole. In this regard, observational approaches constitute a fundamental tool to constrain models. In particular, detailed studies on nearby Seyfert galaxies are valuable for obtaining a better understanding of AGN feedback and kinematics in Seyfert galaxies at higher redshift, since they are expected to share same properties \citep[e.g.,][]{Humire2018}. Recently, a correlation was reported between the amount of asymmetry in the [OIII] line and the black hole mass in a sample of 28 NLS1 galaxies exhibiting [OIII] blueshifted wings \citep{Schmidt2018}. This result indicates that in galaxies displaying asymmetric profiles, the larger the black hole mass, the more extended the blue wings of the [OIII] line are.\\

Although many works in the literature have addressed the study of outflows in Seyfert 1 galaxies, most of these have included few objects or have considered samples of galaxies in the northern hemisphere. Thus, comparable studies in southern galaxies are still relatively scarce. In the present study, we aimed to investigate the main characteristics of the [OIII] emission profiles in a sample of 45 nearby (z$<$ 0.06) broad line Seyfert 1 galaxies from the southern hemisphere, most of which have been little studied. We analysed the properties of these emission lines and their possible connection with the central engine. In addition, we examined and compared the black hole mass distribution in BLS1 galaxies either exhibiting or lacking asymmetric profiles. Most of our results were then compared with those previously found for narrow line AGNs.
 The paper is organized as follows: In section $\S$2 we present the galaxy sample and the data we worked with; in section $\S$3, we describe the emission line measurement procedure; in section $\S$4 we analyse the nuclear kinematics; and finally, in section $\S$5 and $\S$6, we discuss the main results and present the final remarks.

\section{Galaxy sample and data}
\label{sec:sample}

\subsection{The broad line Seyfert 1 sample}
\label{subsec:sample}

We selected 58 Seyfert 1 galaxies from the sample presented by \cite{Tremou2015}, all of which have been observed by the 6 Degree Field Galaxy survey \citep[6dFGS,][]{Jones2004, Jones2009}. These galaxies were originally drawn from the
Hamburg/ESO QSO Survey \citep[HES,][]{Reimers1996, Wisotzki1996, Wisotzki2000}, and consist of type 1 southern hemisphere AGNs (z $\lesssim$ 3.2) with a brightness limit of B$_{J} \leq$ 17.3 in the $J$ band. Although AGN detection methods might miss luminous objects at low redshift (introducing in turn a distance-dependent incompleteness), the method used by the HES facilitates the inclusion of bright extended targets \citep[see][]{Tremou2015}. The main selection criteria established by \cite{Tremou2015} was to work with galaxies with z$\leq$ 0.06. 

From a total of 58 Seyfert 1 galaxies observed by the 6dFGS, 6 are classified as NLS1, 3 as no broad line emitters, and 4 of these do not have available blue spectra or their spectra are very noisy and do not present [OIII] emission. This left us with a final sample of 45 poorly studied BLS1 galaxies with [OIII] profile information. Table \ref{tabla:sample} lists the main characteristics of the sample, namely, the ID (column 1), the 6dFGS name (column 2), J2000 equatorial coordinates: right ascension and declination (columns 3 and 4, respectively), redshift (column 5) and the distance in Mpc (column 6). 
Considering that in the nearby universe it is more appropriate to use a value of H$_{0}$ locally determined through supernovae \citep[e.g.,][]{Dhawan2018}, then in order to calculate the galaxies distance we adopted H$_{0} =$ 74.03 km s$^{-1}$ Mpc$^{-1}$ \citep{Riess2019}.

\begin{figure*}
   \includegraphics[width=88mm, height=88mm]{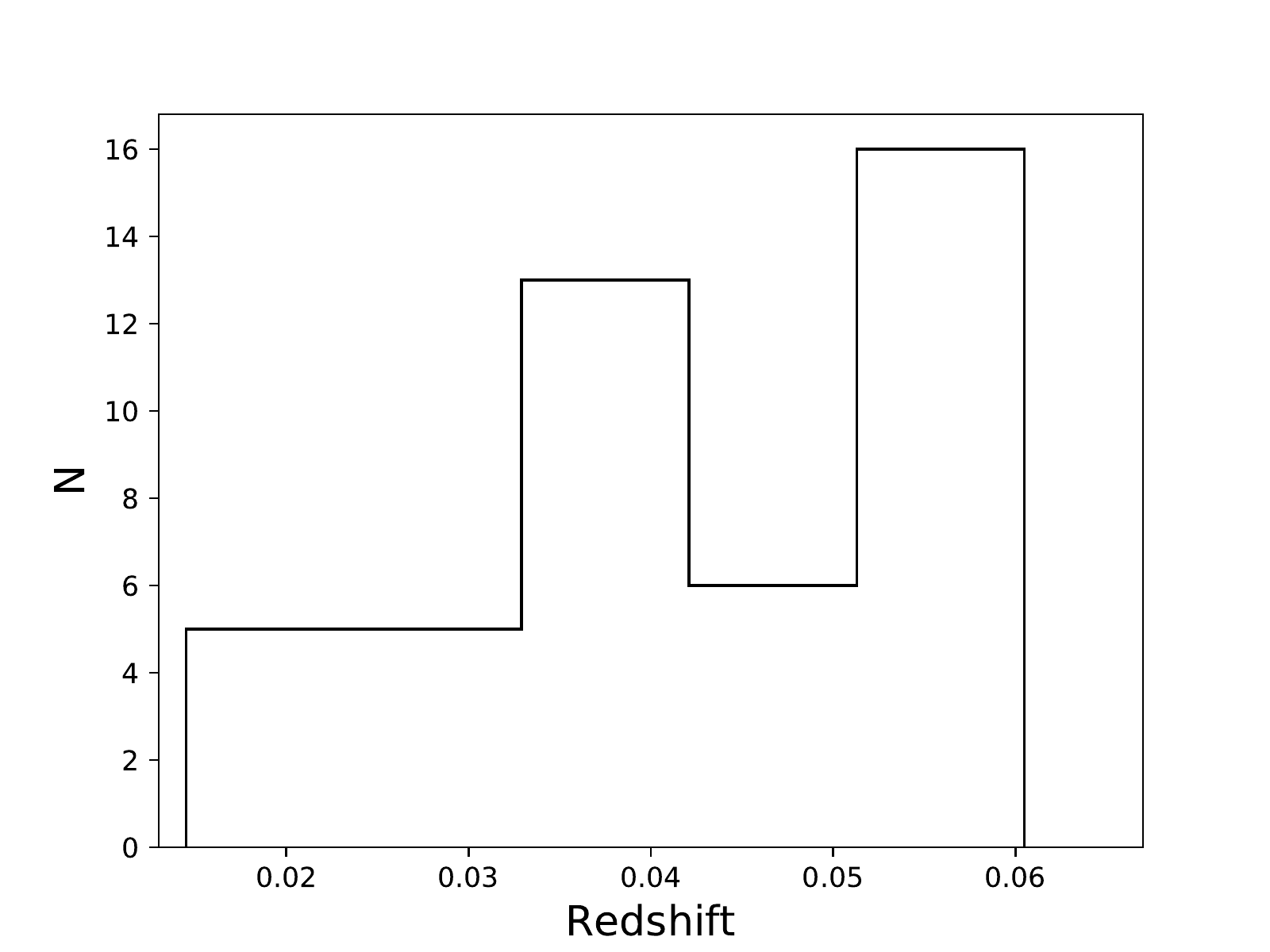}
   \includegraphics[width=88mm, height=88mm]{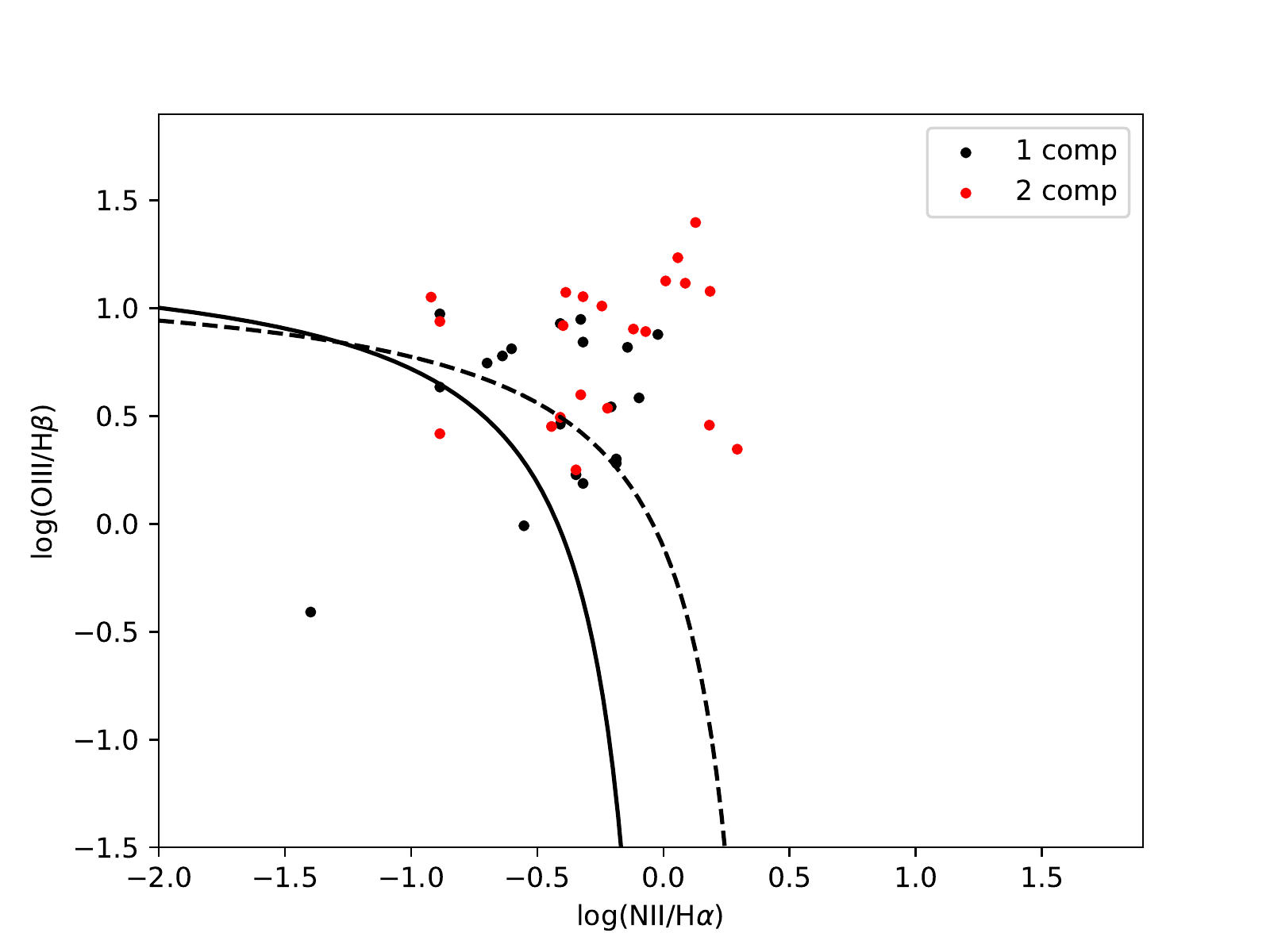}
    \caption{Redshift distribution and BPT diagram for the galaxies of our sample are shown in the left and right panels, respectively. Demarcation lines in the diagnostic diagram are taken from \citet{Kauffmann2003} (solid line) and \citet{Kewley2001} (dashed line). Black and red circles represent galaxies whose [OIII] profiles were fitted using 1 and 2 Gaussian functions, respectively. The data of both plots were taken from \citet{Tremou2015}.}
     \label{fig:sample}
\end{figure*}

In Figure \ref{fig:sample}, we present the redshift distribution and the diagnostic BPT diagram \citep{Baldwin1981} for our sample (left and right panels, respectively). It can be seen that the redshift distribution ranges from 0.016 to 0.059, with a mean z $=$ 0.042. Regarding the BPT diagram, the line ratios are those reported by \cite{Tremou2015} and correspond to the 40 galaxies that have both line ratios measured. The demarcation lines are taken from \cite{Kauffmann2003} (solid line) and \cite{Kewley2001} (dashed line). In addition, from the spectral decomposition that we performed (see Section \ref{sec:measurement}), galaxies whose [OIII] profiles were fitted using 1 and 2 Gaussian functions are shown in black and red circles, respectively.
It can be seen from the BPT diagram that most galaxies of the sample lie in the AGN region, as expected. There are 4 galaxies in the star-forming domain and 6 objects in the intermediate region, delimited by both lines. 
This suggests that the ionization in the NLR could be due to nuclear activity plus a stellar contribution, as indicated by the evidence of hidden starburst activity reported in the BLS1 galaxies \citep{RA2003}. In addition, one object can be observed that is located in the star-forming region, away from the remaining galaxies in the BPT diagram. This object corresponds to 6dF J0230055-085953 (ID$=$12), which has [NII]/H$\alpha$ and [OIII]/H$\beta$ ratios of 0.04 and 0.39, respectively \citep[see Table 4 of][]{Tremou2015}. However, it exhibits a very broad component of H$\beta$ and FeII emission (see Figure \ref{fig:fits1}), which are characteristics of an intense AGN. \\

Certain considerations regarding the BPT diagram should also be taken into account.
As \cite{Tremou2015} did not perform spectral decomposition to the [OIII] profiles, the [OIII]/H$\beta$ ratio corresponds to the [OIII] (total emission) rather than the core component only. Bearing this in mind, the galaxies presenting wings (red dots) would be shifted downward in the plot. In this regard, it can be seen from Fig. \ref{fig:fwhm} (right panel) that most of these galaxies have a wing-to-core line intensity ratio of between 0.2 $-$ 0.6, and therefore, the calculated shift would be of about 0.08~dex $-$ 0.2~dex. Moreover, for a mean distance of 186 Mpc and taking into account the fiber diameter (6.7 arcsec), the obtained spectra correspond to a 3 kpc radius projected distance. Assuming that the stellar contribution of the host galaxy in broad line type 1 AGNs could be around 10$\%$  \citep[e.g.,][]{Vanden-Berk2001,Vanden-Berk2006, Donoso2018, Oio2019}, we estimate an uncertainty of 0.04~dex for all the galaxies in the BPT diagram. In conclusion, most of the objects will still remain in the AGN region, as expected.

 \begin{table*}
 \center
 \caption{The sample of broad line Seyfert 1 galaxies. Data were obtained from Tremou et al. (2015). Distance (D) was calculated using H$_{0} =$ 74.03  km s$^{-1}$ Mpc$^{-1}$.}\label{tabla:sample}
  \begin{tabular}{lcccccc}
    \hline
   ID & 6df Name  & RA (J2000)      & DEC (J2000)   & Redshift   &  D\\
      &           & [$^{\circ}$]    & [$^{\circ}$]  &            &  [Mpc]\\
    \hline
  1    &    g0023554-180251 &    5.98042   &    -18.0472 &    0.0532 &     215.59 \\
  2    &    g0025013-452955 &    6.255     &    -45.4983 &    0.056  &     226.94 \\
  3    &    g0042369-104922 &    10.6533   &    -10.8225 &    0.042  &     170.20 \\
  4    &    g0053544-240437 &    13.4767   &    -24.0767 &    0.056  &     226.94 \\
  5    &    g0111143-161555 &    17.8092   &    -16.265  &    0.052  &     210.73 \\
  6    &    g0113499-145057 &    18.4575   &    -14.8492 &    0.0527 &     213.56 \\
  7    &    g0129067-073830 &    22.2775   &    -7.64167 &    0.056  &     226.94 \\ 
  8    &    g0151419-361116 &    27.9246   &    -36.1878 &    0.0335 &     135.76 \\
  9    &    g0206160-001729 &    31.5662   &    -0.29139 &    0.0424 &     171.82 \\
  10   &    g0214336-004600 &    33.64     &    -0.76667 &    0.0264 &     106.98 \\
  11   &    g0226257-282059 &    36.6071   &    -28.3497 &    0.0605 &     245.17 \\
  12   &    g0230055-085953 &    37.5225   &    -8.99806 &    0.0164 &     66.46 \\
  13   &    g0234378-084716 &    38.6571   &    -8.78778 &    0.043  &     174.25 \\
  14   &    g0259305-242254 &    44.8771   &    -24.3817 &    0.035  &     141.83 \\
  15   &    g0333078-135433 &    53.2821   &    -13.9094 &    0.04   &     162.10 \\
  16   &    g0334245-151340 &    53.6021   &    -15.2281 &    0.035  &     141.83 \\
  17   &    g0345032-264820 &    56.2633   &    -26.8053 &    0.058  &     235.04 \\
  18   &    g0351417-402759 &    57.9233   &    -40.4664 &    0.0582 &     235.85 \\
  19   &    g0400407-370506 &    60.1696   &    -37.0853 &    0.051  &     206.67 \\
  20   &    g0401462-383320 &    60.4425   &    -38.5558 &    0.059  &     239.09 \\
  21   &    g0405017-371115 &    61.2567   &    -37.1878 &    0.0552 &     223.69 \\
  22   &    g0414527-075540 &    63.7192   &    -7.92806 &    0.0379 &     153.59 \\
  23   &    g0431371-024124 &    67.9042   &    -2.69028 &    0.041  &     166.15 \\
  24   &    g0436223-102234 &    69.0925   &    -10.3758 &    0.0355 &     143.86 \\
  25   &    g0856178-013807 &    134.07401 &    -1.63528 &    0.0597 &     241.93 \\
  26   &    g0952191-013644 &    148.07899 &    -1.61222 &    0.0197 &     79.83 \\
  27   &    g1014207-041841 &    153.586   &    -4.31139 &    0.0586 &     237.47 \\
  28   &    g1031573-184633 &    157.989   &    -18.7761 &    0.0404 &     163.72 \\
  29   &    g1110480-283004 &    167.7     &    -28.5008 &    0.024  &     97.26 \\
  30   &    g1138510-232135 &    174.713   &    -23.36   &    0.027  &     109.42 \\
  31   &    g1251324-141316 &    192.88499 &    -14.2214 &    0.0145 &     58.76 \\
  32   &    g1313058-110742 &    198.274   &    -11.1283 &    0.034  &     137.78 \\
  33   &    g1321582-310426 &    200.492   &    -31.0736 &    0.0448 &     181.55 \\
  34   &    g1331138-252410 &    202.808   &    -25.4028 &    0.026  &     105.36 \\
  35   &    g1332391-102853 &    203.16299 &    -10.4814 &    0.0225 &     91.18 \\
  36   &    g1349193-301834 &    207.33    &    -30.3097 &    0.0161 &     65.24 \\
  37   &    g1356367-193145 &    209.153   &    -19.5289 &    0.0349 &     141.43 \\
  38   &    g2130499-020814 &    322.70801 &    -2.1375  &    0.0528 &     213.97 \\
  39   &    g2132022-334254 &    323.009   &    -33.715  &    0.0293 &     118.74 \\
  40   &    g2207450-323502 &    331.93799 &    -32.5839 &    0.0594 &     240.71 \\
  41   &    g2214420-384823 &    333.67499 &    -38.8067 &    0.0398 &     161.29 \\
  42   &    g2234409-370644 &    338.67099 &    -37.1122 &    0.043  &     174.25 \\
  43   &    g2253587-330014 &    343.49399 &    -33.0036 &    0.056  &     226.94 \\
  44   &    g2326376-610602 &    351.65701 &    -61.1003 &    0.0413 &     167.36 \\
  45   &    g2340321-263319 &    355.133   &    -26.5553 &    0.0496 &     201.00 \\
\hline
  \end{tabular}
 \end{table*}

\subsection{Six-Degree Field Galaxy Survey data}
\label{subsec:data}

To perform the spectral analysis of our sample of BLS1 galaxies, we worked with the processed spectroscopic data available in the 6dFGS data release 3 (DR3) archive. This survey measures redshifts and peculiar velocities of a sample of galaxies selected from the Two Micron All Sky Survey (2MASS; \citealt{Skrutskie2006}). Spectral observations were carried out with the Six-Degree Field multiobject fiber spectrograph facility \citep{Parker1998}, located at the 1.2 m Anglo-Australian Observatory's UK Schmidt Telescope (UKST). See \cite{Jones2004} for a full description of the survey.\\
\indent The galaxies were observed with a $6\farcs7$ diameter fiber, using two separate V and R gratings, covering a spectral range of $\sim$4000~\r{A} to $\sim$7500~\r{A}. The spectral resolution is R $\sim$ 1000, equivalent to an approximate value of 300 km~s$^{-1}$ in FWHM. A detailed description of the spectral processing can be found in \cite{Jones2004, Jones2009}.


\section{Emission line measurement}
\label{sec:measurement}

We analysed the [OIII] profiles in order to study the wing components and possible blueshifts in the BLS1 galaxies. Emission lines in type-1 AGNs can be represented by a single or a combination of Gaussian functions \citep[e.g.,][]{RA2006, Schmidt2016}, and provide important information on the processes taking place in the AGN. To perform a detailed study of the [OIII] emission, we used the LINER routine developed by \cite{Pogge1993}, which is a $\chi^2$ minimisation algorithm able to fit several Gaussian functions to a line profile. In this section, we describe the spectral analysis procedure adopted for the 45 galaxies of our sample.

We first carried out a careful fitting of the continuum emission in the vicinity of [OIII]$\lambda$5007 and [OIII]$\lambda$4959 lines, and then both profiles were fitted, as a first approximation, with a single Gaussian each. Since both the [OIII]$\lambda$5007 and [OIII]$\lambda$4959 lines are emitted from the same region of the AGN, they have the same FWHM \citep[e.g.,][]{Villar-Martin2011,Schmidt2018}. In addition, there is a flux ratio of 1:3 between [OIII]$\lambda$4959 and [OIII]$\lambda$5007 according to the theoretical value \citep{Osterbrock1989}. Subsequently, in order to evaluate the fitting quality, we conducted a thorough inspection of residuals. For this purpose, the adopted criteria was to obtain fitting residuals similar to the noise level in the proximities of the line \citep{Schmidt2016}. Then, we applied a double-Gaussian fitting with the same constrains where necessary \citep[e.g.,][]{Cracco2016}. In fact, a double-Gaussian decomposition (core and wings) was necessary in 22 cases, while the 23 remaining galaxies were fitted with a single Gaussian function. Uncertainties in the measurements were estimated by fitting the same emission profile 10 times in a few galaxies with different S/N. We assumed that any errors are given by the dispersion in the distribution of these measurements, considering them at 1 $\sigma$. Relative errors in the FWHM and flux of the [OIII]$_{core}$ are of the order of 5$\%$ and 8$\%$, respectively. Regarding wings ([OIII]$_{wing}$), the relative errors are in the range of $\sim$ 5\%$-$10$\%$ for the flux and $\sim$5\%$-$15$\%$ for the FWHM. Since double Gaussian decomposition might be degenerated in low S/N spectra, extra caution must be taken. With regard to this, among the 22 galaxies exhibiting wings, only one revealed a noisy spectrum (S/N$<$10). Therefore, in this particular case (6dF J1321582-310426), we ran a test to evaluate possible spectral fitting degeneration. We analysed fluctuations in the spectral fitting, considering the initial variations in the FWHM of both the core and wing components, and also the different positions of the wing peak relative to the core component. This procedure was repeated 30 times, with variations within 1 $\sigma$ in the core and wing FWHM being 7\% and 13\%, respectively, while a variation of 20\% was found considering the relative position between the core and wing centroids. Thus, any uncertainties derived from the spectral fitting in this low S/N spectrum are of the same order as errors derived from higher S/N spectra. 

 It should be noted that 6 galaxies: the objects 6dF J0230055-085953, 6dF J0436223-102234, 6dF J1014207-041841, 6dF J1110480-283004, 6dF J1321582-310426, and 6dF J1332391-102853 (see Fig. \ref{fig:fits1}), presented a mild FeII emission, which is a common feature in type 1 AGNs \citep[e.g.,][]{Dong2011}. We inspected all spectra presenting this feature and found that the object 6dF J0230055-085953 (ID=12) presented the most conspicuous iron emission compared with the intensity of the [OIII] emission. Thus, in order to quantify the impact of FeII emission on the [OIII] line profile, we considered this object to be an extreme case. For this purpose, we compared the [OIII] line width of this galaxy with and without subtracting the FeII emission. To do this, we subtracted the iron emission using online software \footnote{\url{http://servo.aob.rs/FeII_AGN/}} developed by \citet{Kovacevic2010} and \citet{Shapovalova2012}, and found that the variation of the FWHM in this extreme case is $\sim$17 \%. This procedure has been widely used in many other studies \citep[e.g.,][]{Foschini2015, Berton2016, Cracco2016, Schmidt2018, Oio2019}.

Taking into account that [OIII]$\lambda$4959 is linked to [OIII]$\lambda$5007, we worked with the latter, which has a higher S/N. Henceforth, all the presented values correspond to the core and the wing components of [OIII]$\lambda$5007.

\begin{figure*}
    \includegraphics[width=58mm, height=60mm]{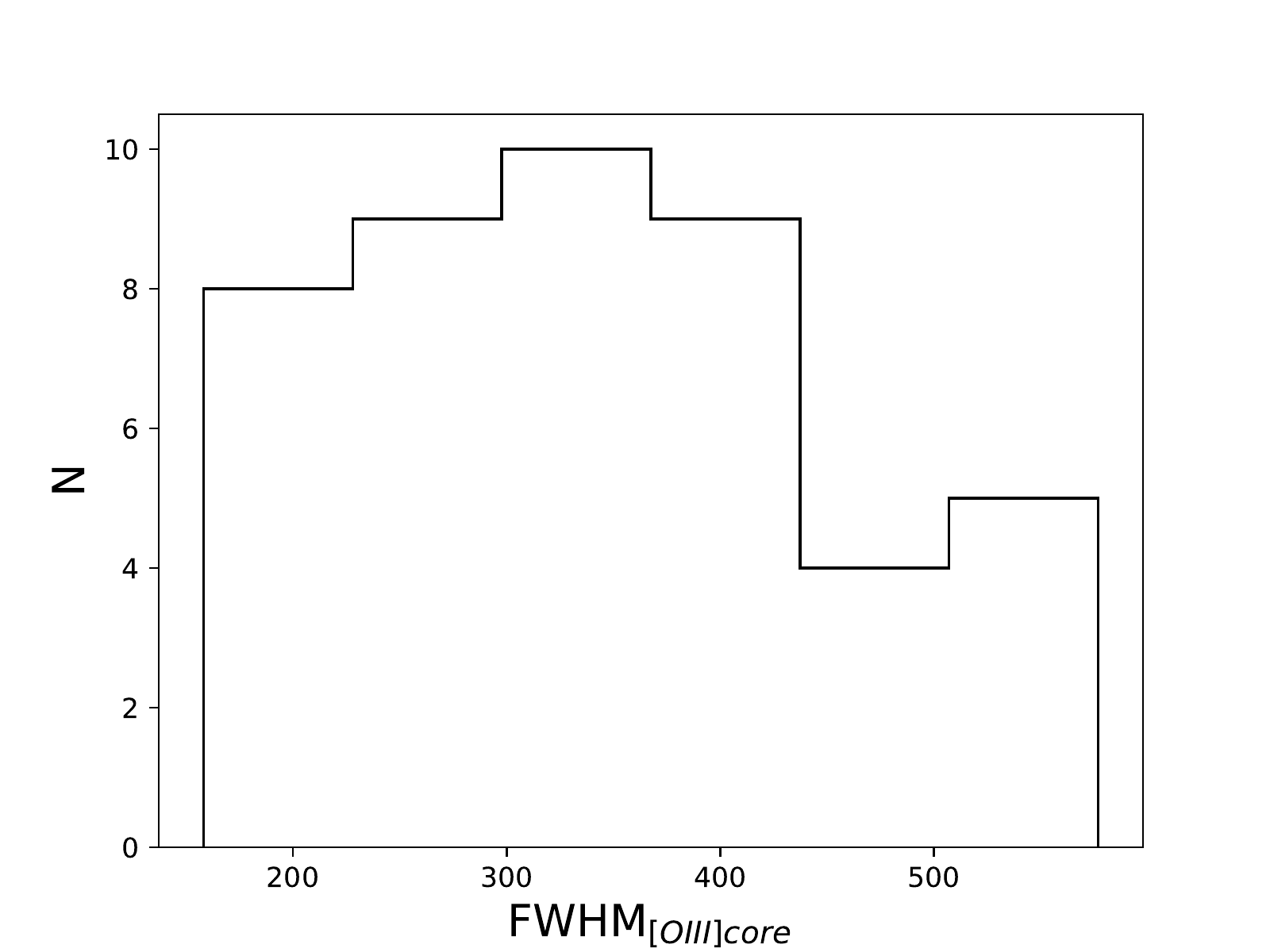}
    \includegraphics[width=58mm, height=60mm]{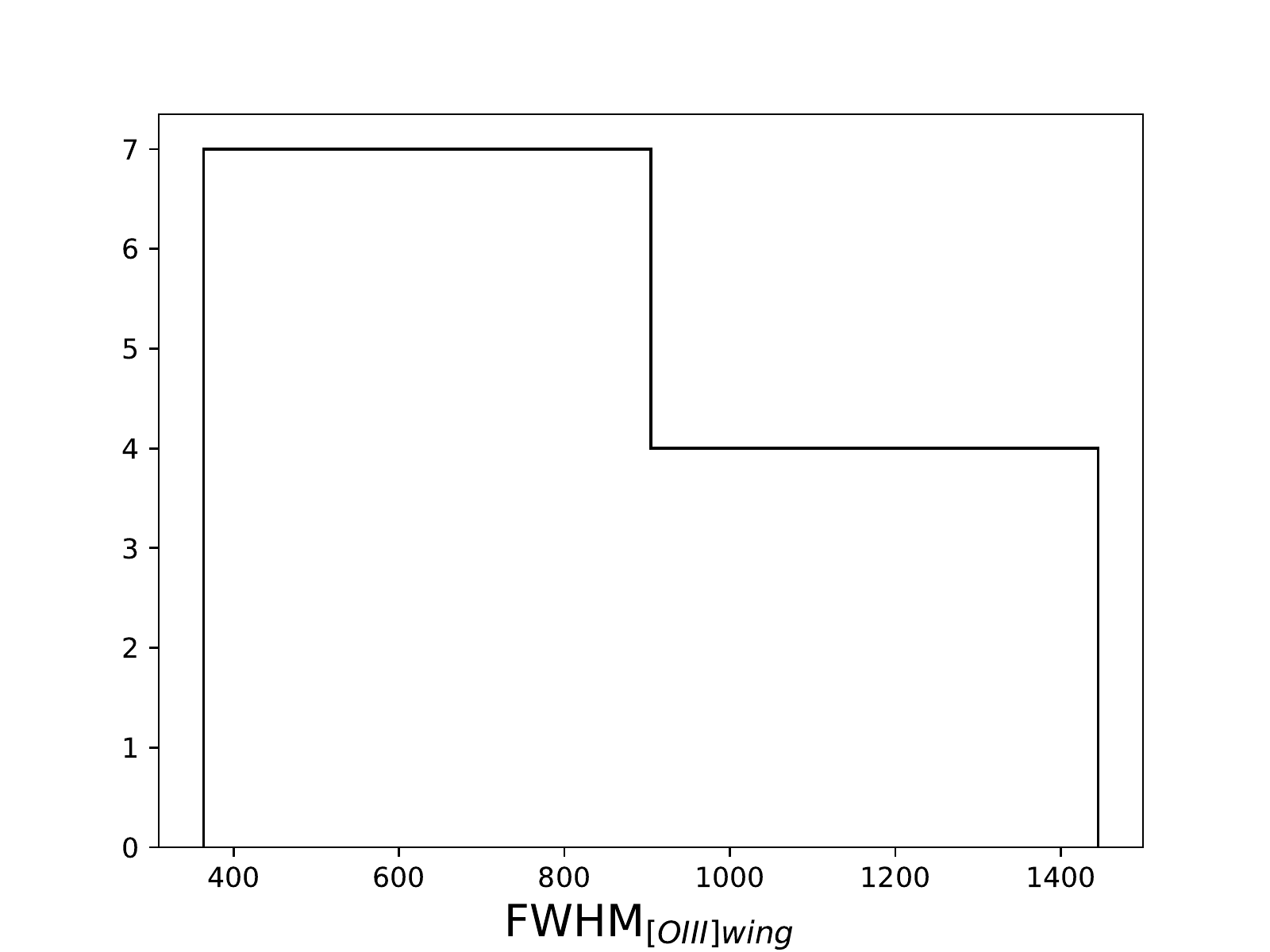}
    \includegraphics[width=58mm, height=60mm]{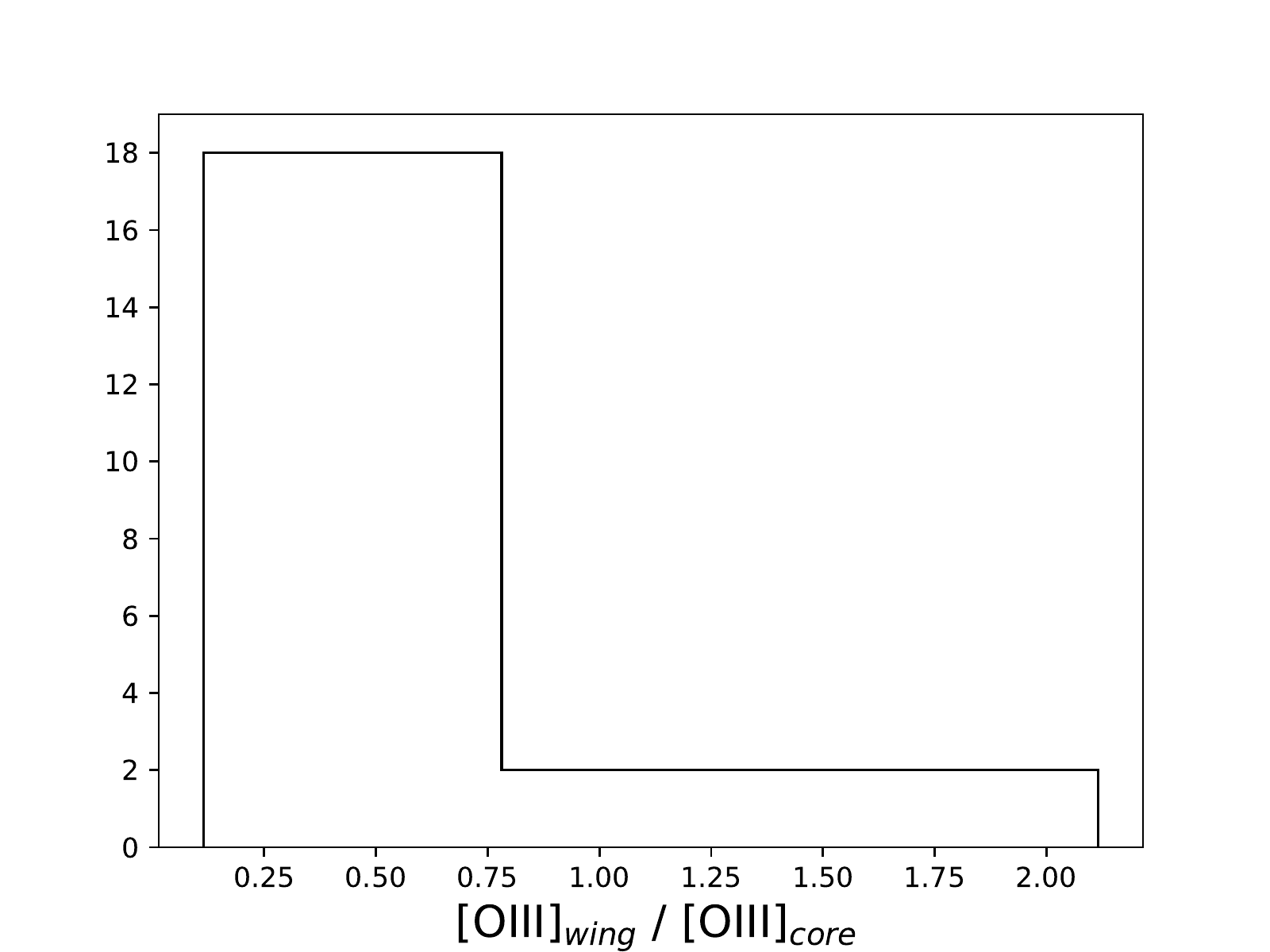}
    \caption{FWHM distributions of the emission lines in km s$^{-1}$. From left
to right : FWHM of [OIII]core and [OIII]wing and the ratio between I$_{[OIII]wing}$ and I$_{[OIII]core}$. The left panel considers the whole sample of 45 galaxies, while the 22 galaxies presenting [OIII] wings are shown in the central and right panels.}
    \label{fig:fwhm}
\end{figure*}

In Fig. \ref{fig:fwhm}, we show the FWHM distribution of the [OIII]$_{core}$ and [OIII]$_{wing}$, and the line intensity ratio between both components, i.e. I([OIII]$_{wing}$)/I([OIII]$_{core}$). The FWHM of [OIII]$_{core}$ was in the range of 156 $-$ 568 km s$^{-1}$, in agreement with previous results considering lines emitted from the NLR \citep[e.g.,][]{Osterbrock1989, RA2000, Boroson2005, Schmidt2016}. The mean value of the [OIII]$_{core}$ FWHM distribution was found to be 288 km s$^{-1}$, with a standard deviation of 129 km s$^{-1}$. Considering the [OIII]$_{wing}$, the FWHM was in the range of 358 $-$ 1423 km s$^{-1}$, which is in agreement with previous studies of blue wings of the [OIII] line in type-1 AGNs \citep{Bian2005,Zakamska2016}. The [OIII]$_{wing}$ FWHM distribution had a mean value of 825 km s$^{-1}$, with a standard deviation of 316 km s$^{-1}$. All the FWHM were corrected by the instrumental width as FWHM$^{2}$ $=$ FWHM$_{obs}^{2}$ $-$FWHM$_{inst}^{2}$, where FWHM$_{obs}$ is the measured FWHM, and FWHM$_{inst}$ is the instrumental broadening, which is $\sim$300 km s$^{-1}$. Regarding the line intensity ratio between both components, we found I([OIII]$_{wing}$)/I([OIII]$_{core}$) to be in the range 0.11 $-$ 2.12, with a mean value of 0.37 and a standard deviation of 0.52. In the right panel in Figure \ref{fig:fwhm}, we can observe that, even though the galaxies showed a clear presence of wings, the core component of the emission was, in general terms, more conspicuous than the wing.

It is worth noting that we are probably missing a fraction of galaxies with [OIII] wings due to the spectral resolution we worked with (R $\sim$ 1000). If this is the case, some galaxies showing symmetric [OIII] lines fitted with only one component might actually exhibit blue wings that could be detected through a higher spectral resolution. In addition, although the 6dFGS lacks absolute flux calibration \citep[e.g.,][]{Chen2018}, this did not affect the results of our work since we analysed the line intensity ratios, which are independent of the flux calibration.


\section{Nuclear kinematics}
\label{sec:nuclearkinematics}

Super massive black holes are well known to play an essential role in the study of active galaxies, as they govern the nuclear kinematics and constitute a fundamental parameter in the understanding of the mechanisms involved in the inner regions of AGNs. In this section we analyse the [OIII] emission profile features for the BLS1 galaxy sample we are studying, and their connection with the black hole masses.

\subsection{Black hole masses}
\label{subsec:BHM}

It is well known that the velocity dispersion ($\sigma_{\star}$) of stars in the bulge of galaxies correlates with the mass (M$_{BH}$) of the black hole they host \citep{Ferrarese2000,Gebhardt2000,Tremaine2002}. This empirical correlation, usually known as the M$_{BH}-\sigma_{\star}$ relation, has been widely invoked to estimate black hole masses. According to \cite{Tremaine2002}, it can by parameterized as:

\begin{equation}
\label{eq:tremaine}
 log\left(\frac{M_{BH}}{M_{\odot}}\right) = (8.13 \pm 0.06) + (4.02 \pm 0.32)log\left(\frac{\sigma_{\star}}{200~ {\rm km~s}^{-1}}\right)
 \end{equation}

As measuring the velocity dispersion of stars in the bulge is a difficult task, the emission line widths can be used as a surrogate \citep[e.g.,][]{Mathur2001,Bian2004,Oio2019}, considering that in a Gaussian function the FWHM is 2.3548 $\sigma$.

It has been found that BLS1 galaxies also follow the M$_{BH}- \sigma_{\star}$ relation \citep[e.g.,][]{Grupe2004,Komossa2008}. Likewise, \cite{Nelson2004} determined the stellar velocity dispersion in the bulge and calculated the black hole masses through the reverberation mapping technique, revealing that these objects follow the M$_{BH}- \sigma_{\star}$ relation as normal galaxies. In addition, \cite{Komossa2007} studied a sample of 39 BLS1 galaxies, and found that the FWHM of the [SII] emission line and the core component of the [OIII] line are the best available surrogates for stellar velocity dispersion, as well as verifying the M$_{BH}- \sigma_{\star}$ relation for Seyfert 1 galaxies.

In order to study the possible relation between [OIII] profile wings and the central engine, we first need to estimate the black hole masses for the galaxies of our sample. 
Despite broad H$\beta$ being the most commonly used for black hole mass estimate in AGNs \citep[e.g.,][]{greene, Oio2019, Torres-Papaqui2020}, the 6dF survey lacks absolute flux calibration \citep[e.g.,][]{Chen2018}. In addition, some galaxies present broad components of H$\beta$ buried in the data noise, with certain objects exhibiting very irregular profiles (see Fig. \ref{fig:fits1}). Consequently, any fitting to these irregular and noise-contaminated profiles would be unreliable. Therefore, in this work, we relied on the M$_{BH}- \sigma_{\star}$ relation using the FWHM of [OIII]$_{core}$ as a surrogate of the stellar velocity dispersion. In this way, we computed the FWHM of the [OIII]$\lambda$5007 core component, and estimated the black hole mass from Equation \ref{eq:tremaine}. The uncertainties in these estimates are $\sim$0.1 dex and are given by the errors in the measurements (see Sect. \ref{sec:measurement}).

\begin{figure}
    \centering
    \includegraphics[scale=0.52]{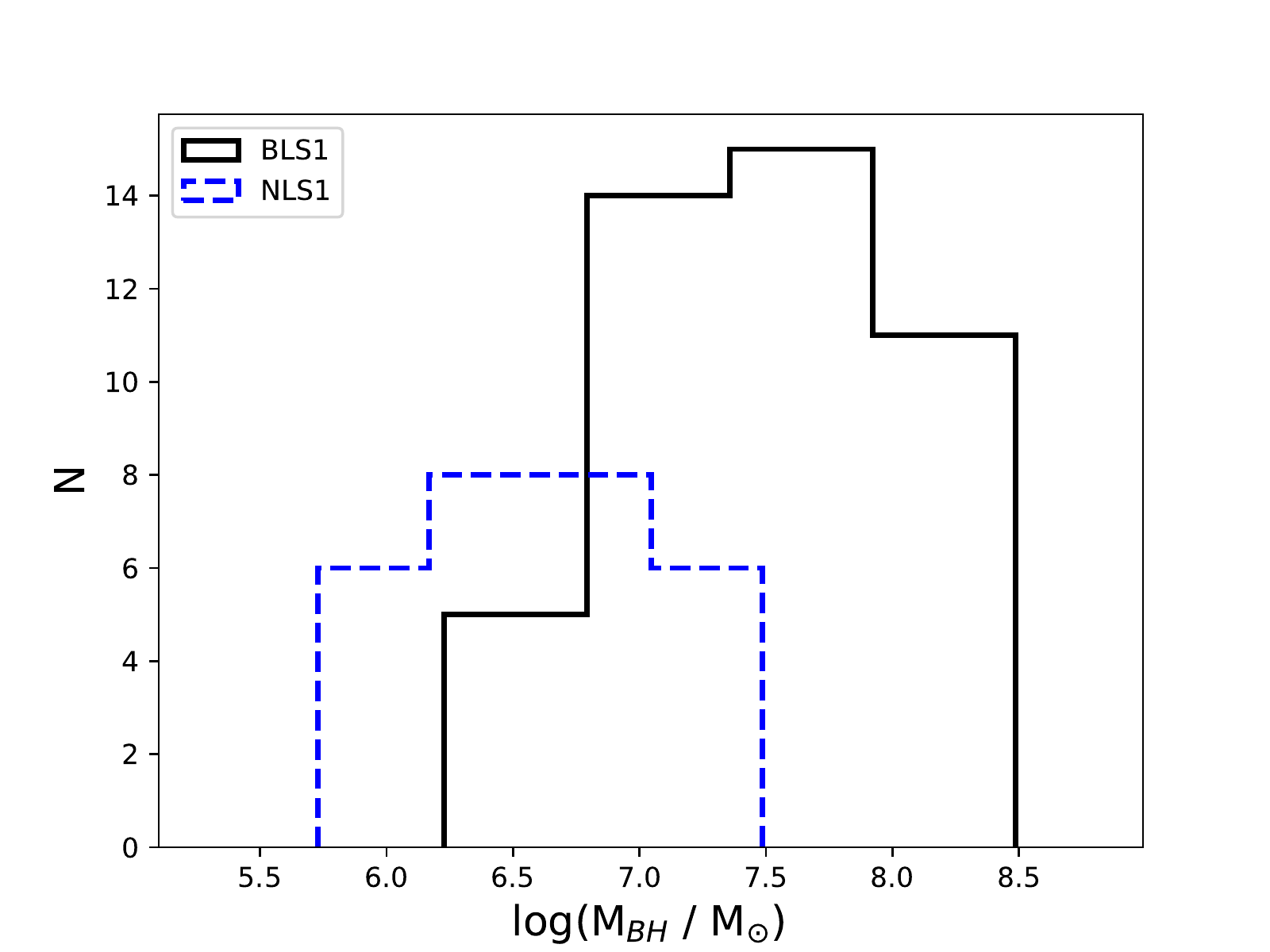}
   \caption{Black hole mass distribution for our BLS1 galaxy sample (black solid line) and the NLS1 sample studied by Schmidt2018 (blue dashed line).}
              \label{fig:bhmass}%
    \end{figure}

In Figure \ref{fig:bhmass}, we show the black hole mass distribution for the BLS1 galaxies of our sample by the black solid line. In addition and for comparison, the blue dashed line indicates the black hole mass distribution for the NLS1 sample studied by \cite{Schmidt2018}. For our sample of galaxies, the masses are in the range of log(M$_{BH}$/M$_{\odot}$)$=$ 6.2$-$8.5, with a median value of log(M$_{BH}$/M$_{\odot}$)$=$ 7.6 and a standard deviation of 0.6. which are typical values for BLS1s \citep[e.g.,][]{Grupe2004,Komossa2007}. As expected, the BLS1 galaxies revealed higher black hole masses than the NLS1s, in agreement with previous studies \citep[e.g.,][]{Komossa2007}. 


Nevertheless, even though the [OIII] core is known to be a good proxy for $\sigma_{\star}$, this might not be true for galaxies with strong linear radio emission \citep[e.g.,][]{Nelson1996}. In this context, the presence of radio jets could influence the nuclear region, which are able to generate kinematic disturbances of the [OIII] line emitting gas. This indicates that galaxies with radio jets can produce broader [OIII] emission lines than if they were gravitationally generated \citep[e.g.,][]{Whittle1992}. To examine this possibility, we carried out an inspection of the NASA/IPAC Extragalactic Database (NED) \footnote{The NASA/IPAC Extragalactic Database (NED) is operated by the Jet Propulsion Laboratory, California Institute of Technology, under contract with the National Aeronautics and Space Administration.} and the Centre de Données astronomiques de Strasbourg \citep{Wenger2000}. We found that only 19 galaxies from the 45 members of the entire sample have been observed by radio frequency surveys, of which 18 were observed with the NVSS (NRAO VLA Sky Survey, \citealt{Condon1998}) at 1.4 GHz, and one with the SUMSS (Sydney University Molonglo Sky Survey) at 843 MHz (object ID 44), with no evidence of radio observations being found for the remaining 26 objects. All galaxies observed with NVSS presented luminosities L$_{\rm 1.4~GHz}\sim$ 10$^{22}$~W~Hz$^{-1}$, indicating they might be radio-quiet according to the criterion adopted by \cite{Best2005} (L$_{\rm 1.4~GHz}<$ 10$^{23}$~W~Hz$^{-1}$). On the other hand, considering that most radio-loud AGNs are found in galaxies with black hole masses $\gtrsim$ 10$^{9}$M$_{\odot}$ \cite[e.g.,][]{Laor2000,Dunlop2003,McLure2004}, the remaining galaxies in our sample would also probably be radio-quiet. 
Nevertheless, a recent study of \citet{Foschini2020} states that there is no mass threshold in jetted NLS1 galaxies with respect to the generation of relativistic jets. In this scenario, we also inspected the radio continuum emission of each galaxy to search for evidence of the presence of jets, and found only one galaxy that had a clear jet-like structure at the resolution achieved by the NVSS (45 arcsec), namely 6dF J349193-301834 (ID 36). According to our study, the black hole mass in this galaxy is log(M$_{BH}$/M$_{\odot}$)$=$ 8.4, which is in agreement with different estimations from the literature which also reported a value of log(M$_{BH}$/M$_{\odot}$)$\sim$8. For example, \cite{DeMarco2013} determined the black hole mass through the stellar velocity dispersion and found log(M$_{BH}$/M$_{\odot}$)$=$ 8.34. In addition, \cite{Markowitz2009} calculated the black hole mass of this galaxy through X-rays observations and reported log(M$_{BH}$/M$_{\odot}$)$=$ 8.11, and \cite{Khorunzhev2012} estimated log(M$_{BH}$/M$_{\odot}$)$=$ 7.98 using the correlation with the bulge luminosity. In this particular case, the black hole mass we estimated represents an upper limit. We indicate this specific galaxy with a red plus sign in Figures \ref{fig:deltav_mass} and \ref{fig:blue_emission_mass}.

    

As previously mentioned in Section \ref{sec:sample}, few galaxies lie in the limit between the star-forming and AGN domain (see the BPT diagram in the right panel of Fig. \ref{fig:sample}). Thus, these galaxies might be affected by stellar contribution. If this is indeed the case, contamination by HII region emission would produce lines to be narrower than if they were AGN dominated, thereby yielding an underestimation in the BH mass determination. Assuming that the stellar contribution of the host galaxy in broad line type 1 AGNs is about 10\% \citep[e.g.,][]{Vanden-Berk2001, Vanden-Berk2006, Donoso2018, Oio2019}, we estimate there is an approximately 5\% decrease in the FWHM of [OIII]$_{core}$, resulting in turn in an underestimation of $\sim$0.1 dex in black hole masses. In addition, as mentioned in Sect. \ref{sec:measurement}, 6 galaxies presented a mild FeII emission, which in the extreme case of 6dF J0230055-085953 (ID=12) produces a variation of 17 \% in the FWHM of the [OIII] line. This effect generates an uncertainty of 0.3 dex in the black hole mass.

\subsection{Emission profile of [OIII]$\lambda$5007 }
\label{sub:wing}

According to the spectral fitting of [OIII]$\lambda\lambda$4959,5007 performed in Section \ref{sec:measurement}, double-Gaussian decomposition (core and wings components) was necessary for 22 out of the 45 BLS1 galaxies. For these objects, we computed the difference between core [OIII]$_{core}$ and wing [OIII]$_{wing}$ line centroids, defined as $\Delta$v, to quantify the wing shift relative to the core component. In this way, we determined that 18 galaxies show blue wings ($\Delta$v $<$ 0), 2 objects present practically symmetrical wings relative to the core component ($\Delta$v$\sim$ 0), and 2 of the targets exhibit red wings ($\Delta$v $>$ 0).

The distribution of $\Delta$v in the sample of BLS1 was characterized by a mean value of $-$219 km s$^{-1}$ with a standard deviation of 192 km s$^{-1}$. Moreover, the $\Delta$v values ranged from $-$670 to 124 km s$^{-1}$, in agreement with previous results \citep[e.g.,][]{VeronCetty2001, Cracco2016, Cooke2020}, with typical uncertainties in the determination of this parameter being around 15$-$20$\%$. 

\begin{figure}
    \centering
    \includegraphics[scale=0.52]{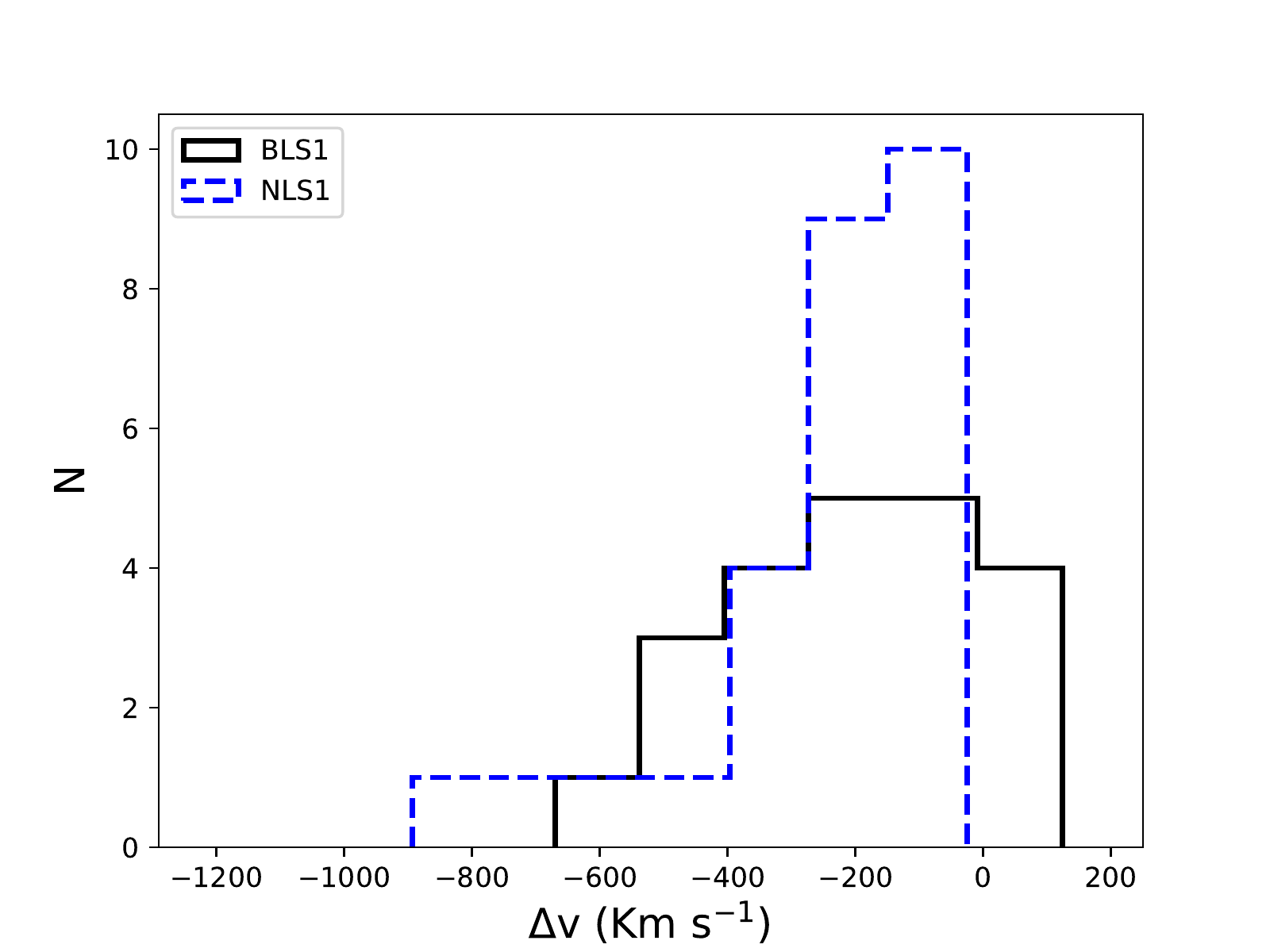}
    \caption{$\Delta$v distribution for our BLS1 galaxy sample (black solid line), and the NLS1 sample studied by \citet{Schmidt2018} (blue dashed line).}
              \label{fig:hist_deltav}%
    \end{figure}

In Figure \ref{fig:hist_deltav}, we compare the $\Delta$v distribution obtained (solid line) with that reported by \cite{Schmidt2018} for a sample of NLS1 galaxies (dashed line). It can be observed that even though all of the NLS1 galaxies have blue wings ($\Delta$v $<$0), in general terms, both samples revealed similar distributions in $\Delta$v.


\begin{figure}
    \centering
   \includegraphics[scale=0.52]{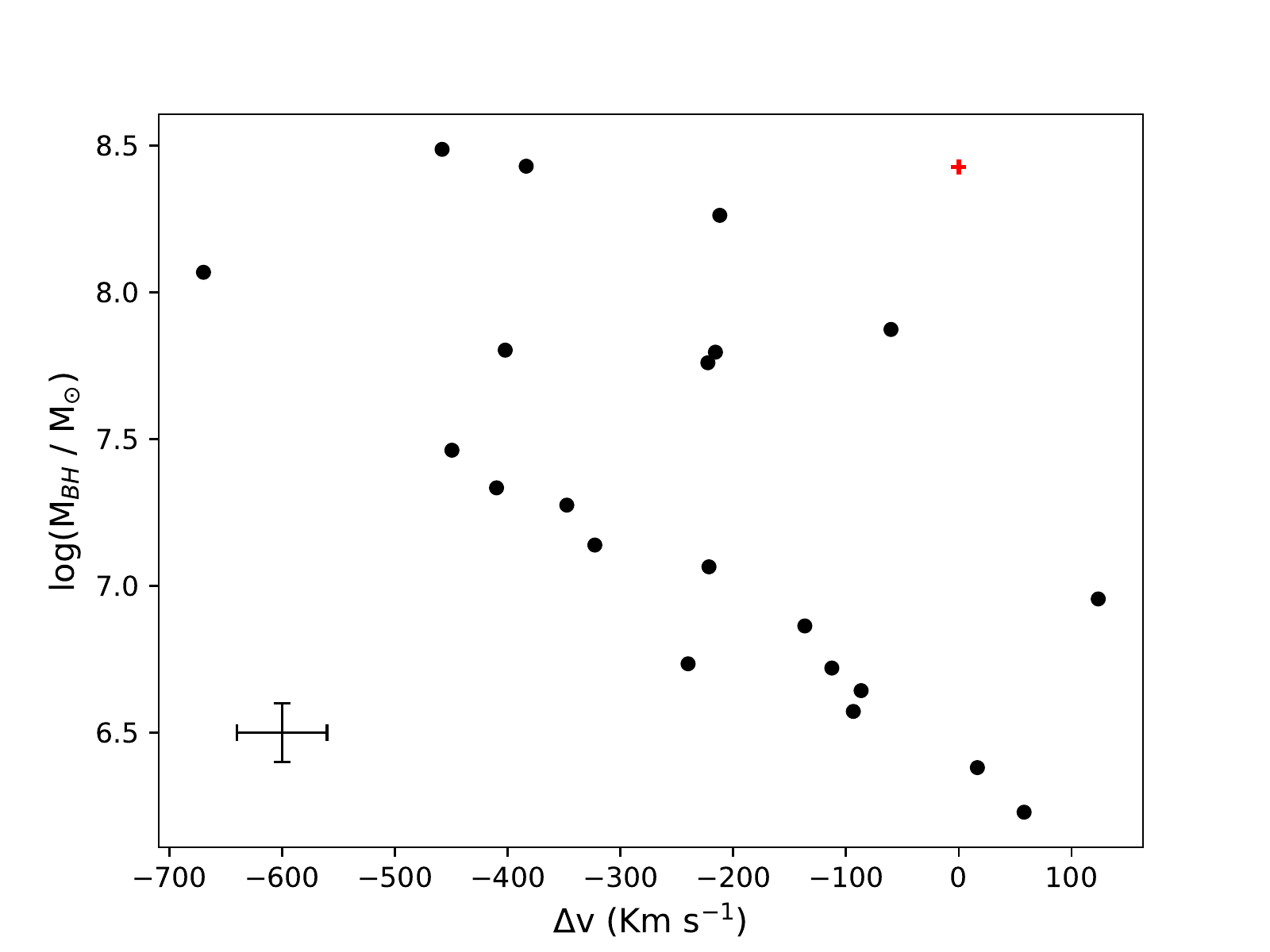}
   \caption{Relation between the black hole mass and $\Delta$v for the 22 galaxies that exhibit [OIII] wings. The red plus sign indicates the galaxy 6dF J349193-301834 (ID 36), which presents a radio jet. A typical error bar is shown in the bottom-left corner.}
              \label{fig:deltav_mass}%
    \end{figure}

 For those galaxies exhibiting wings, we also analysed the relation between $\Delta$v and the FWHM of the [OIII] core component. We found that these two parameters correlated with a Pearson coefficient of r$_{p}=-$0.45 and a p-value of 3$\times$10$^{-2}$. Thus, galaxies with larger FWHM tend to show more blueshifted wings, in agreement with previous studies \citep[e.g.,][]{Bian2005,Ludwig2012, Berton2016}. Based on this result, we also expected a correlation between $\Delta$v and the black hole mass. In Figure \ref{fig:deltav_mass}, we show the behaviour of these two variables, where a Pearson coefficient was obtained of r$_{p}=-$0.51 and a p-value of 1$\times$10$^{-2}$. Although galaxies with $\Delta$v $\sim$ 0 presented a wide range of black hole mass, there is a zone lacking galaxies with $|\Delta$V| $<$ 300 km s$^{-1}$ and log(M$_{BH}$/M$_{\odot}$) $>$ 7.1. Regarding this, more blueshifted wings appeared to be associated with more massive black holes.
 

\subsubsection{The wing blue-end}
\label{sub:total}

\begin{figure}
    \centering
    \includegraphics[scale=0.52]{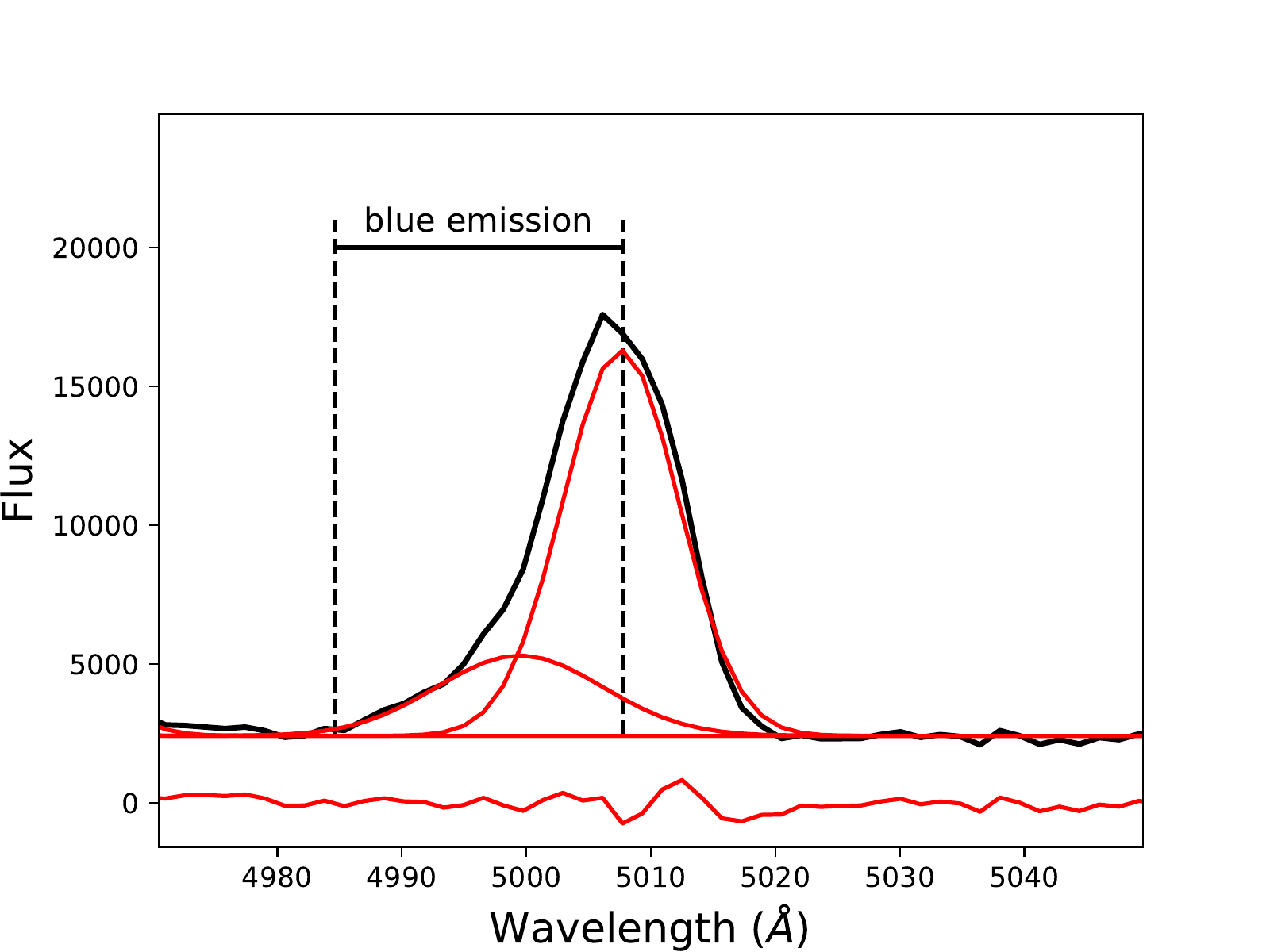}
   \caption{The $\emph{blue emission}$ parameter quantifies the radial velocity difference between the wing blue end and the centroid of the core component.}
                 \label{fig:plot_blue-emission}
    \end{figure}

One of the goals of this work was to study a possible connection between the black hole mass and the [OIII] wings. For this purpose, we defined the parameter \emph{blue emission}, which quantifies the entire blue-side extension of the wing component in all galaxies with [OIII] double-Gaussian decomposition, i.e., those lines showing not only blue but also symmetric and red wings:

\begin{equation}
    {blue \ emission} = \Delta v - \frac{1}{2} {\rm Width}_{{base}[OIII]_{wing}}
    \end{equation}

\noindent where $\Delta$v is the velocity difference between the [OIII]$_{wing}$ and [OIII]$_{core}$ line centroids, and Width$_{{base}[OIII]_{wing}}$ is the total width of the wing at its base.
Thus, defined in this way, $\emph{blue emission}$ represents the wing blue-end and its radial velocity relative to the centroid of the core component. Fig. \ref{fig:plot_blue-emission} shows a representation of this parameter in an illustrative way.

Taking into account that the total width of the wing at its base can be difficult to measure, we assumed it to be twice the FWHM of the wing, i.e., Width$_{base}$[OIII]$_{wing}$ $\sim$ 2 FWHM[OIII]$_{wing}$. Thus, the $\emph{blue emission}$ can be written as:

\begin{equation}
\label{eq:totalasym}
{blue \ emission} =  \Delta v - {\rm FWHM}[OIII]_{wing}
\end{equation}

\noindent According to the estimation of uncertainties given in Section \ref{sec:measurement}, it can be noticed that \emph{blue emission} typical errors ($\sim$10 $-$ 15$\%$) are smaller than those associated with $\Delta$V and the FWHM of the wing. Moreover, in some cases, smaller FWHMs of the wing component were found to correspond to higher $\left | \Delta V \right |$ (or vice versa). Consequently, blue emission constitutes a more stable parameter, which can be more accurately determined than $\Delta$v and the FWHM of the wings when considered separately.

\begin{figure}
    \centering
    \includegraphics[scale=0.52]{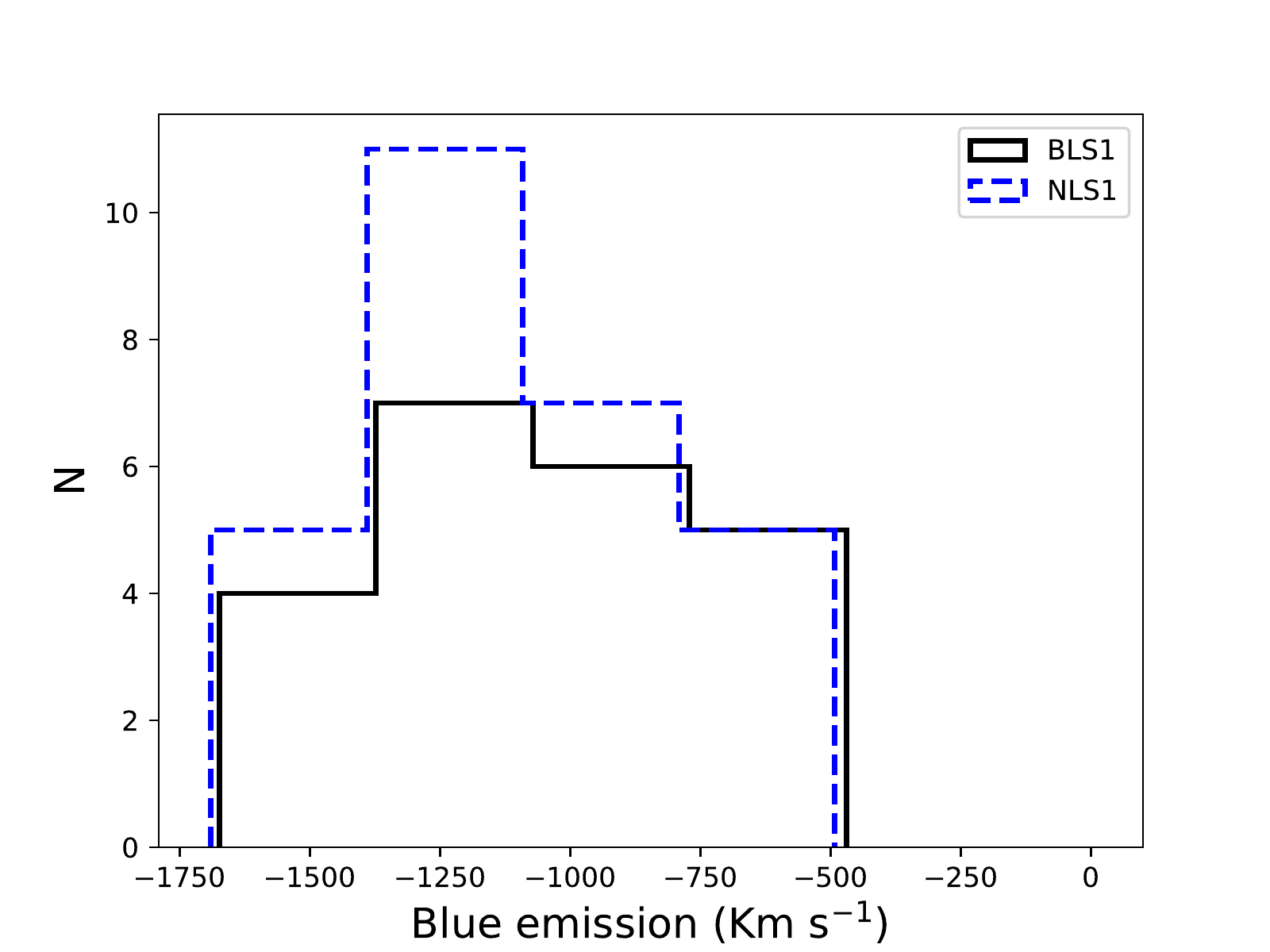}
   \caption{Distribution of the $\emph{total blue emission}$. The solid black line indicates the BLS1 galaxies of our sample and the dashed blue line represents the NLS1 galaxies from \citet{Schmidt2018}.}
                 \label{fig:hist_blue_emission}
    \end{figure}

This parameter has already been studied in a sample of 28 NLS1 galaxies with blue asymmetric [OIII] profiles \citep{Schmidt2018}, and we compared the results we obtained for our sample of BLS1s with those reported by these authors for NLS1s. In Figure \ref{fig:hist_blue_emission}, we show the \emph{blue emission} distribution for the 22 BLS1 galaxies exhibiting [OIII] wings (black solid line), which ranged from $-$1674 to $-$469 km s$^{-1}$, with a mean value of $-$1086 km s$^{-1}$ and a standard deviation of 351 km s$^{-1}$. These values are consistent with those reported by \citealt{Schmidt2018} for NLS1s (blue dashed line) with higher spectral resolution (R=1600).

\begin{figure}
    \centering
    \includegraphics[width=100mm, height=100mm]{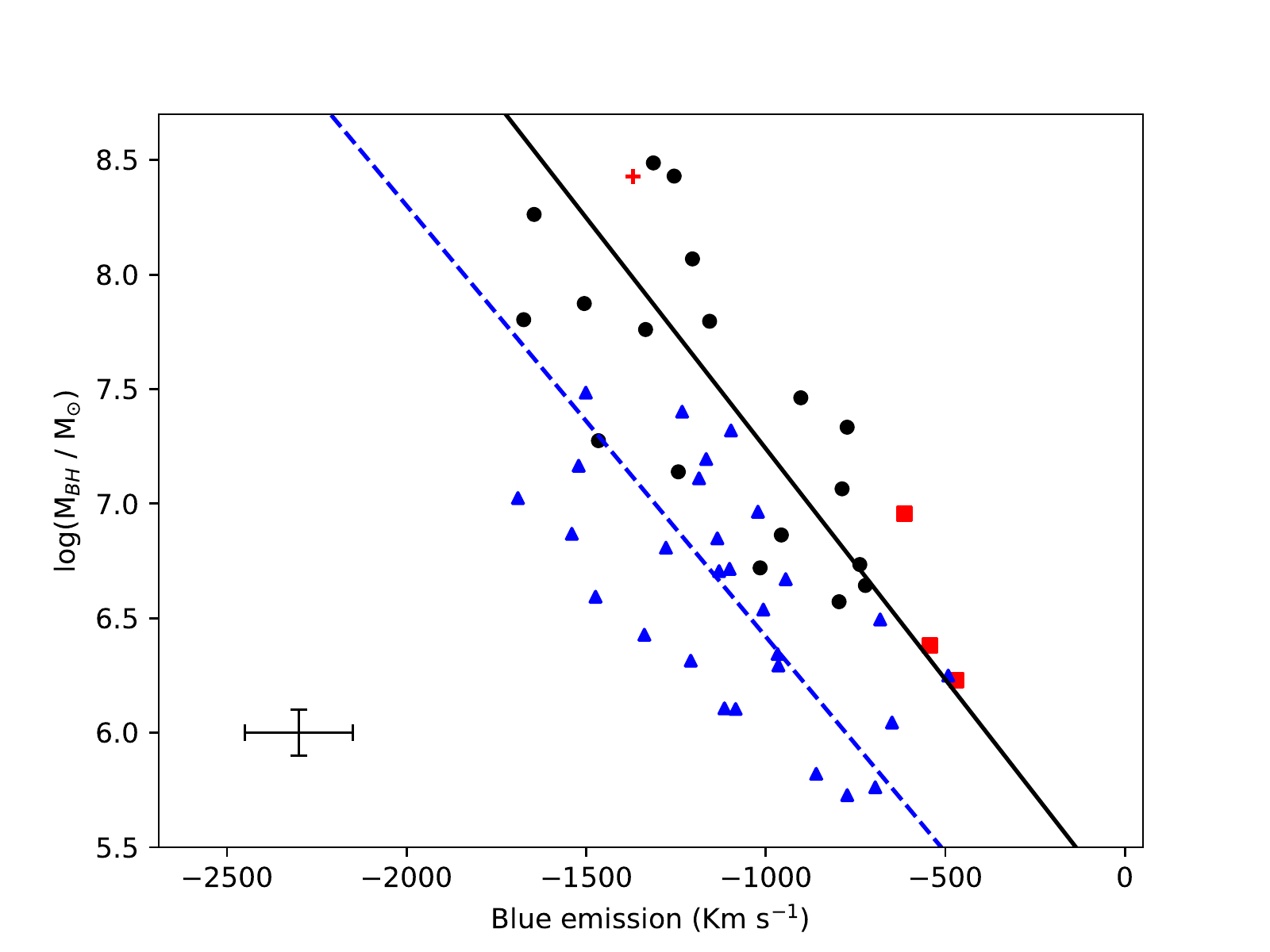}
   \caption{Relation between the $\emph{blue emission}$ and the black hole mass, expressed as log(M$_{BH}$/M$_{\odot}$). Black circles represent the BLS1 galaxies with [OIII] blue wings, while red symbols (squares and plus sign) indicate the BLS1 galaxies with symmetrical or red [OIII] wings ($\Delta$v $\geq$ 0). In addition, the red plus sign indicates the galaxy 6dF J349193-301834 (ID 36), which presents a radio jet. Finally, blue triangles represent the NLS1 galaxies from \citet{Schmidt2018}. The black solid line and blue dashed lines represent the best fit for the BLS1 and NLS1 data, respectively.  A typical error bar is shown in the bottom-left corner.}
              \label{fig:blue_emission_mass}%
    \end{figure}

We also studied a possible relation between the $\emph{blue emission}$ and the black hole mass. For BLS1 galaxies, these two quantities were found to be strongly correlated, with a Pearson coefficient of r$_{p}= -$0.79 and a p-value of 1$\times$10$^{-5}$. In Figure \ref{fig:blue_emission_mass}, we show the correlation between the $\emph{blue emission}$ and the black hole mass for our sample of BLS1 galaxies. Black circles correspond to galaxies presenting [OIII] blueshifts ($\Delta$v $<$ 0), while galaxies exhibiting symmetric ($\Delta$v $\sim$ 0) and red wings ($\Delta$v $>$ 0) are marked with red squares. In addition, the red plus sign corresponds to the galaxy ID 36 (6dF J349193-301834), which presents a radio jet, and the NLS1 galaxies studied by \cite{Schmidt2018} are shown with blue triangles. It can be observed that both NLS1s and BLS1s follow the same trend, with the correlation being stronger for BLS1s (since the correlation coefficient for NLS1s is r$_{p}= -$0.63). By performing an ordinary least squares (OLS) bisector fit to our data, we obtained a slope of $-$0.0020 and a zero point of 5.22 (black solid line). These values are quite similar to those found in NLS1 galaxies, with a slope of $-$0.0019 and a zero point of 4.54 (blue dashed line). However, a slight difference in the zero point is observed, as expected, considering that black holes in BLS1 are systematically more massive than in NLS1s galaxies (see Fig. \ref{fig:bhmass}).

These results indicate that objects presenting more extended wings towards the blue side of the spectrum also correspond to larger black hole masses. Furthermore, we noticed that galaxies with symmetric wings, and even those presenting red-shifted wings (red squares in Fig. \ref{fig:blue_emission_mass}), also seemed to follow the trend observed for galaxies with blueshifted asymmetric profiles (black dots). Although most of galaxies associated with outflows exhibit blueshifted asymmetries, there are few cases where red wings have been reported \citep[e.g.,][]{Cracco2016}. Therefore, the blue emission parameter represents a suitable tool for studying asymmetric profiles and wings in general.

Regarding the detection of the \textit{blue emission} parameter, it is worth noting that a fraction of galaxies with presence of [OIII] wings might be missed due to the spectral resolution of our data (see Sec. \ref{sec:measurement}). Since the lower bound of the \textit{blue emission} values we detected is $\sim$ 500~km s$^{-1}$, we would not expect that those galaxies with [OIII] single-Gaussian fit having a FWHM$\gtrsim$ 500~km s$^{-1}$ would exhibit wings at a higher resolution. Thus, narrower [OIII] profiles, which in turn are associated with smaller BH masses, might result more affected by this effect. If this is the case, the wing component (hypothetically detected with higher resolution) could not extend more than the width of the line measured with a lower resolution, following the trend (Fig. \ref{fig:blue_emission_mass}) and starting from the lower black hole masses.

In Table \ref{tab:oiii}, we list the [OIII] emission line properties and the estimated black hole masses.
Column 1 presents the galaxy ID, column 2 and 3 correspond to the FWHM of the core and wing components, respectively, column 4 shows the I([OIII]$_{wing}$)/I([OIII]$_{core}$) line intensity ratio, column 5 presents the $\Delta$v, column 6 the $\emph{blue emission}$, and finally, column 7 corresponds to the black hole mass.

\begin{table*}
 \center
 \caption{[OIII] emission line properties and black hole mass estimation.}
  \label{tab:oiii}
  \begin{tabular}{lcccccc}
    \hline
    ID & FWHM$_{core}$  & FWHM$_{wing}$   & I([OIII]$_{wing}$)/I([OIII]$_{core}$)  & $\Delta V$         & Blue emission  &  log (M$_{BH}$/M$_{\odot}$)  \\
   {} & [km s$^{-1}$]  & [km s$^{-1}$]   &              {}                    & [km s$^{-1}$]      & [km s$^{-1}$]&    \\
   \hline
       1   &   228    &  820   &   0.5   &   -137  &  -957  &  6.9   \\
       2   &   366    &  --      &   --    &   --      &  --      &  7.7 \\
       3   &   201    &  636   &   0.4   &   -87   &  -723  &  6.6 \\
       4   &   211    &  498   &   0.2   &   -240  &  -738  &  6.7 \\
       5   &   298    &  364   &   0.1   &   -410  &  -774  &  7.3 \\
       6   &   383    &  --      &   --    &   --      &  --      &  7.8 \\
       7   &   288    &  1118  &   0.2   &   -348  &  -1466 &  7.3 \\
       8   &   346    &  --      &   --    &   --      &  --      &  7.6 \\
       9   &   463    &  --      &   --    &   --      &  --      &  8.1 \\
      10   &   388    &  941   &   0.3   &   -216  &  -1157 &  7.8 \\
      11   &   357    &  --      &   --    &   --      &  --      &  7.6 \\
      12   &   566    &  --      &   --    &   --      &  --      &  8.5 \\
      13   &   406    &  1445  &   0.2   &   -60   &  -1505 &  7.9 \\
      14   &   272    &  --      &   --    &   --      &  --      &  7.2 \\
      15   &   321    &  453   &   0.1   &   -449  &  -903  &  7.5 \\
      16   &   253    &  --      &   --    &   --      &  --      &  7.0 \\
      17   &   360    &  --      &   --    &   --      &  --      &  7.7 \\
      18   &   430    &  --      &   --    &   --      &  --      &  8.0 \\
      19   &   365    &  --      &   --    &   --      &  --      &  7.7 \\
      20   &   373    &  --      &   --    &   --      &  --      &  7.7 \\
      21   &   380    &  1112  &   0.2   &   -223  &  -1335 &  7.8 \\
      22   &   158    &  527   &   0.7   &   58    &  -469  &  6.2 \\
      23   &   294    &  --      &   --    &   --      &  --      &  7.3 \\
      24   &   558    &  872   &   0.8   &   -384  &  -1255 &  8.4 \\
      25   &   371    &  --      &   --    &   --      &  --      &  7.7 \\
      26   &   390    &  1272  &   1.0   &   -402  &  -1674 &  7.8 \\
      27   &   193    &  703   &   2.1   &   -94   &  -796  &  6.6 \\
      28   &   541    &  --      &   --    &   --      &  --      &  8.4 \\
      29   &   419    &  --      &   --    &   --      &  --      &  7.9 \\
      30   &   347    &  --      &   --    &   --      &  --      &  7.6 \\
      31   &   301    &  --      &   --    &   --      &  --      &  7.3 \\
      32   &   328    &  --      &   --    &   --      &  --      &  7.5 \\
      33   &   507    &  1433  &   1.9   &   -212  &  -1645 &  8.3 \\
      34   &   577    &  855   &   0.3   &   -458  &  -1313 &  8.5 \\
      35   &   210    &  903   &   0.7   &   -113  &  -1015 &  6.7 \\
      36   &   558    &  1370  &   0.6   &   0     &  -1370 &  8.4 \\
      37   &   266    &  921   &   0.3   &   -323  &  -1244 &  7.1 \\
      38   &   259    &  --      &   --    &   --      &  --      &  7.1 \\
      39   &   240    &  737   &   0.7   &   124   &  -614  &  7.0 \\
      40   &   243    &  --      &   --    &   --      &  --      &  7.0 \\
      41   &   221    &  --      &   --    &   --      &  --      &  6.8 \\
      42   &   173    &  560   &   0.5   &   17    &  -543  &  6.4 \\
      43   &   500    &  --      &   --    &   --      &  --      &  8.2 \\
      44   &   454    &  535   &   0.1   &   -670  &  -1204 &  8.1 \\
      45   &   255    &  566   &   0.3   &   -222  &  -788  &  7.1 \\
    \hline
  \end{tabular}
 \end{table*}


\subsection{Black hole mass distribution}
\label{sub:comp}

Hitherto, O[III] lines exhibiting \textit{blue emission} were observed to be associated with both low and high BH mass objects (see Fig. \ref{fig:blue_emission_mass}). For instance, as previously mentioned, \cite{Schmidt2018} reported blueshifted asymmetric O[III] profiles in a sample of NLS1 galaxies, which have systematically lower black hole masses than BLS1s (see Fig. \ref{fig:bhmass}). However, on comparing our $\Delta$v and \textit{blue emission} distributions with the results presented by these authors, it can be observed that NLS1s and BLS1s have values within the same range (see Fig. \ref{fig:hist_deltav} and Fig. \ref{fig:hist_blue_emission}, respectively). Regarding this, we now considered the behaviour of the black hole mass distribution for: (a) galaxies exhibiting blueshifted asymmetric O[III] profiles compared with those presenting symmetric O[III] lines, and (b) galaxies with or without the presence of O[III] wings.

\begin{figure}
    \centering
    \includegraphics[width=100mm, height=100mm]{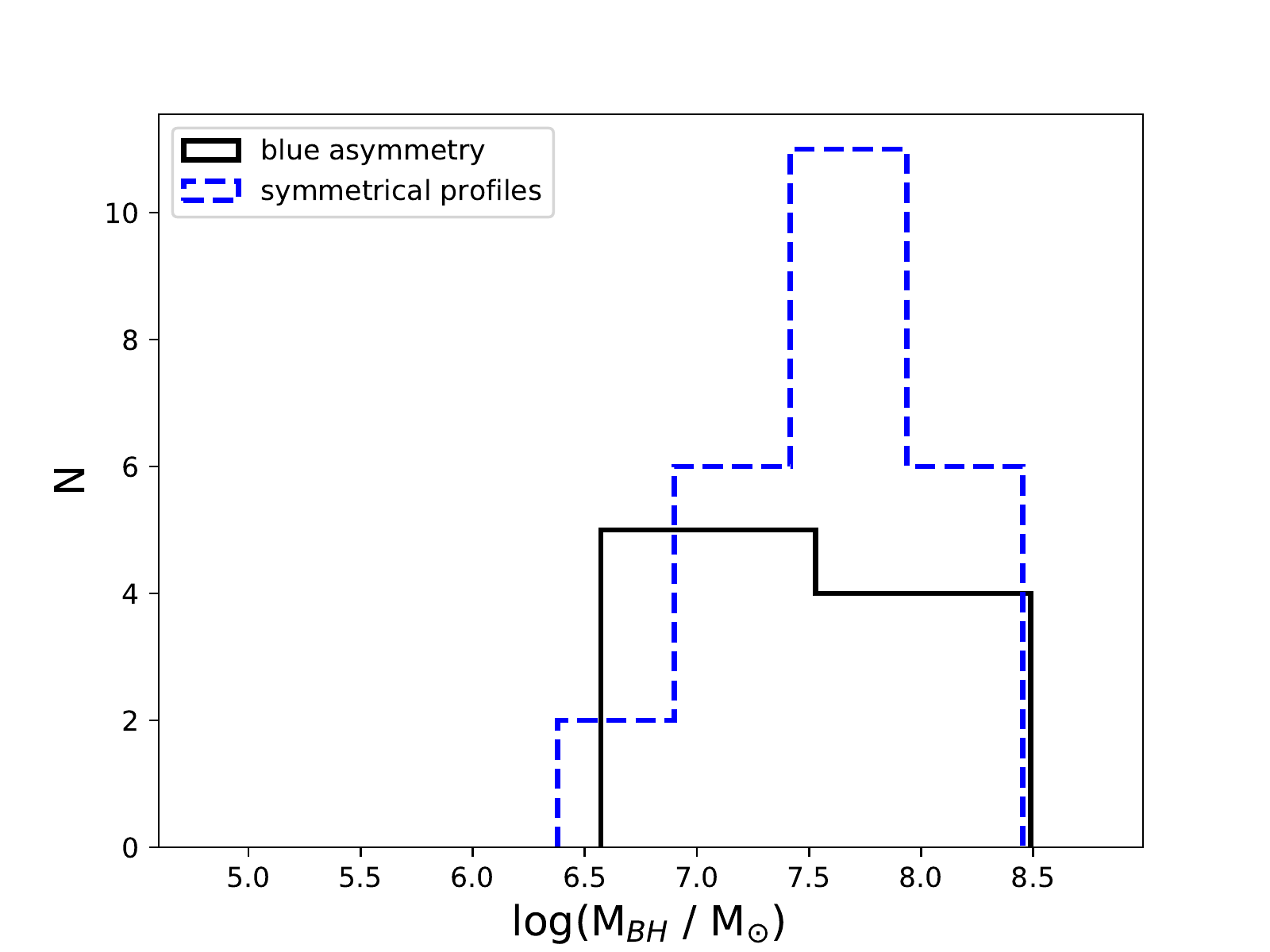}
   \caption{Black hole mass distribution for galaxies with symmetric [OIII] profiles (blue dashed line) and galaxies showing blue asymmetric [OIII] emission (solid black line).}
              \label{fig:blackholemass_comp}
    \end{figure}

In the first case, we computed the black hole mass distribution for 18 galaxies exhibiting blueshifted O[III] profiles ($\Delta$v < 0) and also for 25 objects presenting symmetric O[III] lines, which comprised 23 galaxies with single Gaussian fitting plus 2 galaxies with symmetrical wings ($\Delta$v = 0); see sample details in Sec. \ref{subsec:sample}. In Figure \ref{fig:blackholemass_comp}, we show the black hole mass distribution for these two populations, which are quite similar. Galaxies exhibiting blueshifted wings were found to have a mean value of log(M$_{BH}$/M$_{\odot}$) = 7.4 with a standard deviation of 0.6, while black hole masses ranged from log(M$_{BH}$/M$_{\odot}$) = 6.6 to 8.5. Similarly, the black hole mass distribution for galaxies with symmetric O[III] lines was found to have a mean value of log(M$_{BH}$/M$_{\odot}$) = 7.7 with a standard deviation of 0.5, with black hole masses ranging from log(M$_{BH}$/M$_{\odot}$) = 6.4  to 8.5. According to this, the presence or absence of blue-shifted [OIII] profiles is independent of the black hole mass.

In the second case, we computed the black hole mass distribution for galaxies with or without presence of O[III] wings, i.e., all galaxies with double and single Gaussian fitting, respectively. Galaxies exhibiting wings were found to have a mean value of log(M$_{BH}$/M$_{\odot}$) = 7.3 with a standard deviation of 0.7, while black hole masses ranged from log(M$_{BH}$/M$_{\odot}$) = 6.2 to 8.5. On the other hand, the black hole mass distribution for galaxies that did not exhibit O[III] wings was characterized by a mean value of log(M$_{BH}$/M$_{\odot}$) = 7.7, a standard deviation of 0.4, and with black hole masses ranging from log(M$_{BH}$/M$_{\odot}$) = 6.8 to 8.5. As in the previous case, both these distributions are quite similar. Therefore, since the \emph{blue emission} parameter is defined for all galaxies with double Gaussian decomposition, this result indicates that BLS1 galaxies within a given range of black hole masses might or might not exhibit wings. This, together with the results previously mentioned, suggests that \emph{blue emission} is not biased to higher BH masses or more luminous objects. Nonetheless, according to the analysis performed in Section \ref{sub:total}, whenever O[III] lines do exhibit wings the observed trend indicated that the more extended the \emph{blue emission} was, the more massive the black hole would be.


\section{Discussion}
\label{sec:discussion}

There are different models to explain the origin of outflows. For instance, radiation-pressure \citep[e.g.,][]{Ge2019,Leftley2019} and magnetic driving mechanisms are frequently invoked, and in the latter case, outflows can be generated by the interaction of ionized gas and magnetic fields \citep[e.g.,][]{Bisnovatyi-Kogan2001, King2012}.
These outflows interact with the surrounding medium and transfer kinetic energy to the environment, giving rise to the asymmetries observed in line profiles. In star forming galaxies, the [OIII] emission line is well fitted by a single Gaussian function \citep{Concas2017}. However, based on a study of 600.000 local galaxies, these authors suggest that there is a trend between the need of a second component and the AGN contribution to the spectrum of the galaxy. In agreement with this result, [OIII] asymmetric profiles are usually present in quasars and Seyfert galaxies \citep[e.g.,][]{VeronCetty2001, Komossa2007}. In addition, more type 1 AGNs usually exhibit outflows compared to type 2 galaxies \citep[e.g.,][]{Vaona2012, Woo2016}. 

In the present work, we studied a sample of 45 BLS1 galaxies from the southern hemisphere with available free-access spectra, basing our analysis on the [OIII]$\lambda$5007 emission profile. By conducting a thorough fitting and inspection of [OIII]$\lambda$5007, we found that spectral decomposition was necessary in 22 cases exhibiting core and wing components, while a single Gaussian function was fitted in the remaining 23 galaxies. Among the 22 objects presenting wings, 18 were found to have blueshifted wings and 2 red-shifted wings, while 2 objects exhibited symmetrical wings relative to the core component centroid.

For the whole sample, the core component was found to be in the range 158 km s$^{-1}$ to 577 km s$^{-1}$, with a mean value of 293 km s$^{-1}$. Considering the 22 galaxies showing [OIII] wings, the FWHM of the wing ranged from 364 km s$^{-1}$ to 1445 km s$^{-1}$ (Sect. \ref{sub:wing}). This FWHM difference observed between the core and wing components might imply that they are associated with different emitting regions, in agreement with previous works, \citep[e.g.,][]{Holt2003, Bian2005, Schmidt2018}. The FWHM range of the wings lay between the FWHM of the core component emitted by the NLR (see Sect. \ref{sec:measurement}) and the FWHM of typical broad Balmer emission lines \citep[see for instance][]{Tremou2015}. Thus, considering that gas squared velocity decreases with distance to the AGN centre \citep[e.g.,][]{Schmidt2016}, we infer that wings are most probably originating at the inner layers of the NLR, in agreement with previous studies \citep[e.g.,][]{VeronCetty2001, Bian2005, Schmidt2018}. This region is located between the zone where the [OIII] core component originates and the BLR (i.e., the region where the broad Balmer lines are produced), indicating that inner layers are more turbulent than the external emitting regions of the NLR, as reported by other authors \citep[e.g.,][]{VeronCetty2001, Holt2003,Bian2005}.

Outflows and winds not only manifest themselves through the [OIII] asymmetric profile, but also through asymmetries observed in other high ionization emission lines, such as [Fe VII], [Fe X], [Fe XI] \citep[e.g.,][]{RA2006, Komossa2008}, [Ne III], and [Ne V] \citep[e.g.,][]{Spoon2009}. Moreover, very high ionization lines originating at the BLR, such as C IV$\lambda$1549, have been found to be blueshifted \citep[e.g.,][]{Sulentic2007,Gaskell2013}. However, no asymmetries were found in low ionization emission lines, such as [SII]$\lambda\lambda$6717,6731, each of which are well described by a single Gaussian function \citep[e.g.,][]{Cracco2016,Schmidt2016,Schmidt2019a}. Since low ionization lines originate at the external regions of the NLR, and high ionization lines are emitted from its inner layers (and from the BLR in the case of C~IV$\lambda$1549), we interpret that outflows could be moving throughout a stratified medium ionized by the AGN, in agreement with previous studies \citep[e.g.,][]{Cracco2016}.

Some works have suggested that outflows reduce the line equivalent width and increase the FWHM of the [OIII] profile \citep{Cracco2016}, also in agreement with results reported by \cite{Ludwig2012}, who found that the [OIII] equivalent width correlated with the blueshift of blue wings. These results are consistent with the relation we found between $\Delta$v and the FWHM of the core component of [OIII] (see Sect. \ref{sub:wing}), also studied by other researchers \citep[e.g.,][]{VeronCetty2001,Bian2005}.

We analysed the full extension of [OIII] wings relative to the centroid of the core component using the \emph{blue emission} parameter, and found this parameter to have a strong correlation with black hole mass, characterized by a Pearson coefficient of r$_{p}= -$0.79 and a p-value of 1$\times$10$^{-5}$ (Sect. \ref{sub:total}). This is in agreement with previous results presented by \citet{Schmidt2018} for a sample of 28 NLS1 galaxies exhibiting blueshifted wings (Pearson coefficient of r$_{p}= -$0.63). Considering the BLS1 and NLS1 samples, the OLS bisector fits of the two correlations revealed slopes of $-$0.0020 and $-$0.0019 and zero points of 5.22 and 4.54 \citep{Schmidt2018}, respectively, with both these distributions having very similar parameters, and  covering similar \emph{blue emission} ranges. The slight difference observed in zero points was expected due to the difference in the black hole masses in BLS1 and NLS1 (see Fig. \ref{fig:bhmass}). Thus, for any given value of [OIII] $\emph{blue emission}$, BLS1 galaxies would have higher black hole mass than NLS1s. On the other hand, for a given black hole mass, BLS1 galaxies will exhibit fewer extended wings to the blue-end than NLS1s, with $\sim$ 340 km s$^{-1}$ being the typical velocity difference. Given that NLS1 galaxies would not represent a different type of object \citep[e.g.,][]{Sulentic2000, VeronCetty2001}, it is not clear what could produce this difference. However, it is known that BLS1 galaxies present higher black hole masses and NLS1s show higher Eddington ratios \citep[e.g.,][]{Bian2004, Xu2007, Cracco2016}, so this difference in Eddington ratios might be responsible for NLS1s showing more extended wings to the blue-end than BLS1s, at a given black hole mass. This could indeed be the case, since outflows could be originated by the interaction of ionized gas with magnetic fields \citep{Bisnovatyi-Kogan2001}. In this scenario, the Eddington ratio in NLS1 was found to show a weak correlation with the [OIII] blue emission, with a Pearson coefficient of r$_{p}= -$0.33 \citep{Schmidt2018}. Also, the blue emission seemed to be related with the luminosity of the broad component of H$\beta$, with a Pearson coefficient of r$_{p}= -$0.40 \citep{Schmidt2018}. According to this, NLS1 galaxies showing wings with more an extended blue-end also present higher Eddington ratios and brighter broad H$\beta$ profiles. Nevertheless, according to the results of \citet{Berton2016}, there is no correlation between the velocity of the winds and the Eddington ratio. As these authors found correlation coefficients of r$_{p}= -$0.1 and r$_{p}= -$0.2 in radio quiet and radio loud samples of NLS1 galaxies, respectively, the Eddington ratio is not related to the velocity of the outflows.

According to previous studies, emission profile asymmetries could depend on the intensity of the AGN \citep[e.g.,][]{Mullaney2013}. In this regard, the observed trend followed by the black hole mass with $\Delta$v, \emph{blue emission}, and the luminosities of emission lines such as [SII], [NII] and [OIII] \citep{Schmidt2019a}, are in agreement with the hypothesis that the intensity of the AGN can influence both the shape and asymmetries in emission [OIII] profiles. However, it was reported that asymmetric profiles could be more related to kinematic mechanisms than to photoionization processes in NLS1 galaxies \citep[e.g.,][]{Schmidt2018}.

By comparing the black hole mass distributions in galaxies with [OIII] blueshift and galaxies showing symmetrical [OIII] profiles, we found that both distributions are quite similar (Sect.\ref{sub:comp}). This is in agreement with previous analyse of NLS1 galaxies \citep[e.g.,][]{Schmidt2018}, indicating that both symmetric and blue asymmetric [OIII] profiles can be observed in galaxies with the same black hole mass distribution.

Summing up, the findings of our study of O[III] emission line asymmetries and their relation with the different spectral and physical parameters associated with BLS1 and NLS1 galaxies, are in agreement with a continuous transition scenario between both types of galaxies \citep{Sulentic2000, VeronCetty2001}, since the behaviour we observed in BLS1 galaxies is consistent with that previously reported for NLS1s. Moreover, the \emph{blue emission} parameter was found to be a useful tool for studying asymmetric profiles and wings in general.

\section{Final remarks}
\label{sec:final}

In this work, we performed a spectroscopic study of a sample of 45 near southern broad line Seyfert 1 (BLS1) galaxies from the 6 Degree Field Galaxy survey, basing our analysis on the [OIII]$\lambda$5007 emission profile. Through a careful Gaussian fitting, we found that 23 galaxies were fitted with a single Gaussian function. On the other hand, 22 galaxies needed a double Gaussian decomposition: one for the core emission and one function for the wing. We investigated how line features could be related to the central black hole. All the results were compared to similar studies performed on narrow line Seyfert 1 galaxies, and can be summarized as follows:

    \begin{itemize}
        \item Black hole masses were estimated in the range log(M$_{BH}$/M$_{\odot}$)$=$ 6.2$-$8.5, with a mean value of log(M$_{BH}$/M$_{\odot}$)$=$ 7.6 and a standard deviation of 0.6, in agreement with typical values measured in BLS1 galaxies.\\

        \item We computed the radial velocity difference $\Delta$v between the [OIII] core and wing components. The $\Delta$v distribution ranged from $-$670 to 124 km s$^{-1}$, having a mean value of $-$219 km s$^{-1}$ with a standard deviation of 192 km s$^{-1}$. These values are similar to those reported for NLS1 galaxies \citep[e.g.,][]{Cracco2016,Schmidt2018}.\\
       
        \item By studying $\Delta$v, we found that 18 galaxies showed a blue asymmetric [OIII] profile ($\Delta$v$<$0), 2 objects had symmetrical wings relative to the core component ($\Delta$v $\sim$0) and 2 galaxies presented red wings ($\Delta$v $ >$0).\\
 
        \item We found the parameter $\Delta$v to be more strongly related to the black hole mass (Pearson coefficient r$_{p}=-$0.51 and p-value of 1.5$\times$10$^{-2}$) than to the FWHM of the [OIII] core component (Pearson coefficient r$_{p}=-$0.45 and p-value of 3.5$\times$10$^{-2}$). This last tendency has been observed in NLS1 galaxies \citep[e.g.,][]{Bian2005, Schmidt2018}.\\

        \item The \emph{blue emission} parameter was calculated as the radial velocity difference between the blue-end of the [OIII] wing and the centroid of the [OIII] core component. $\emph{Blue emission}$ values ranged from$-$1674 to $-$469 km s$^{-1}$, having a mean value of $-$1086 km s$^{-1}$ and a standard deviation of 351 km s$^{-1}$. We found the values of the \emph{blue emission} in BLS1 to be similar to those observed in NLS1 galaxies \citep{Schmidt2018}. \\

       \item The \emph{blue emission} parameter correlated strongly with the black hole mass (Pearson coefficient r$_{p}= -$0.79 and a p-value 1$\times$10$^{-5}$), indicating that galaxies with higher black hole masses also present a higher radial velocity difference between the wing blue-end and the core. This correlation was also observed in NLS1 galaxies (with a Pearson coefficient r$_{p}= -$0.63 and p-value of 3$\times$10$^{-4}$ being reported in \citealt{Schmidt2018}). However, we found the $\emph{blue emission}$ to be more strongly correlated with the black hole mass in the case of BLS1 galaxies, as the Pearson coefficient indicated. Also, the slope of the OLS bisector fit in each case was different ($-$0.0020 for BLS1s and $-$0.0019 for NLS1), since BLS1 galaxies host more massive black holes.\\
        
    \item BLS1 galaxies with [OIII] blueshifted wings presented the same black hole mass distribution as BLS1 with symmetrical [OIII] profiles. This indicates that the blue asymmetries in BLS1 can be present in galaxies with different black hole masses, in agreement with the results of \cite{Schmidt2018} in NLS1 galaxies.\\
    
    
      
        
    \end{itemize}
    
In general terms, we found that the BLS1 galaxies presented similar properties regarding the line features of the [OIII] profile compared to NLS1 galaxies. In addition, the correlations between $\Delta$v and the \emph{blue emission} with different parameters in BLS1s resembled those found in NLS1s, indicating similar properties regarding outflows. In this context, our results support the scenario in which BLS1 and NLS1 galaxies are not different type of objects, but with there existing a continuous transition between them \citep[e.g.,][]{Sulentic2000, VeronCetty2001}.

\section*{Acknowledgements}

We want to thank Damian Mast for fruitful discussions. We are also grateful to our referee for helpful comments and suggestions. This work was partially supported by Consejo de Investigaciones Científicas y Técnicas (CONICET) and Secretaría de Ciencia y Técnica de la Universidad Nacional de Córdoba (SecyT). 
This research has made use of the NASA/IPAC Extragalactic Database (NED) which is operated by the Jet Propulsion Laboratory, California Institute of Technology, under contract with the National Aeronautics and Space Administration.

\section*{Data Availability}

The inclusion of a Data Availability Statement is a requirement for articles published in MNRAS. Data Availability Statements provide a standardised format for readers to understand the availability of data underlying the research results described in the article. The statement may refer to original data generated in the course of the study or to third-party data analysed in the article. The statement should describe and provide means of access, where possible, by linking to the data or providing the required accession numbers for the relevant databases or DOIs.



\bibliographystyle{mnras}
\bibliography{Bibliography} 




\appendix

\section{Some extra material}

In this Appendix, we present all the fits to the [OIII] lines for the 45 studied galaxies (Fig. \ref{fig:fits1}).

\begin{figure*}
\begin{centering}
    \includegraphics[width=0.49\textwidth,height=0.25\textwidth]{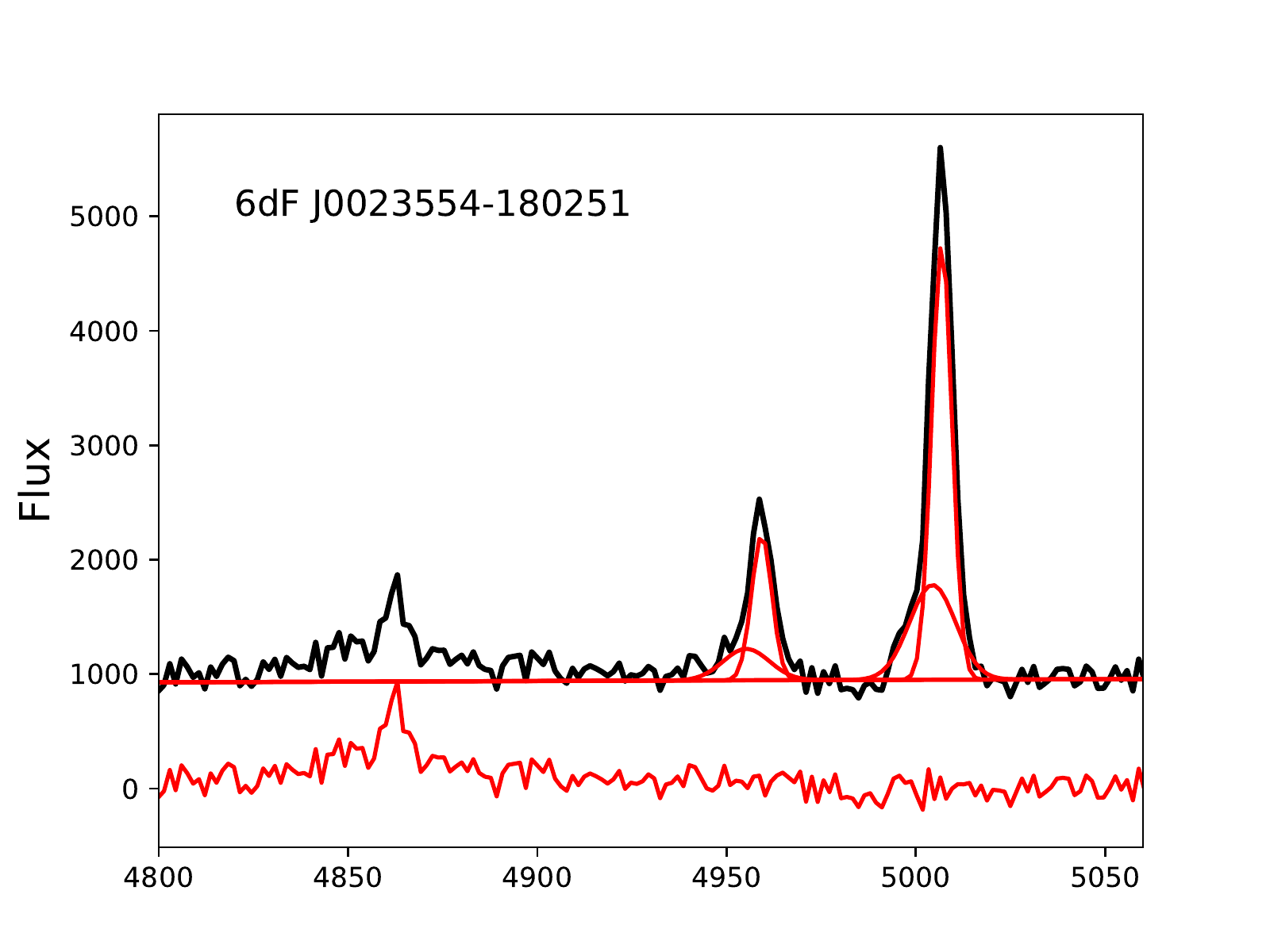}
    \includegraphics[width=0.49\textwidth,height=0.25\textwidth]{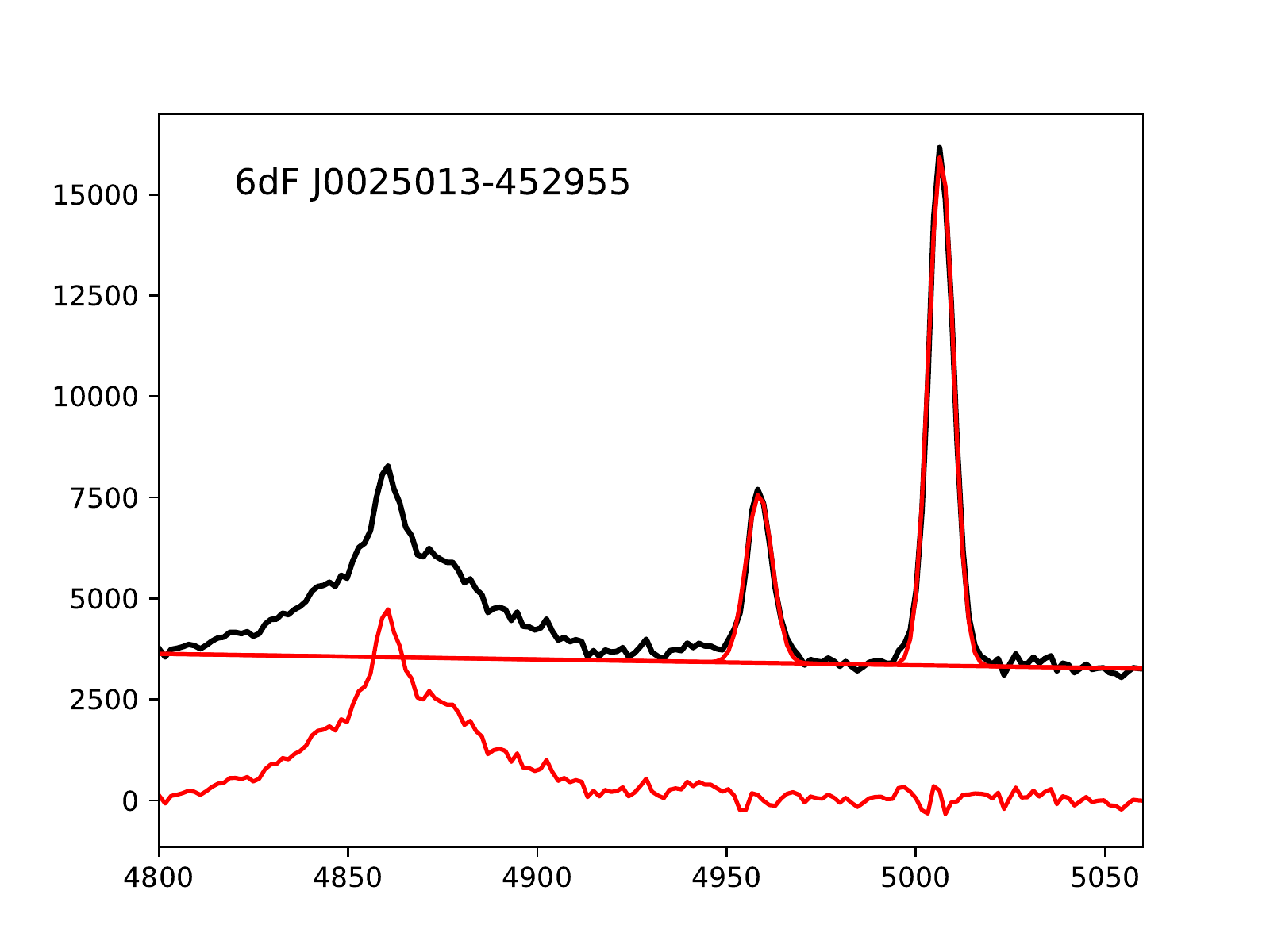}
    \includegraphics[width=0.49\textwidth,height=0.25\textwidth]{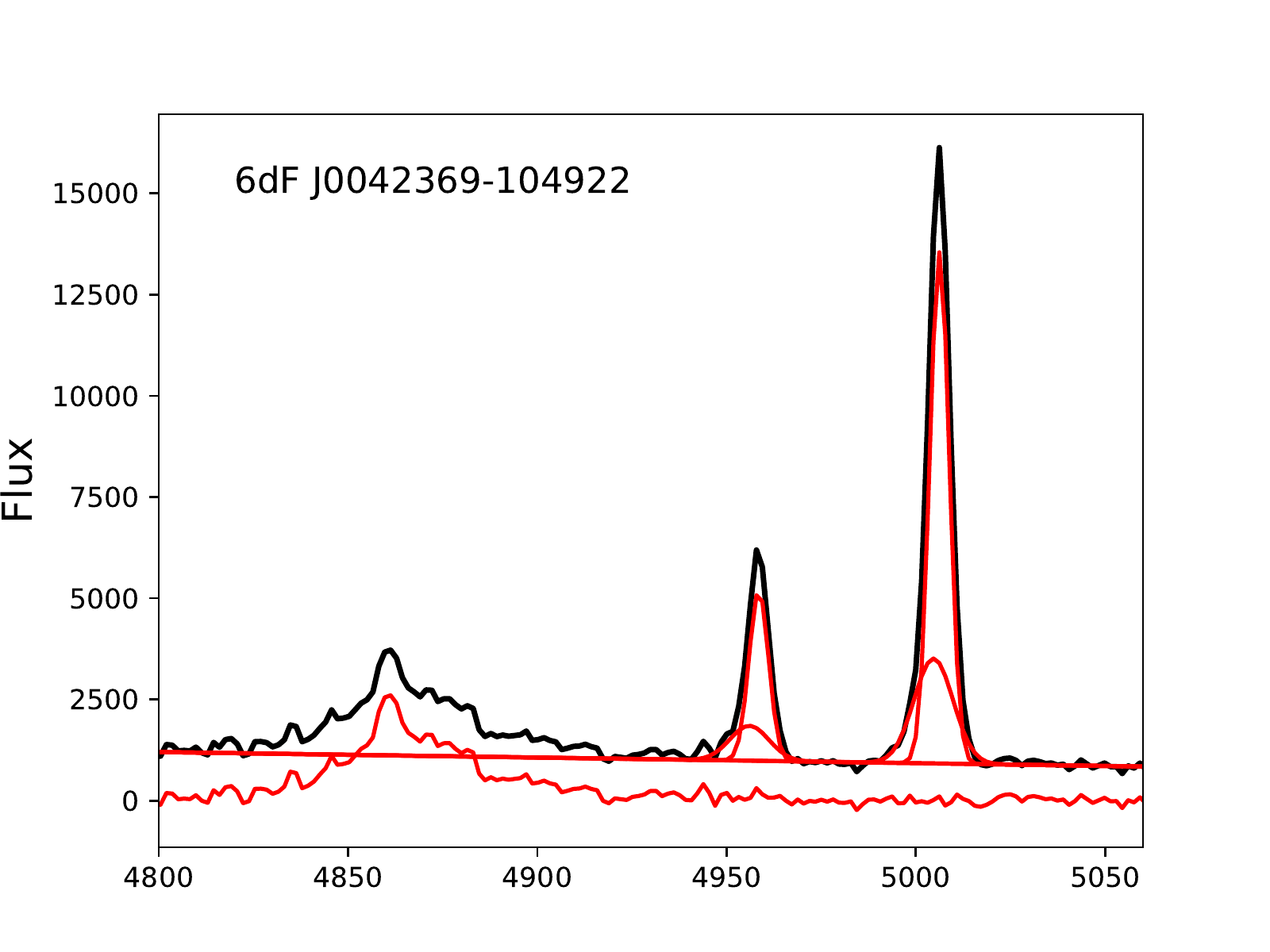}
    \includegraphics[width=0.49\textwidth,height=0.25\textwidth]{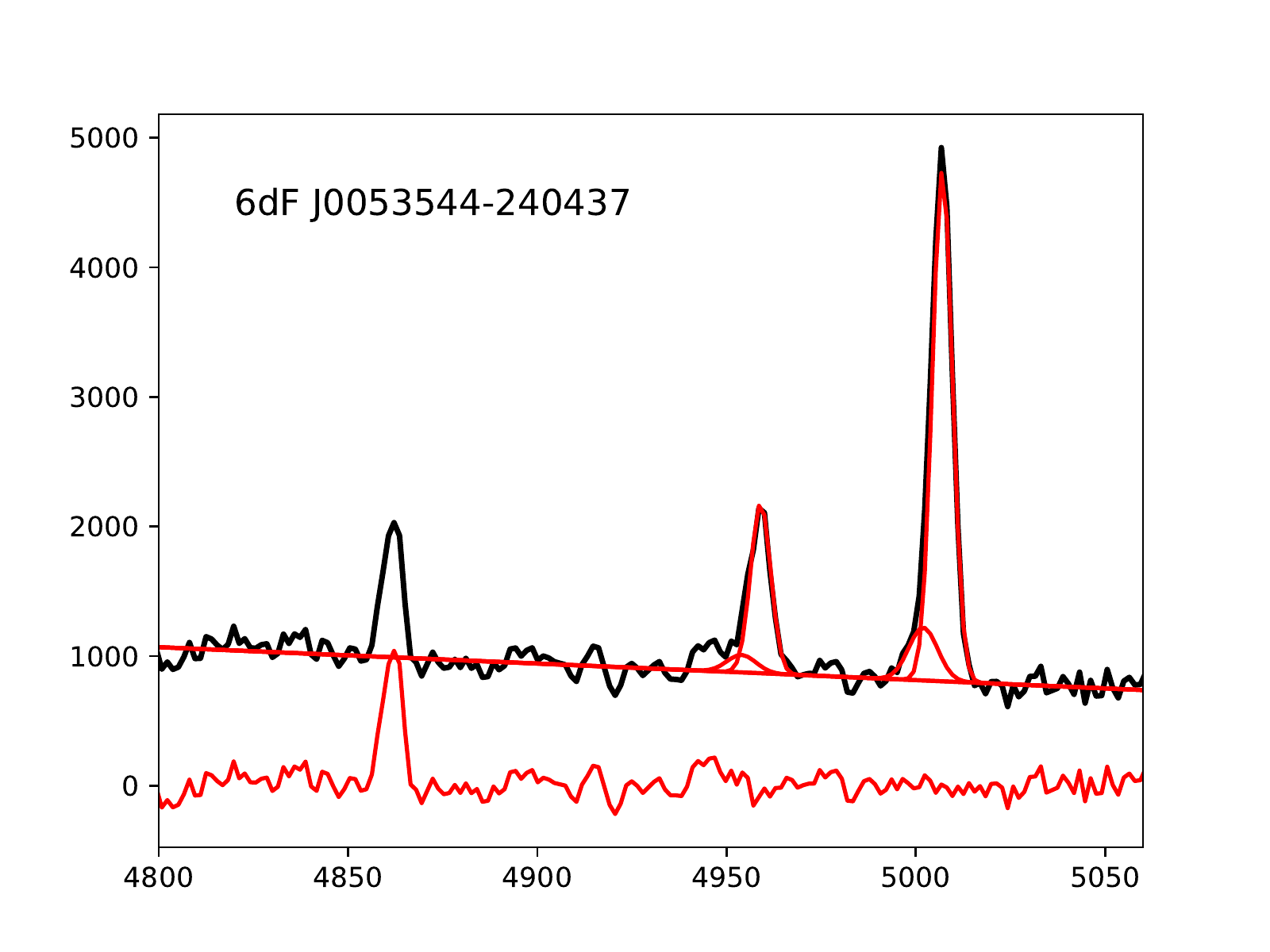}
    \includegraphics[width=0.49\textwidth,height=0.25\textwidth]{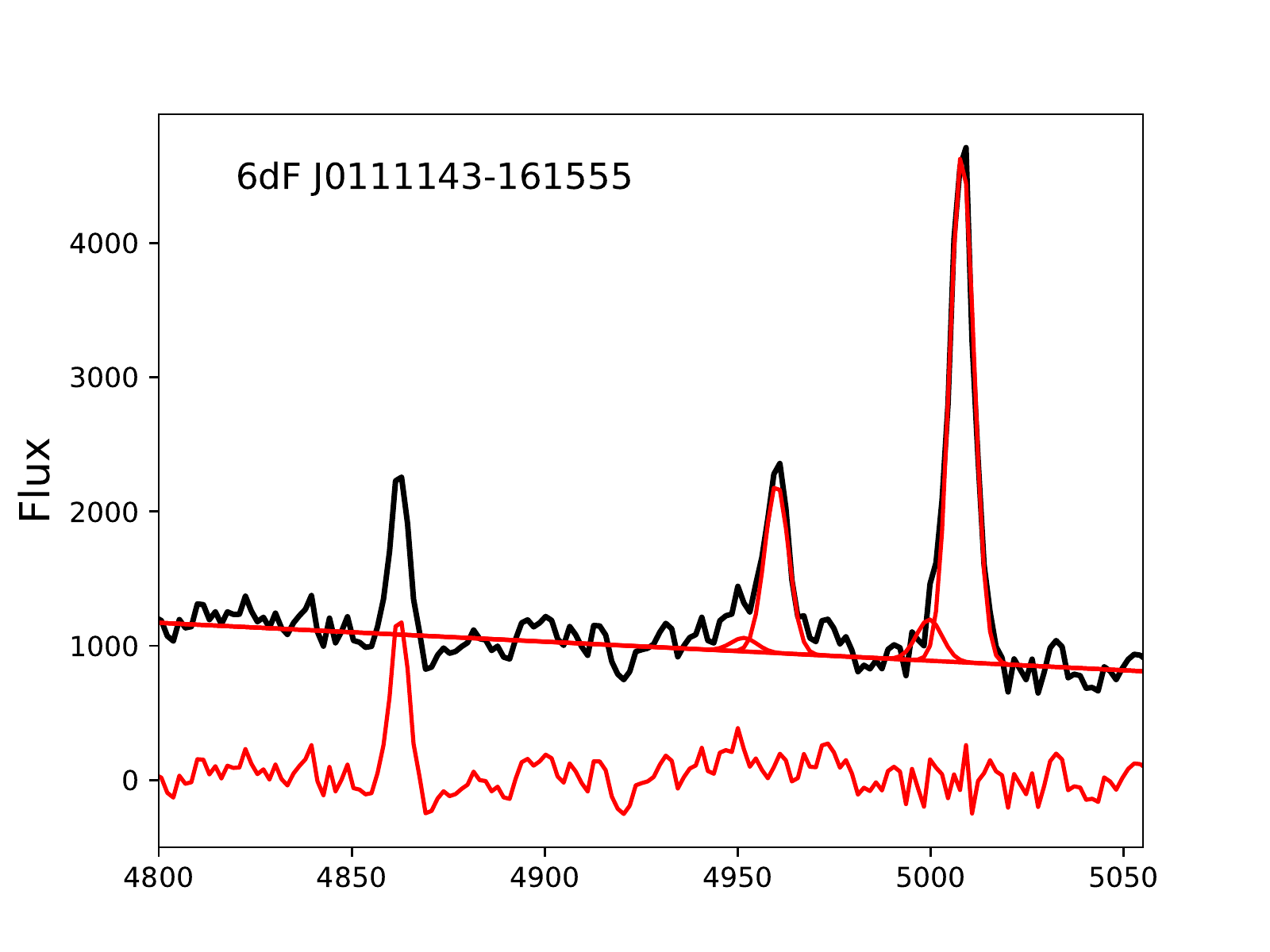}
    \includegraphics[width=0.49\textwidth,height=0.25\textwidth]{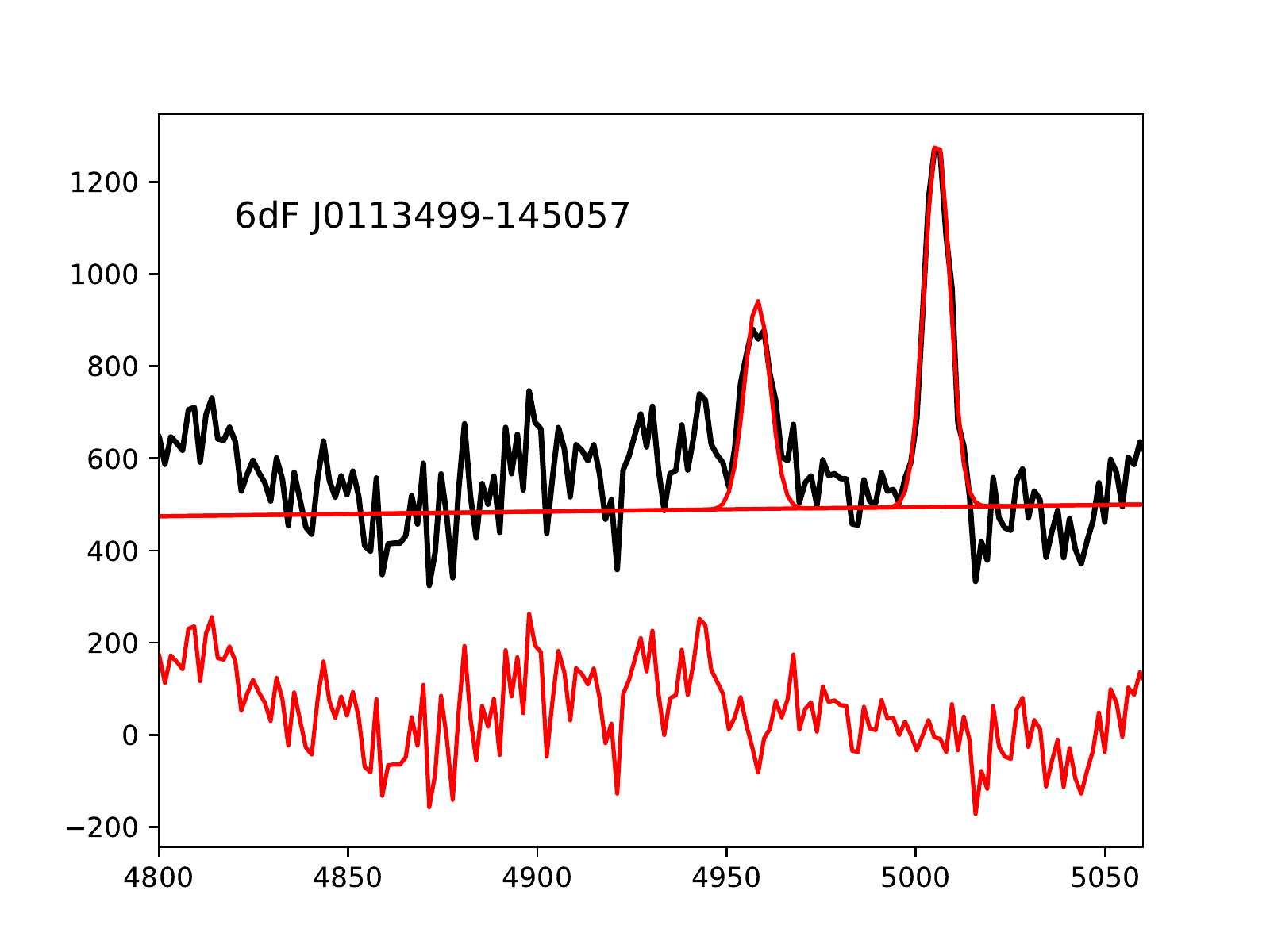}
    \includegraphics[width=0.49\textwidth,height=0.25\textwidth]{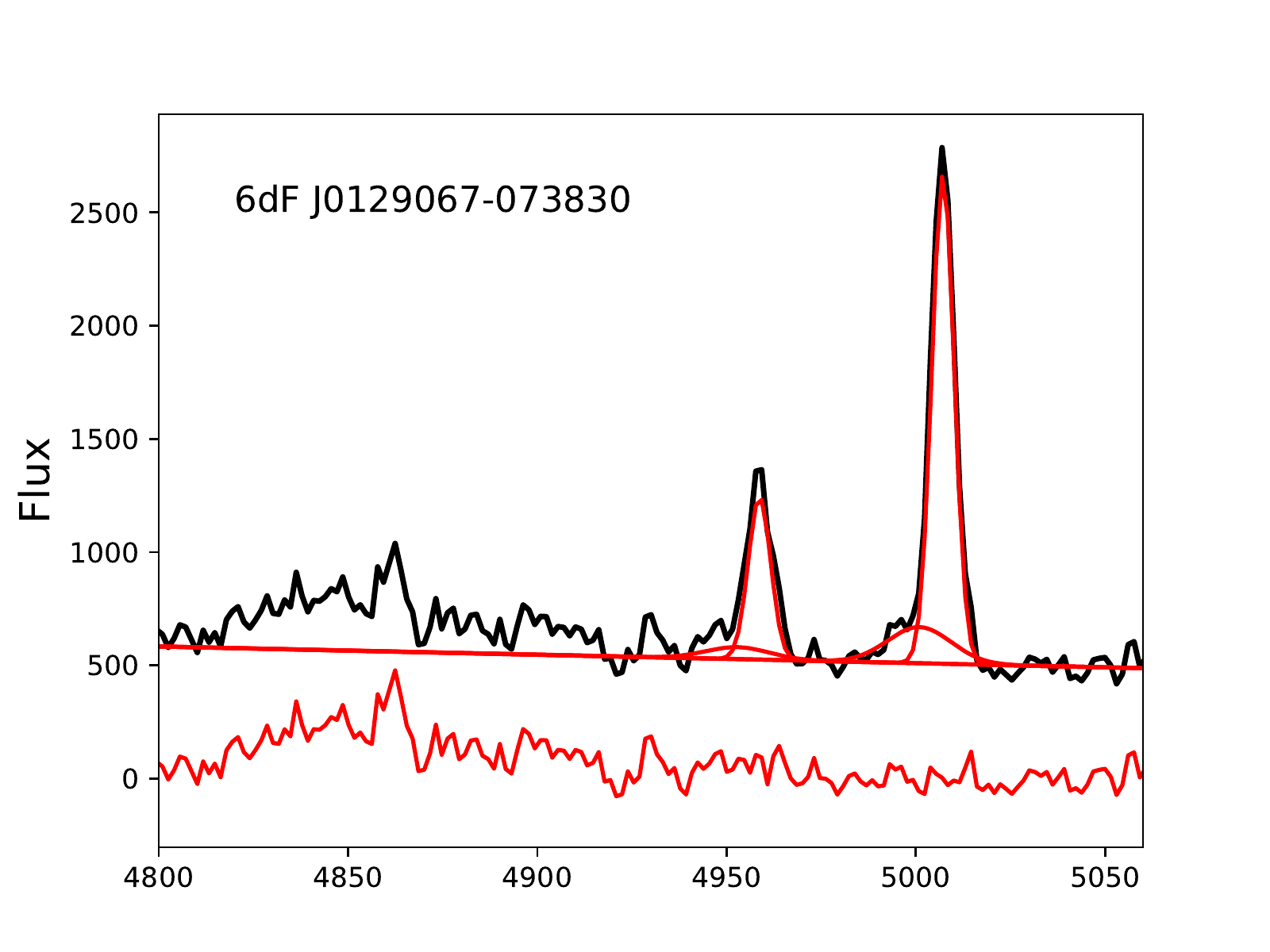}
    \includegraphics[width=0.49\textwidth,height=0.25\textwidth]{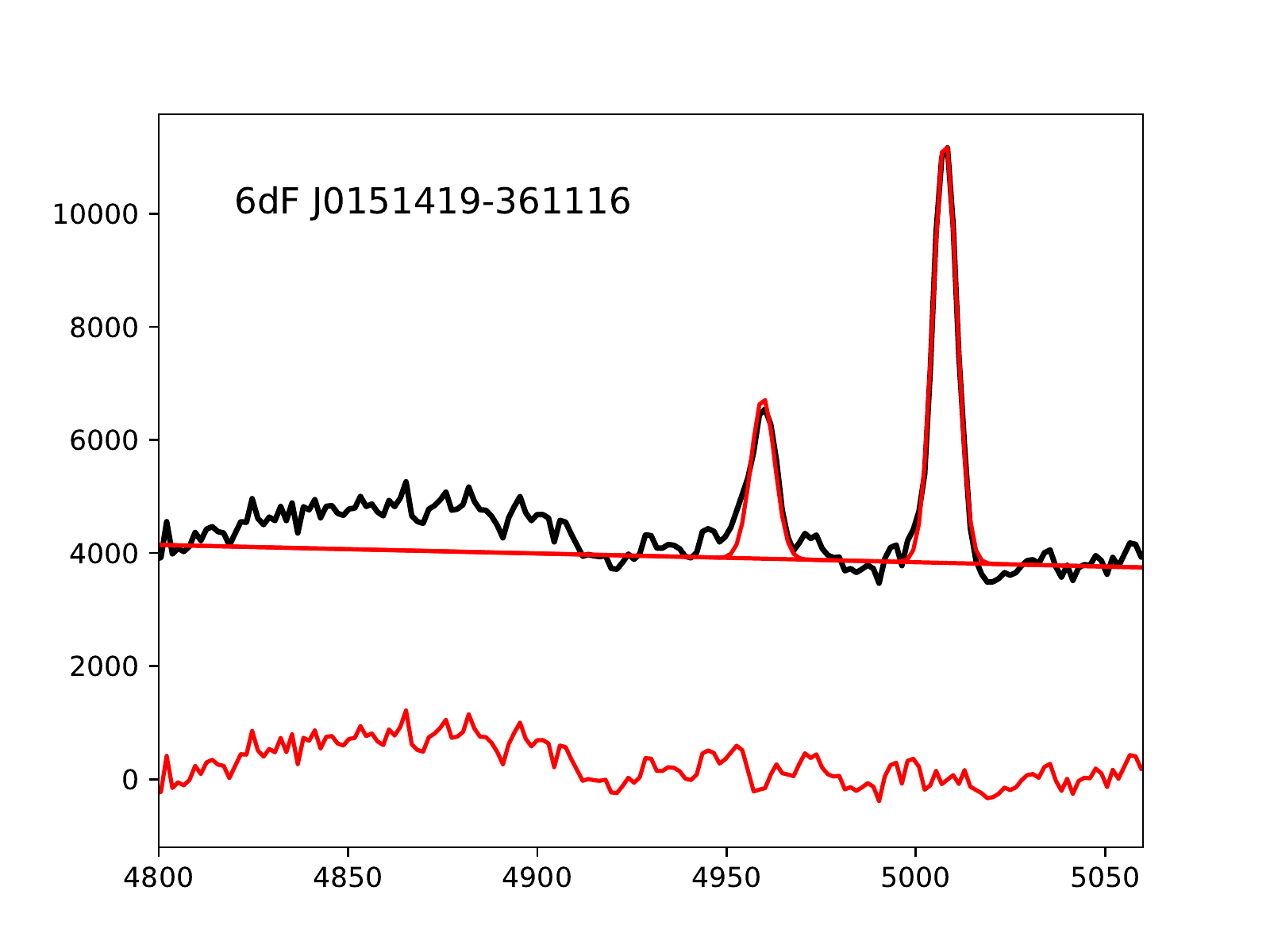}
    \includegraphics[width=0.49\textwidth,height=0.25\textwidth]{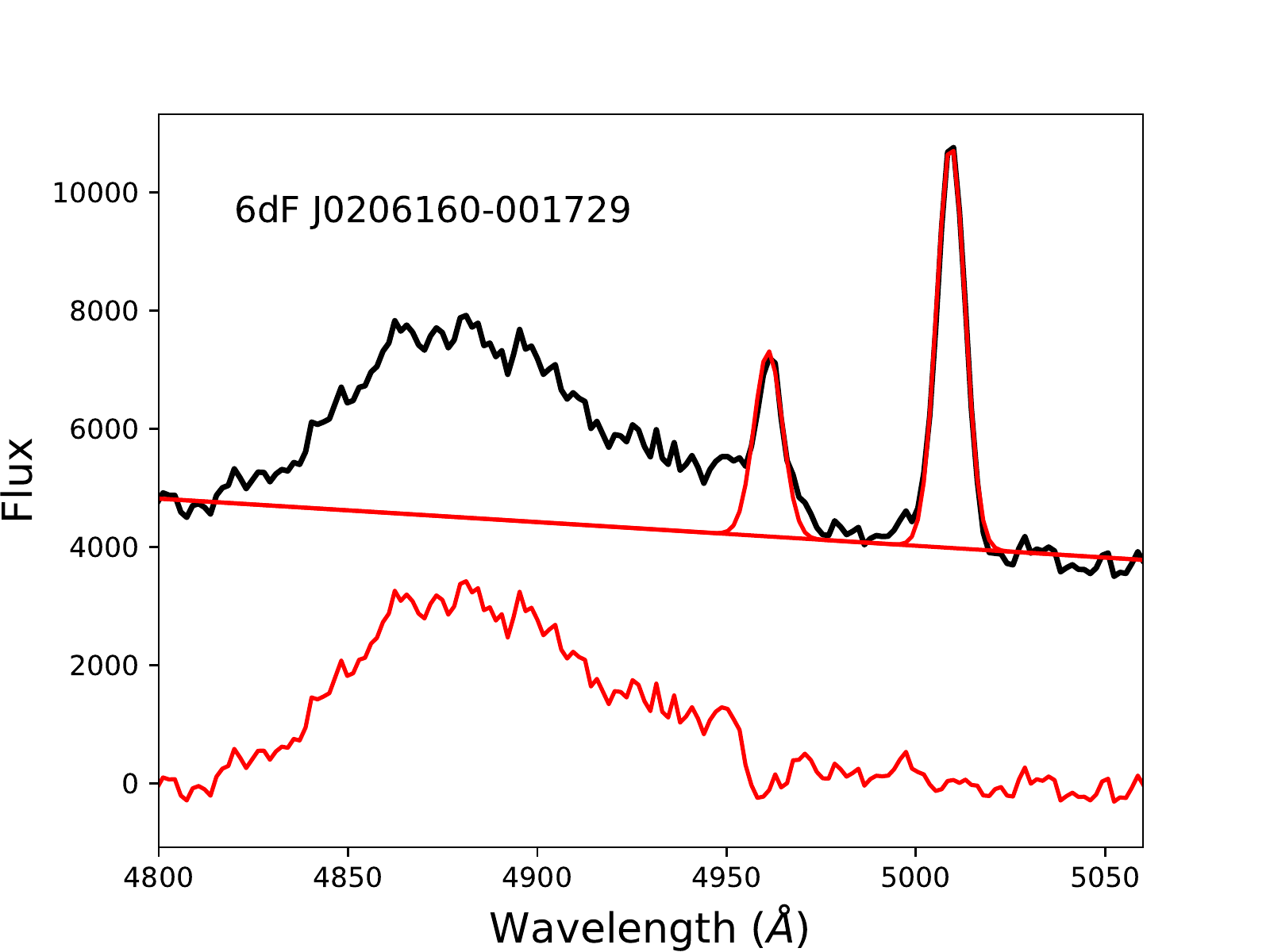}
    \includegraphics[width=0.49\textwidth,height=0.25\textwidth]{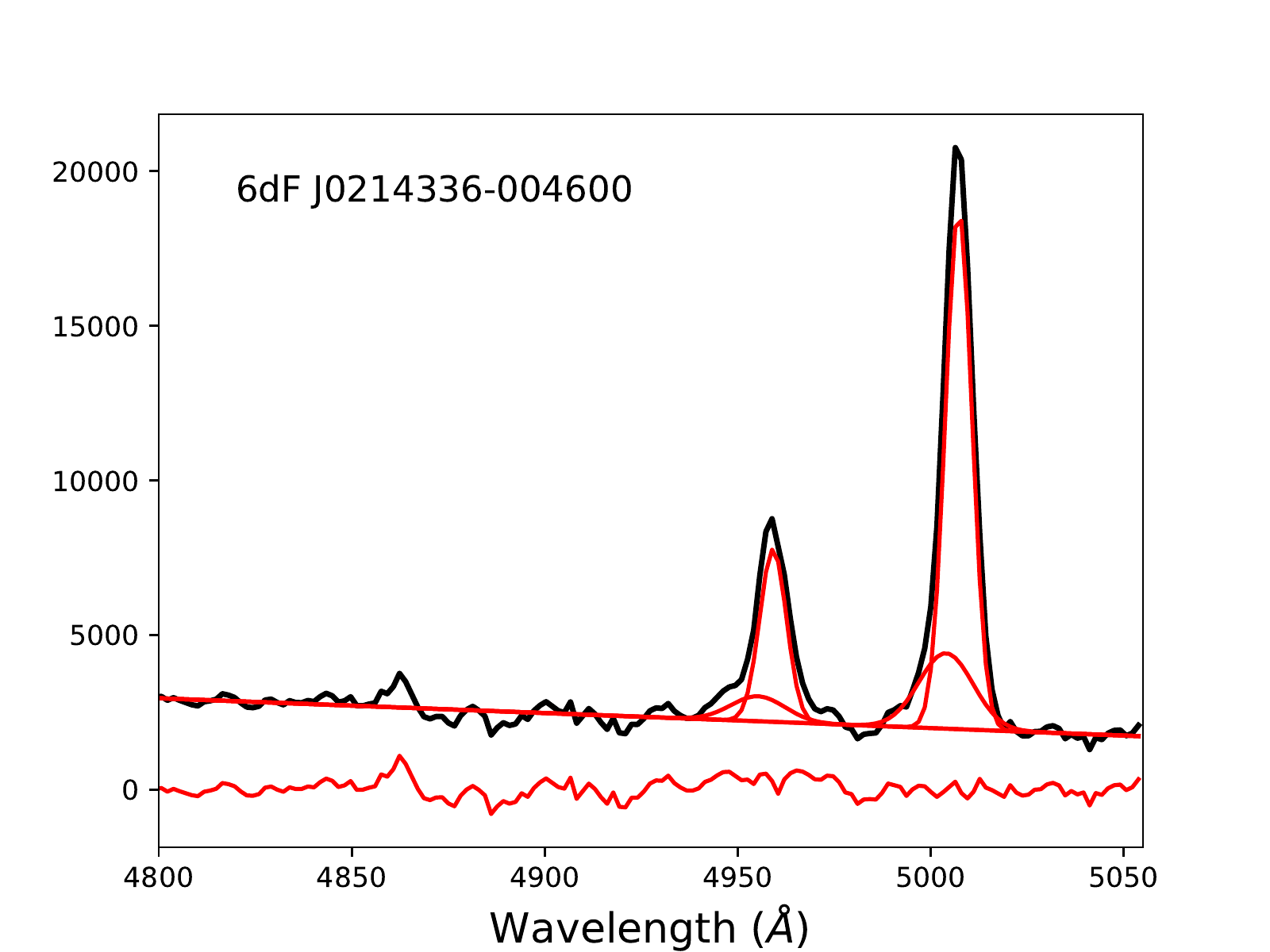}
\caption{Gaussian
decomposition of the [OIII]$\lambda\lambda$4959,5007 emission lines. The individual components are shown in red lines and the residuals are plotted at the bottom of each panel. Fluxes are given in arbitrary units.}
\label{fig:fits1}      
\end{centering}
\end{figure*}

\begin{figure*}
    \includegraphics[width=0.49\textwidth,height=0.25\textwidth]{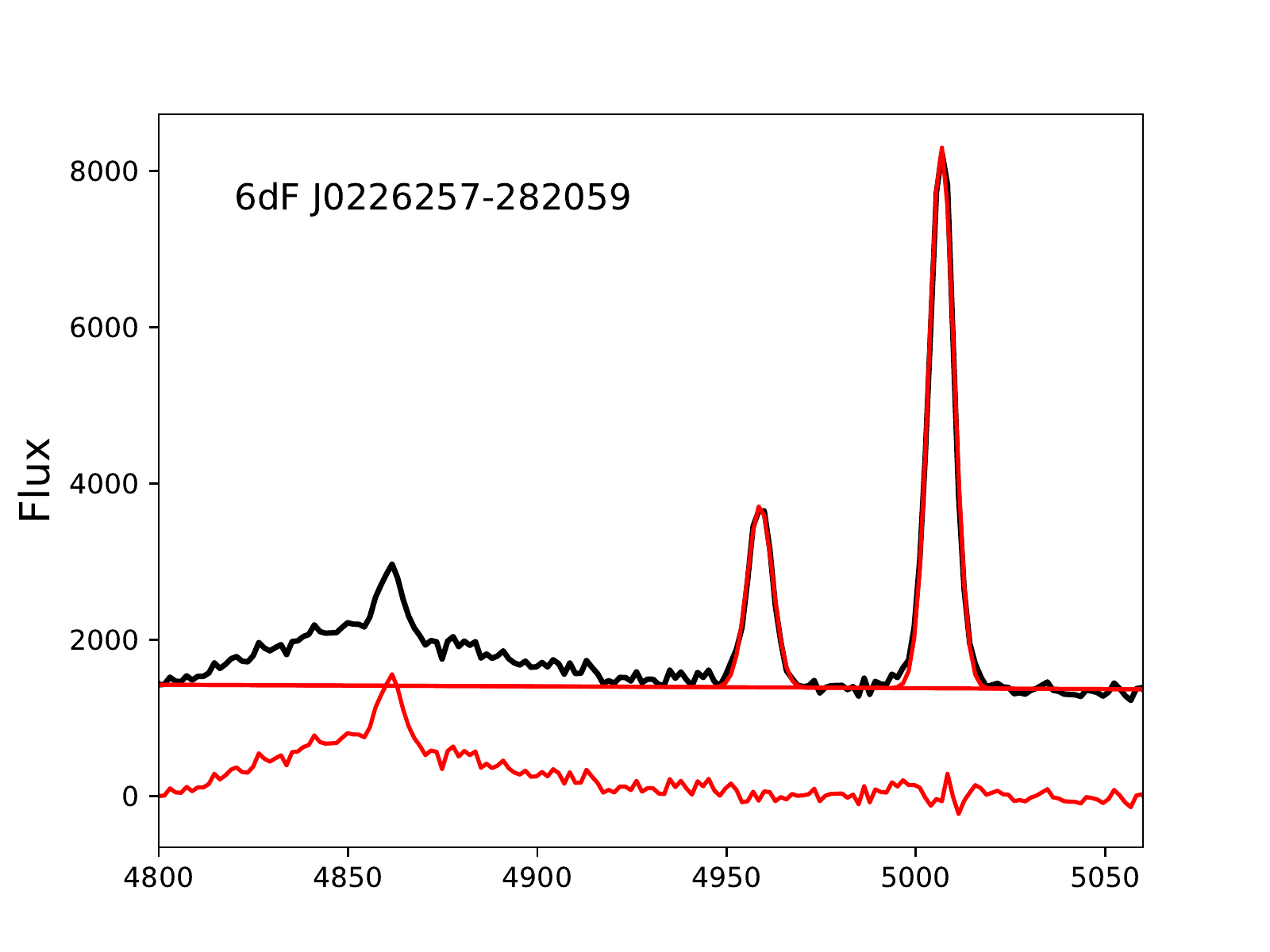}
    \includegraphics[width=0.49\textwidth,height=0.25\textwidth]{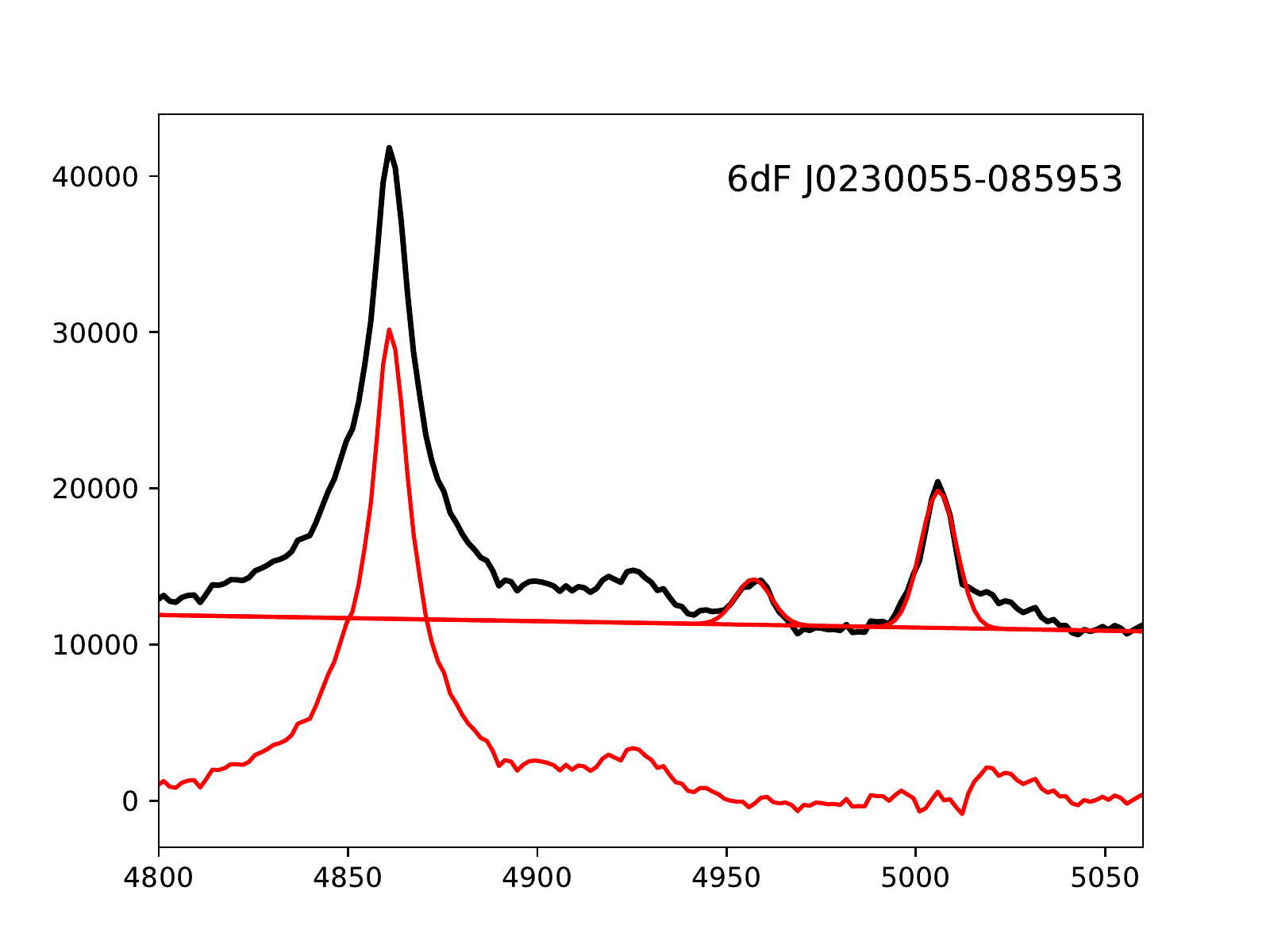}
    \includegraphics[width=0.49\textwidth,height=0.25\textwidth]{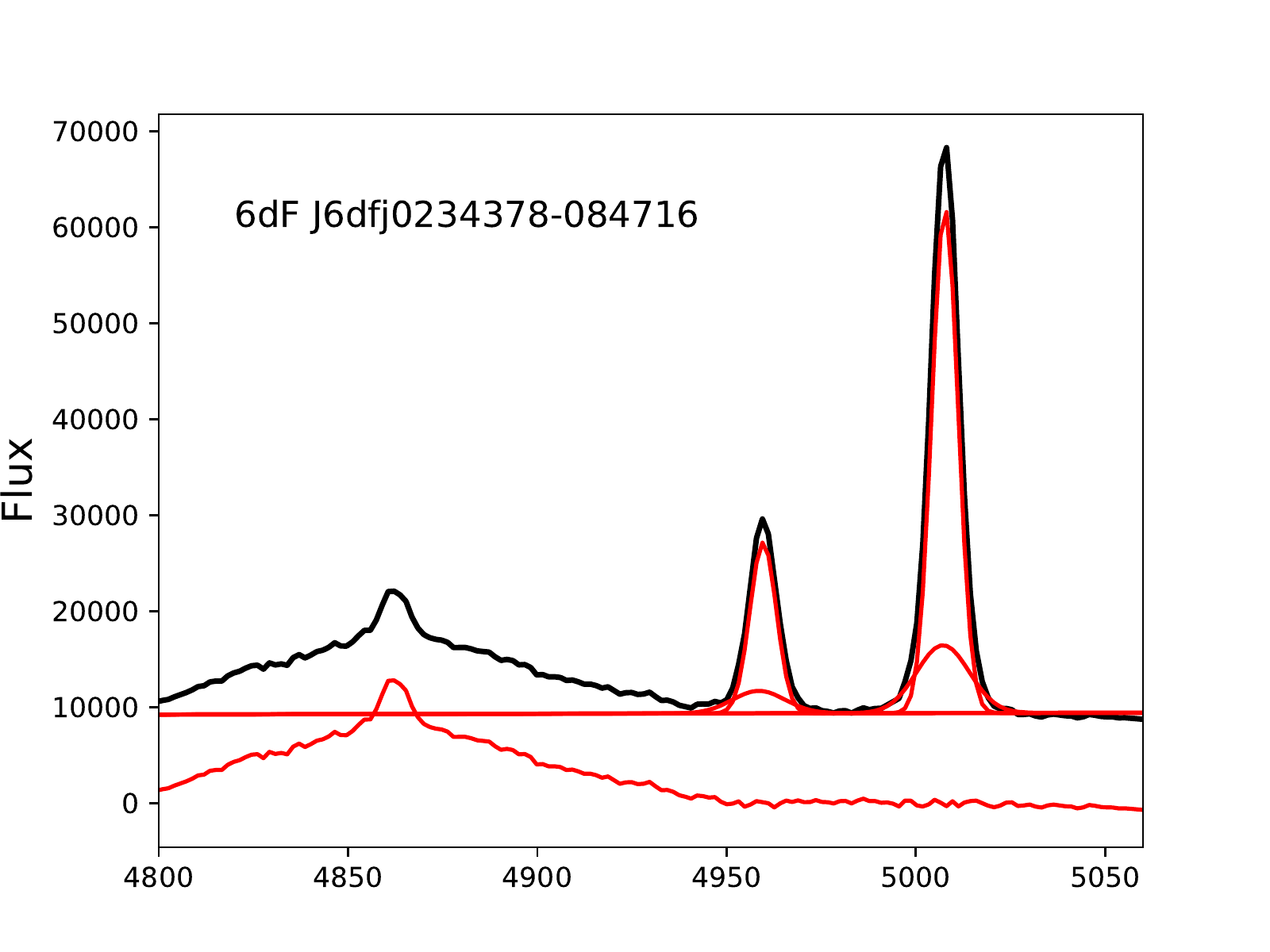}
    \includegraphics[width=0.49\textwidth,height=0.25\textwidth]{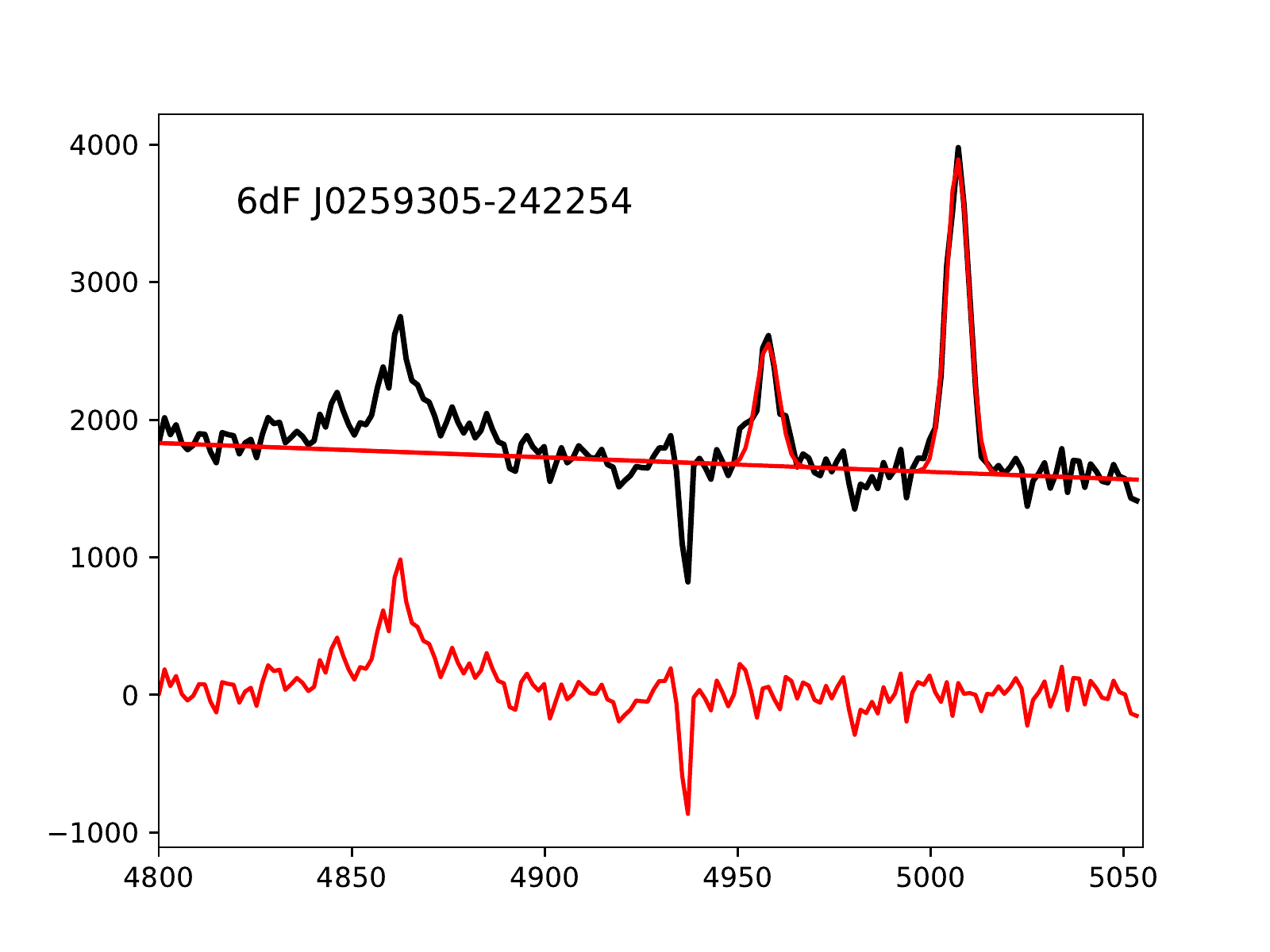}
    \includegraphics[width=0.49\textwidth,height=0.25\textwidth]{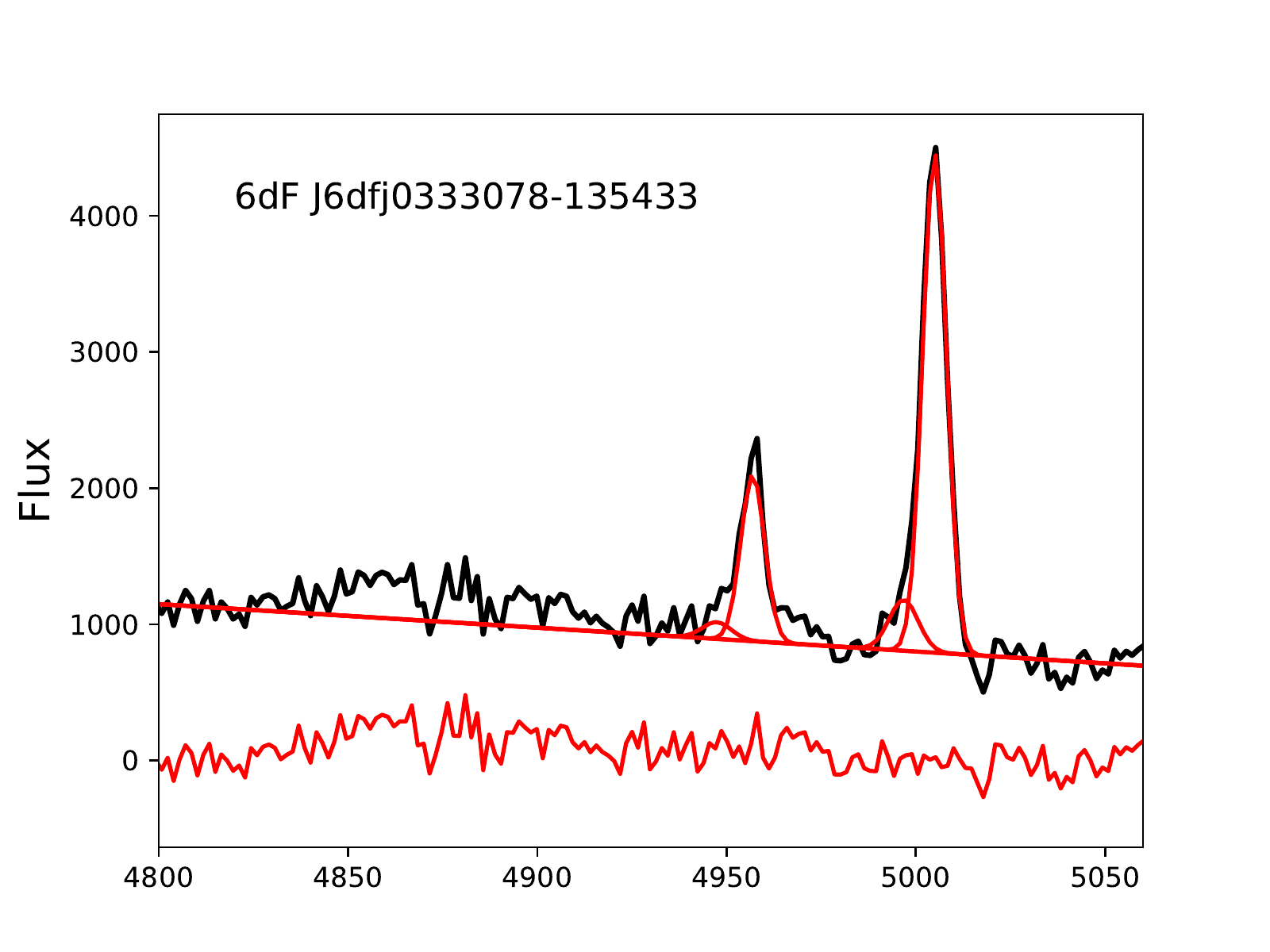}
    \includegraphics[width=0.49\textwidth,height=0.25\textwidth]{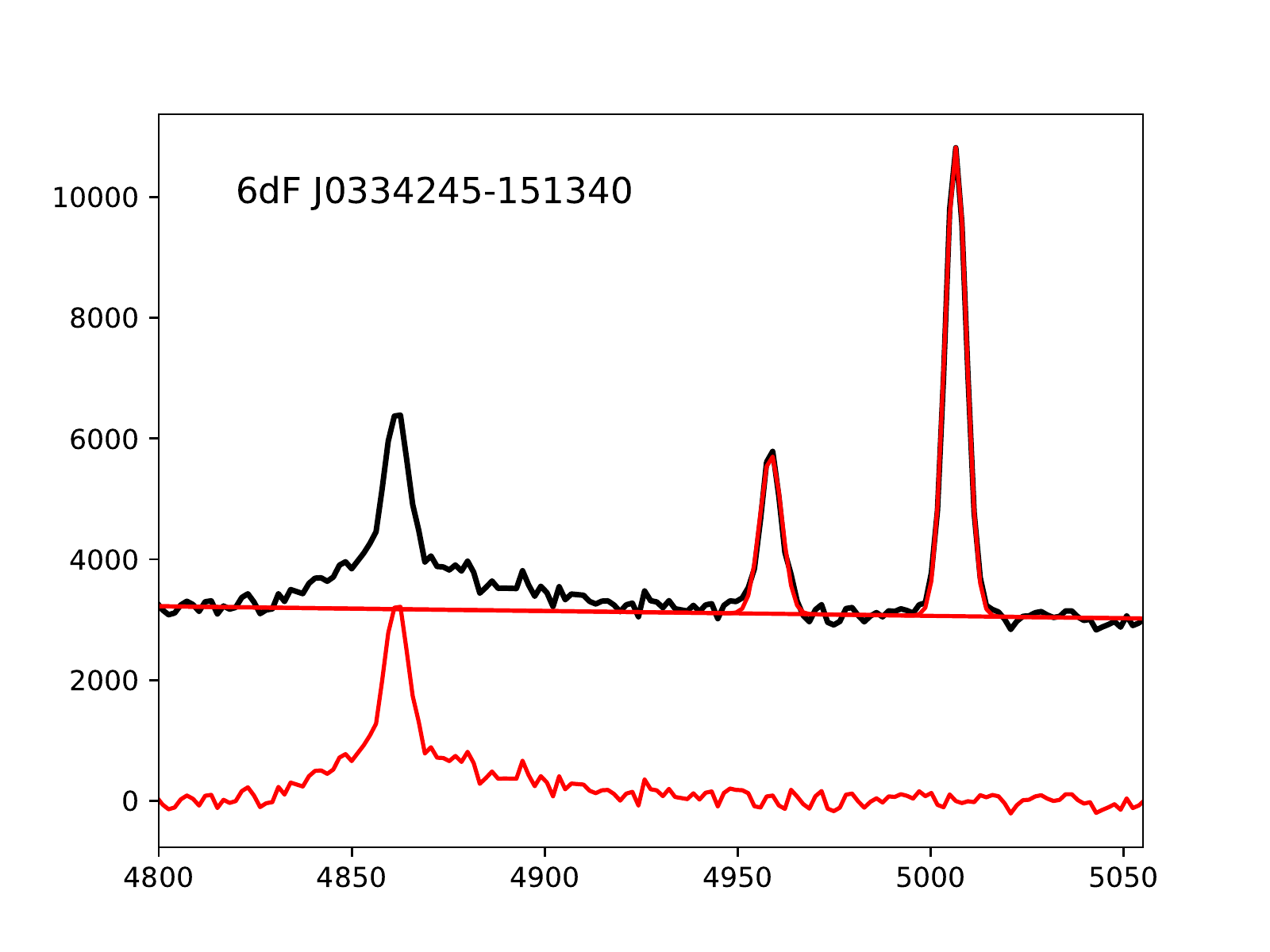}
    \includegraphics[width=0.49\textwidth,height=0.25\textwidth]{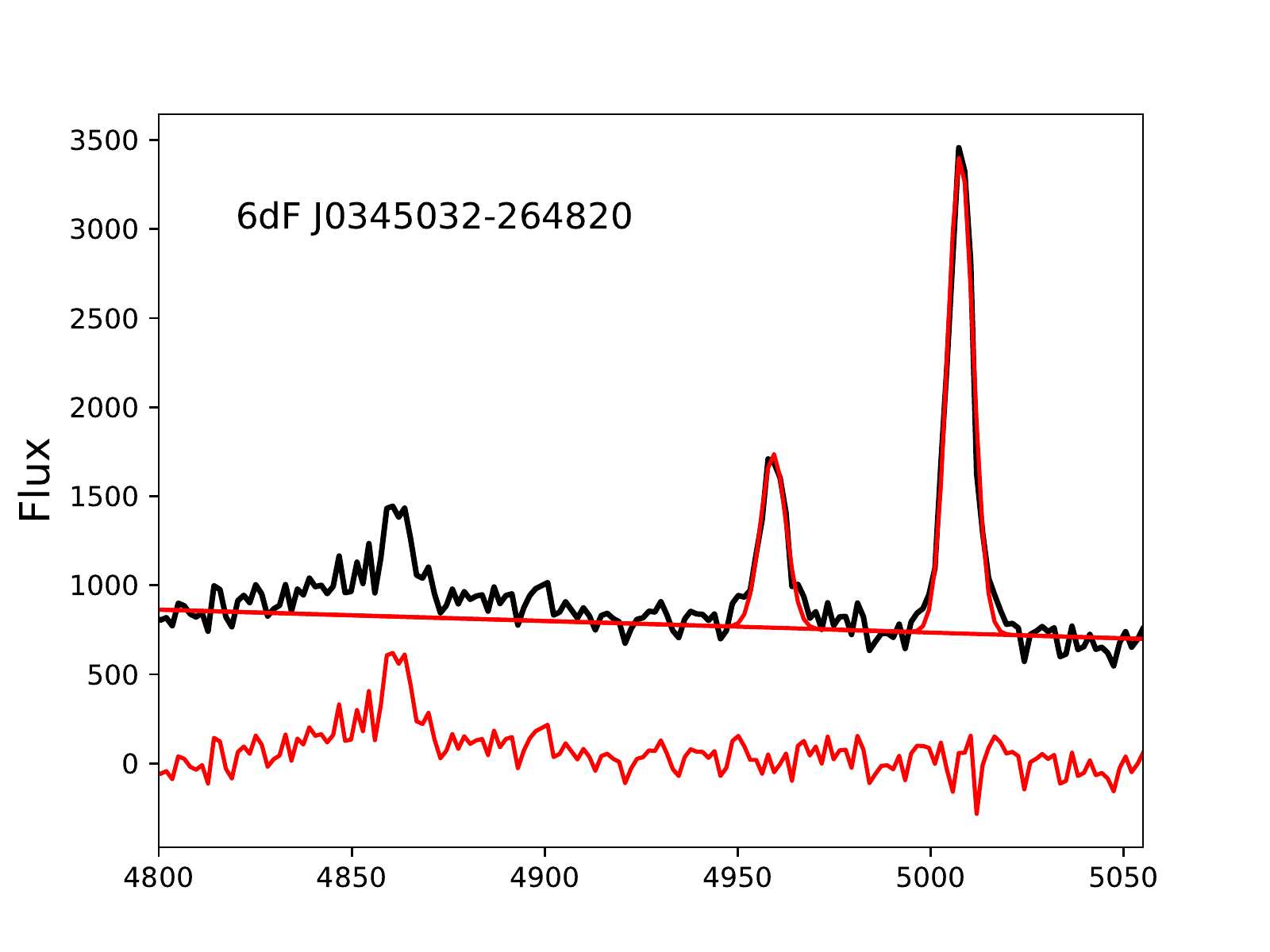}
    \includegraphics[width=0.49\textwidth,height=0.25\textwidth]{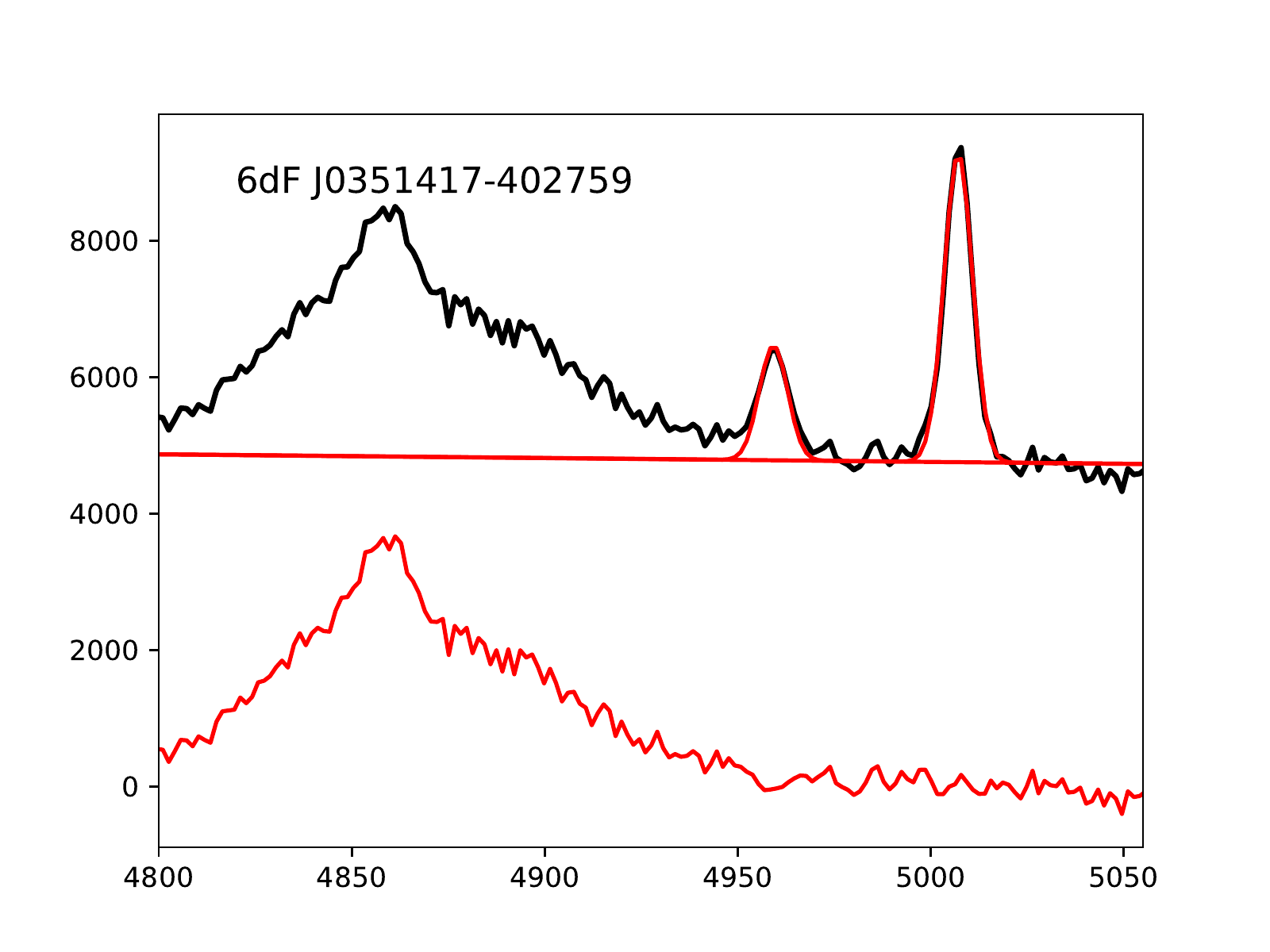}
    \includegraphics[width=0.49\textwidth,height=0.25\textwidth]{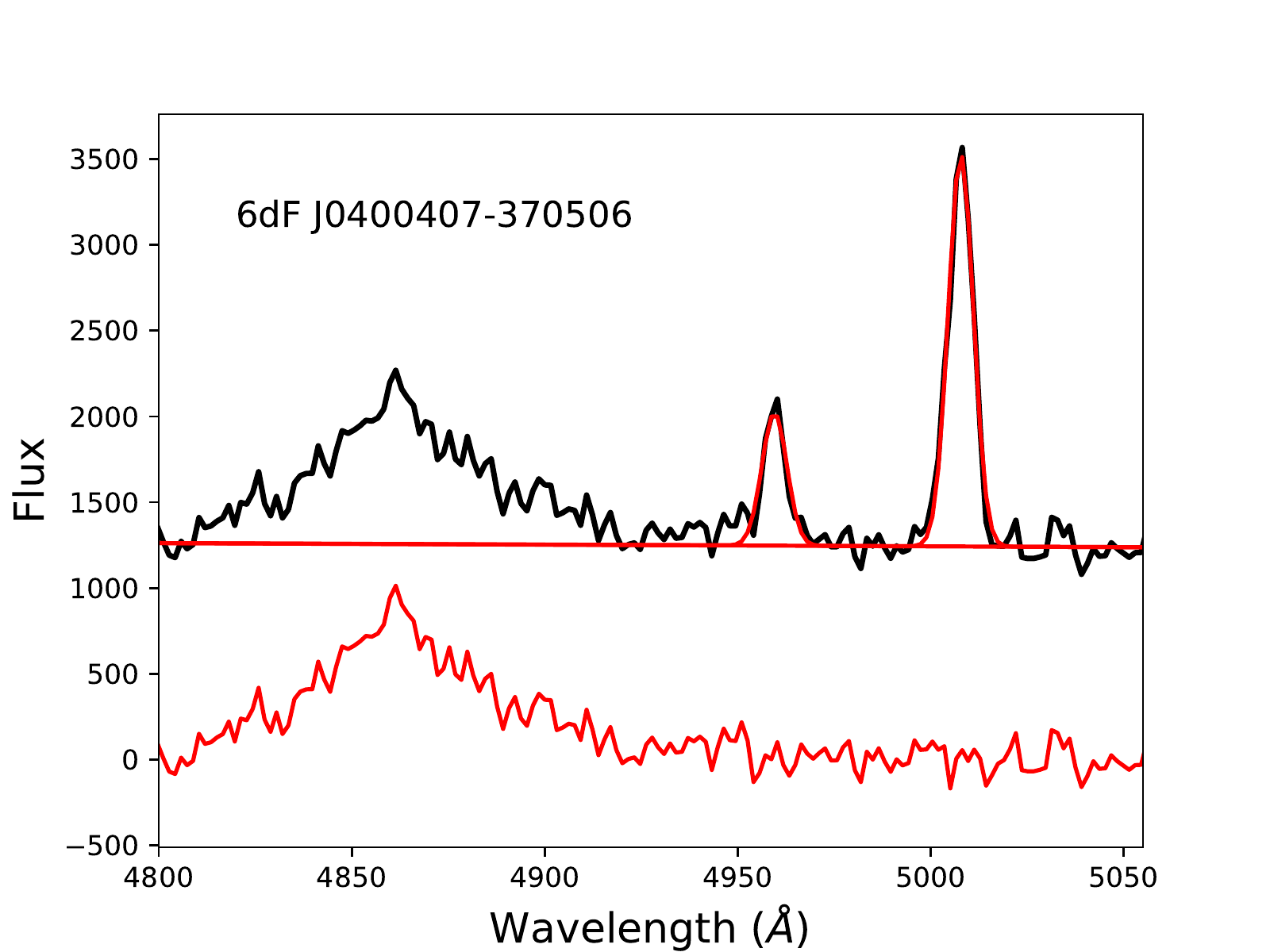}
    \includegraphics[width=0.49\textwidth,height=0.25\textwidth]{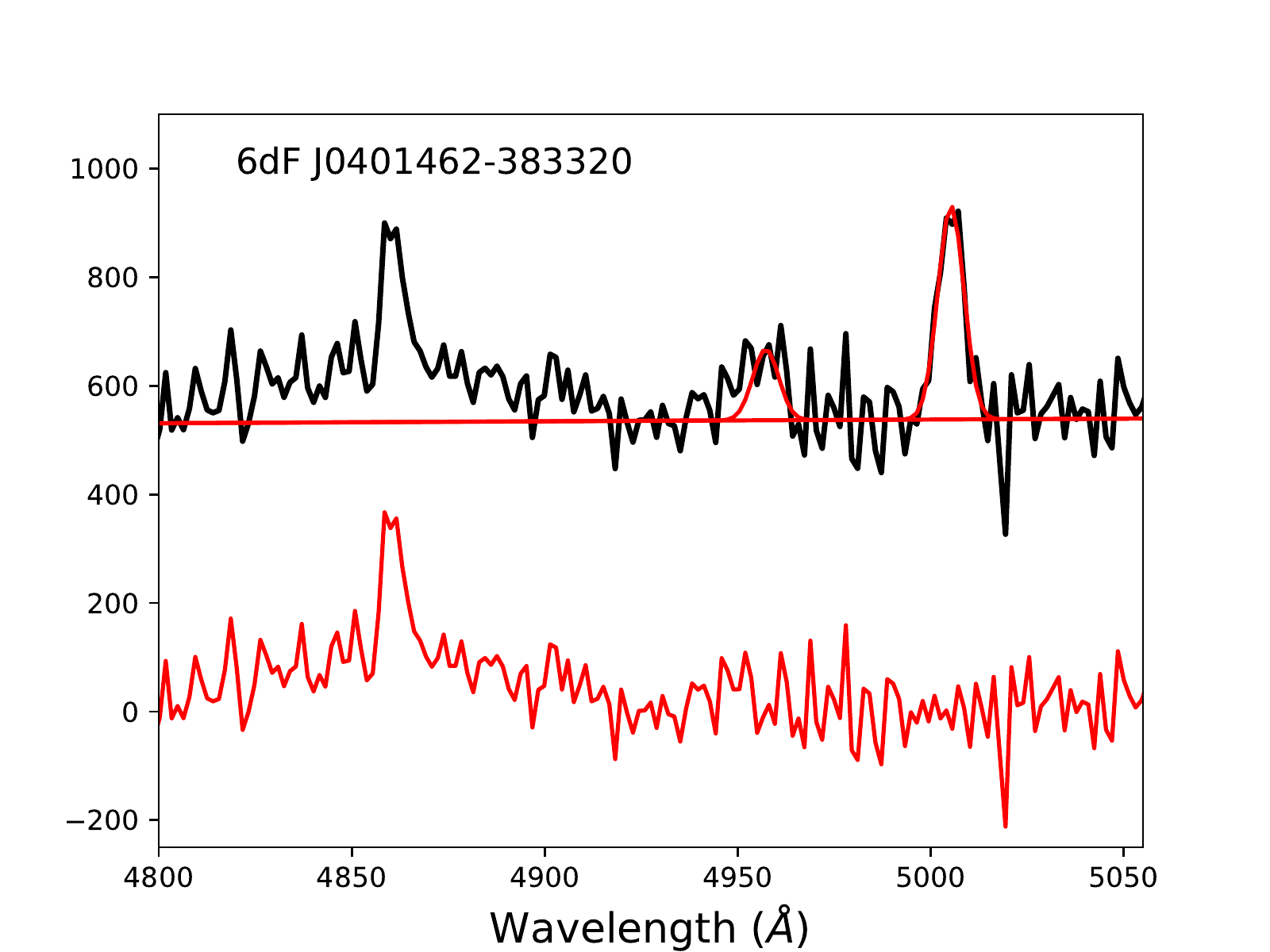}
    \contcaption{}
\label{}
\end{figure*}

\begin{figure*}
    \includegraphics[width=0.49\textwidth,height=0.25\textwidth]{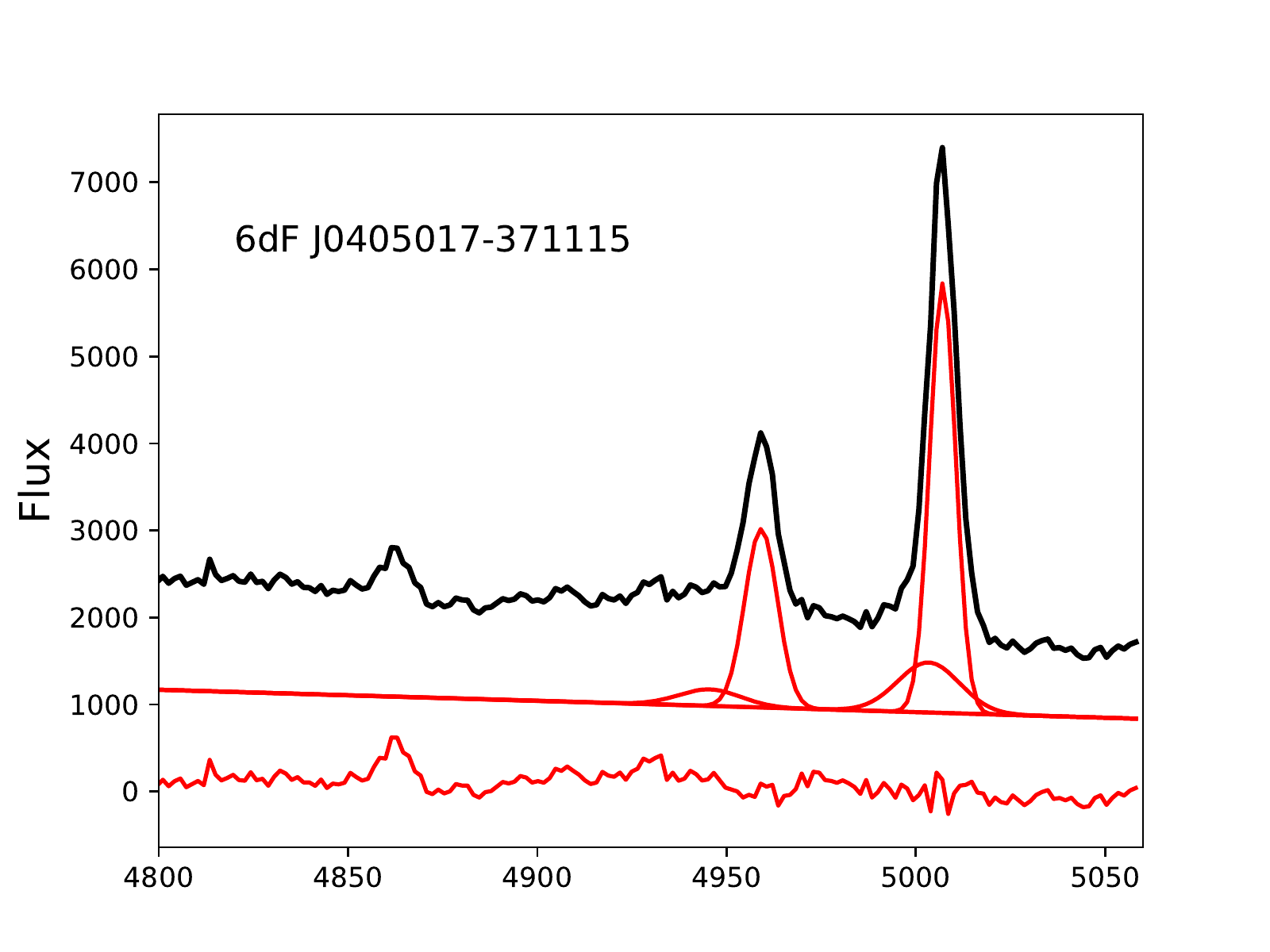}
    \includegraphics[width=0.49\textwidth,height=0.25\textwidth]{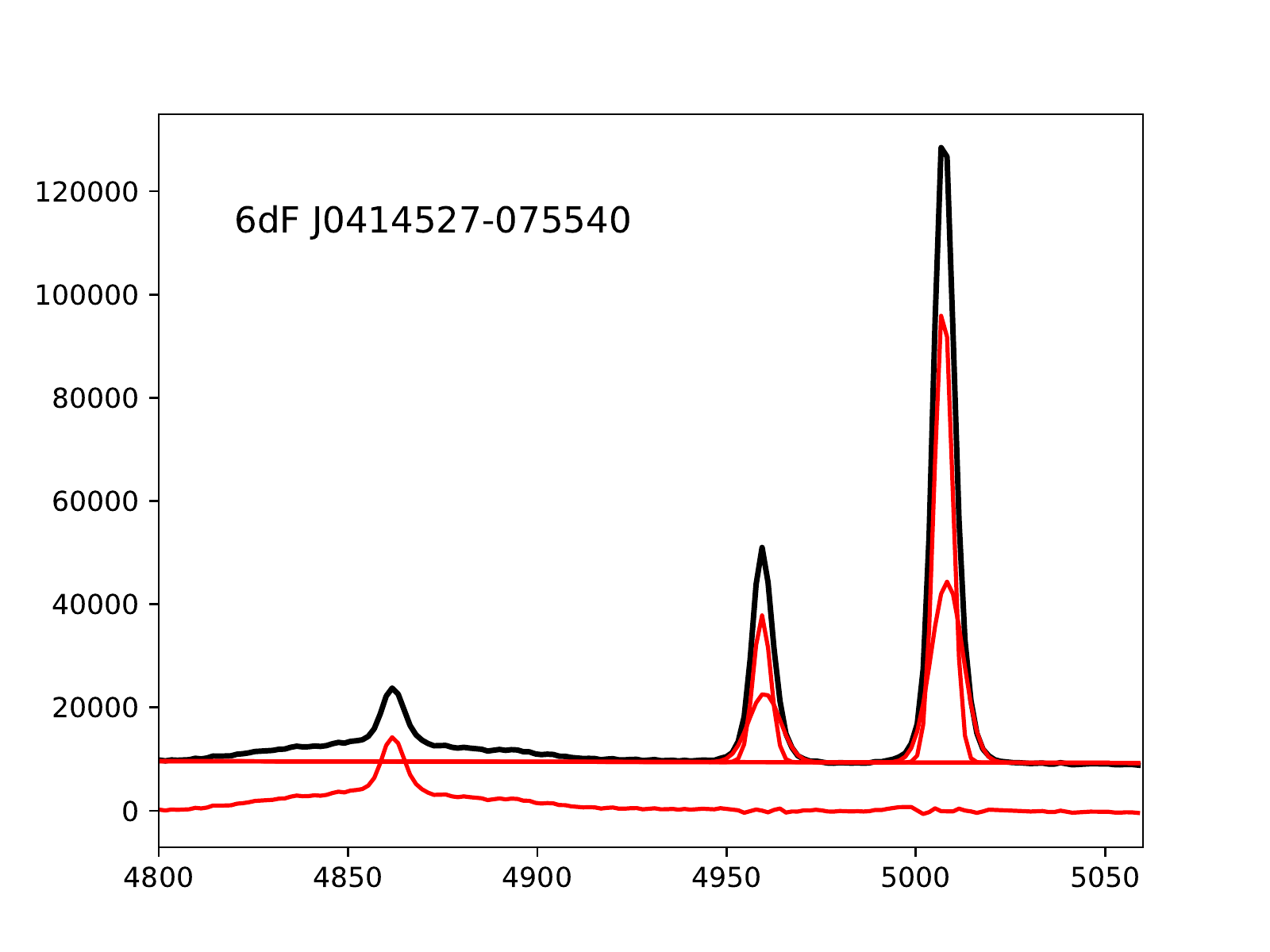}
    \includegraphics[width=0.49\textwidth,height=0.25\textwidth]{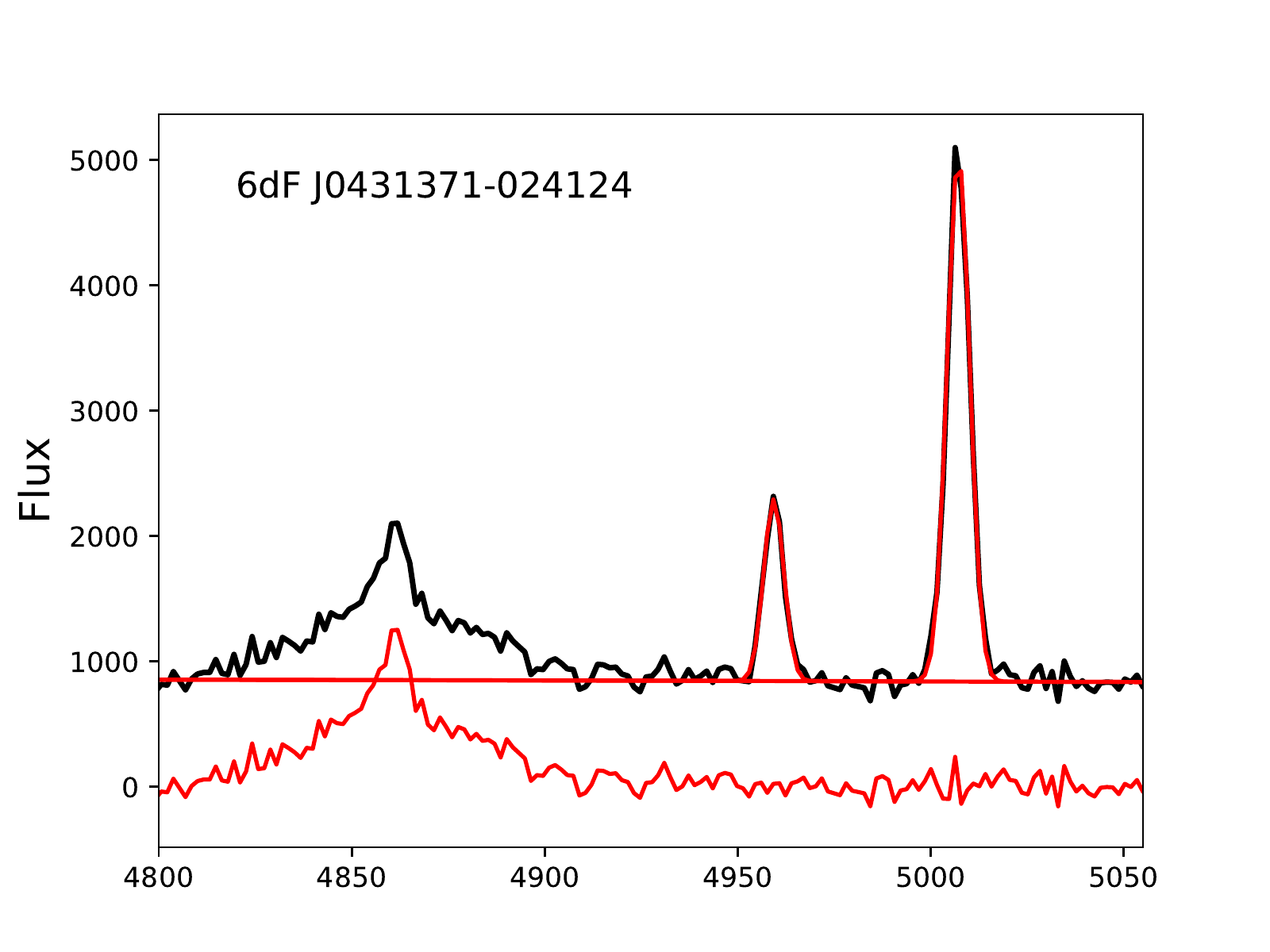}
    \includegraphics[width=0.49\textwidth,height=0.25\textwidth]{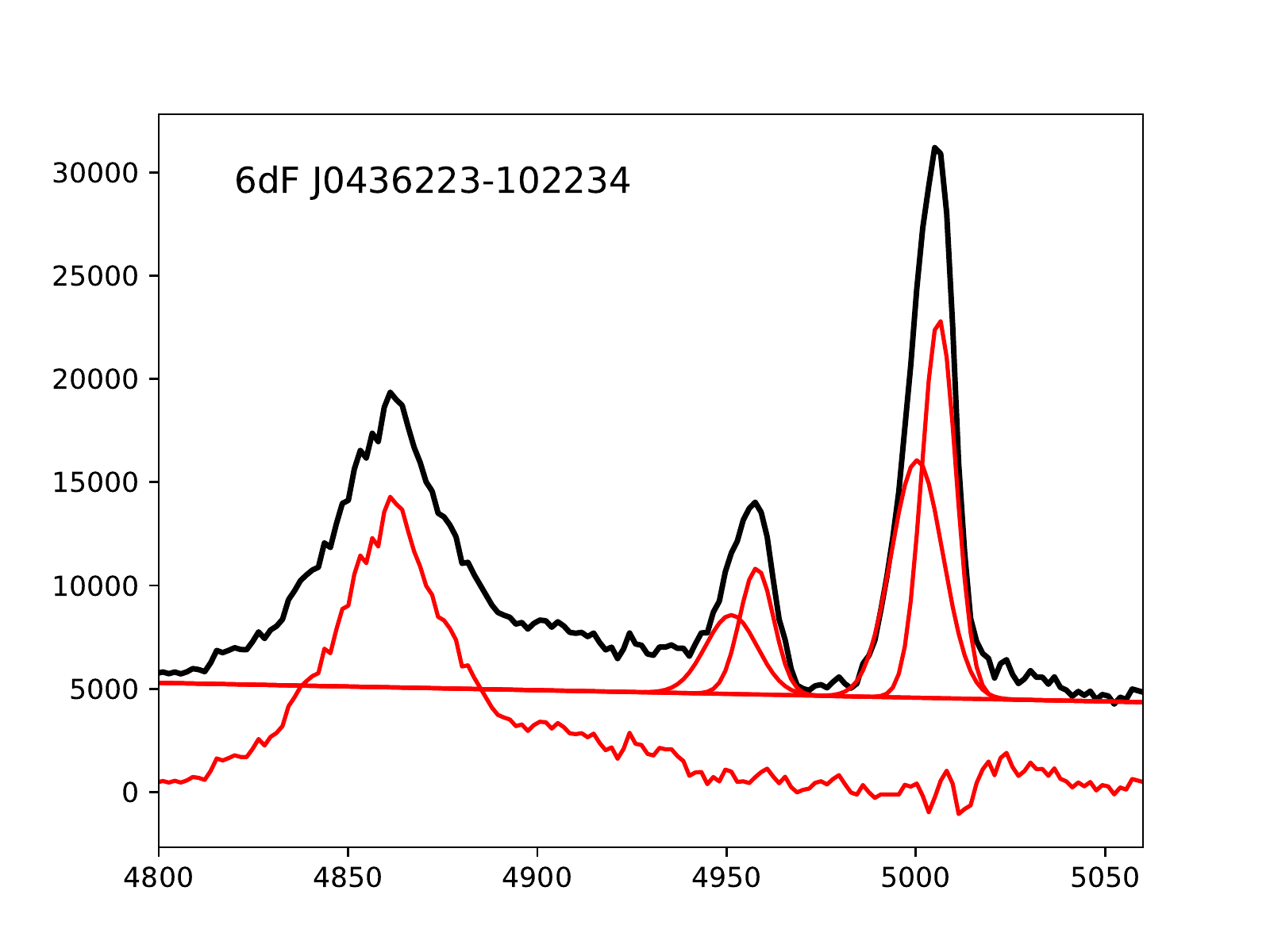}
    \includegraphics[width=0.49\textwidth,height=0.25\textwidth]{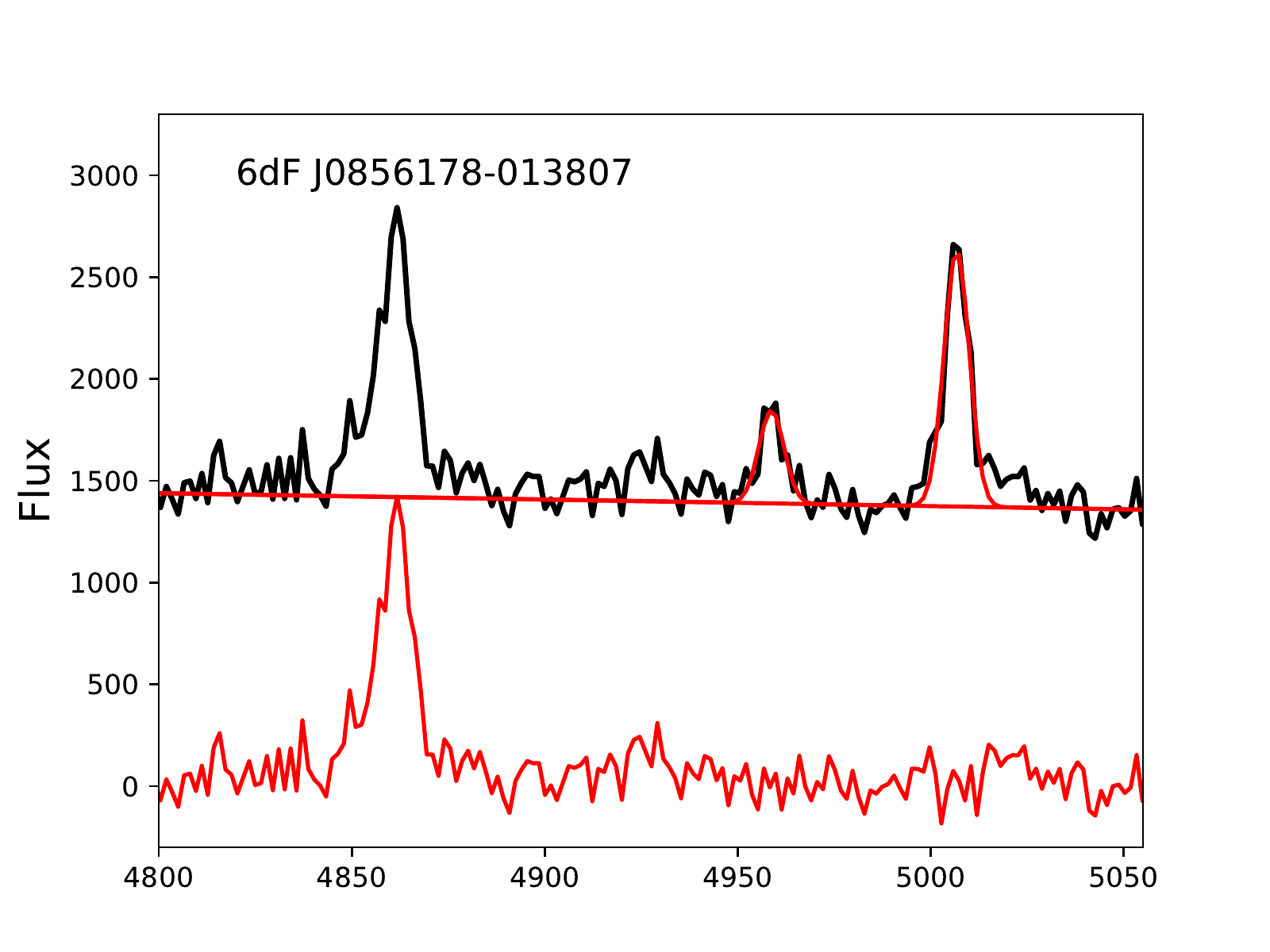}
    \includegraphics[width=0.49\textwidth,height=0.25\textwidth]{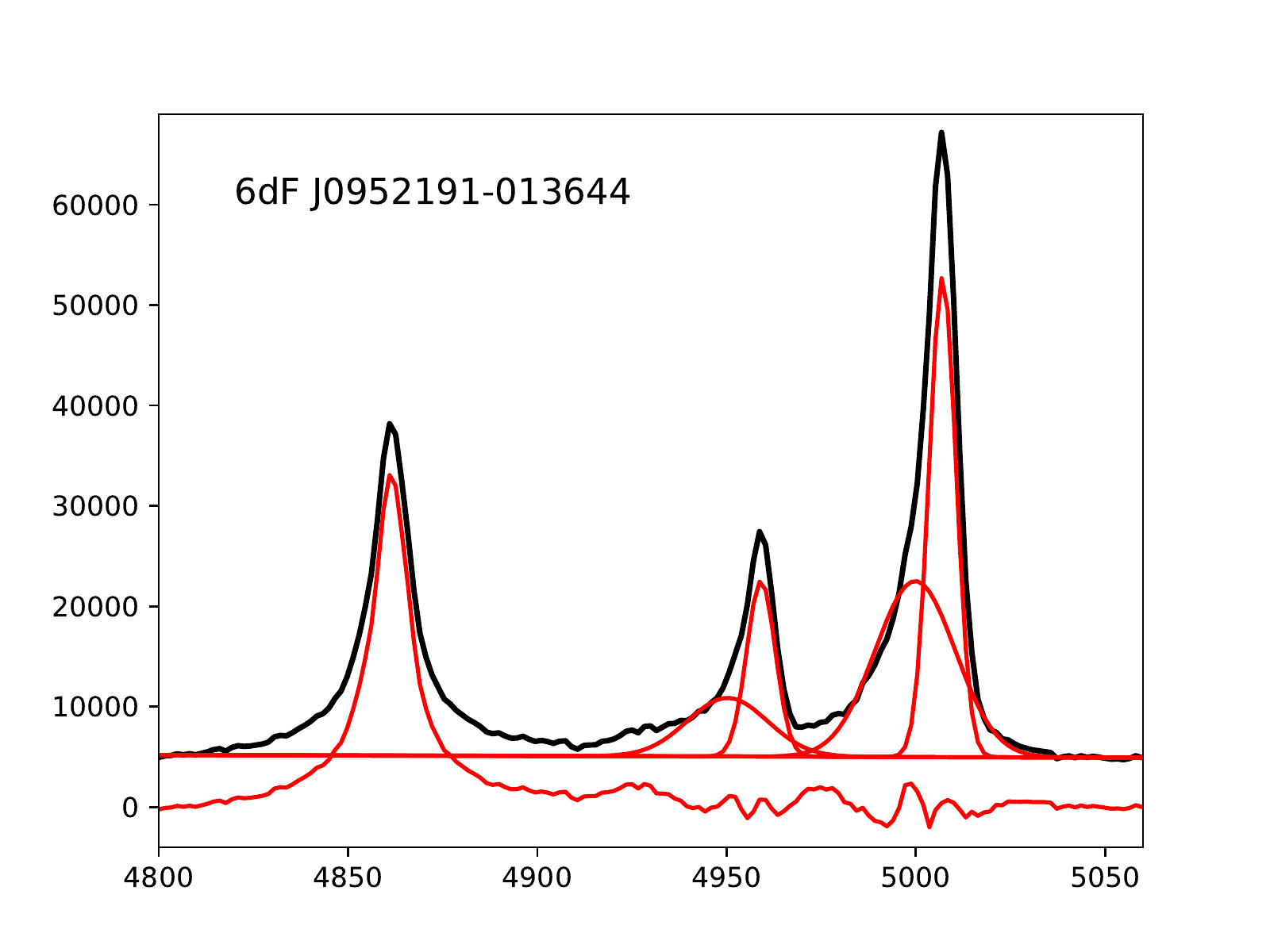}
    \includegraphics[width=0.49\textwidth,height=0.25\textwidth]{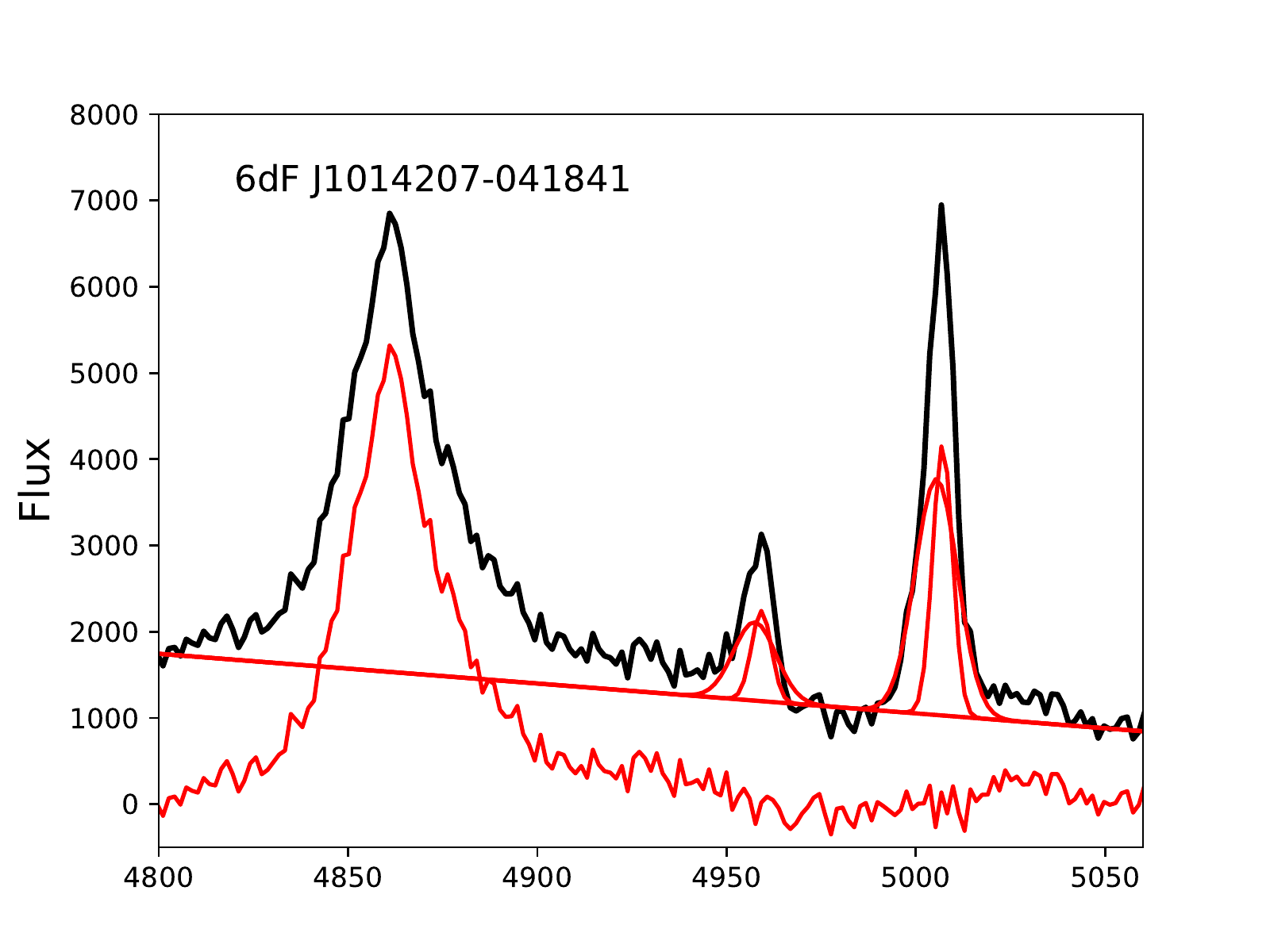}
    \includegraphics[width=0.49\textwidth,height=0.25\textwidth]{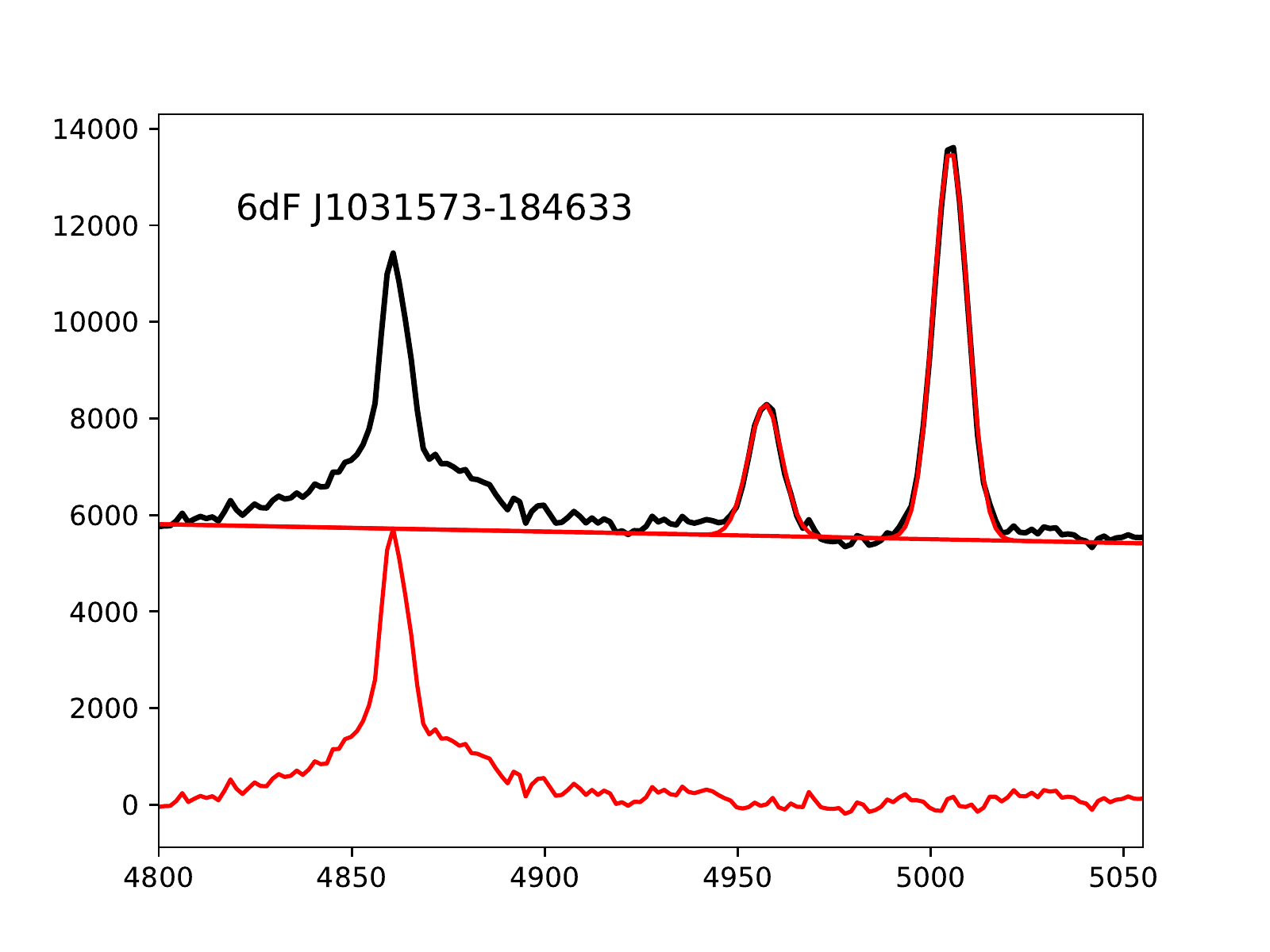}
    \includegraphics[width=0.49\textwidth,height=0.25\textwidth]{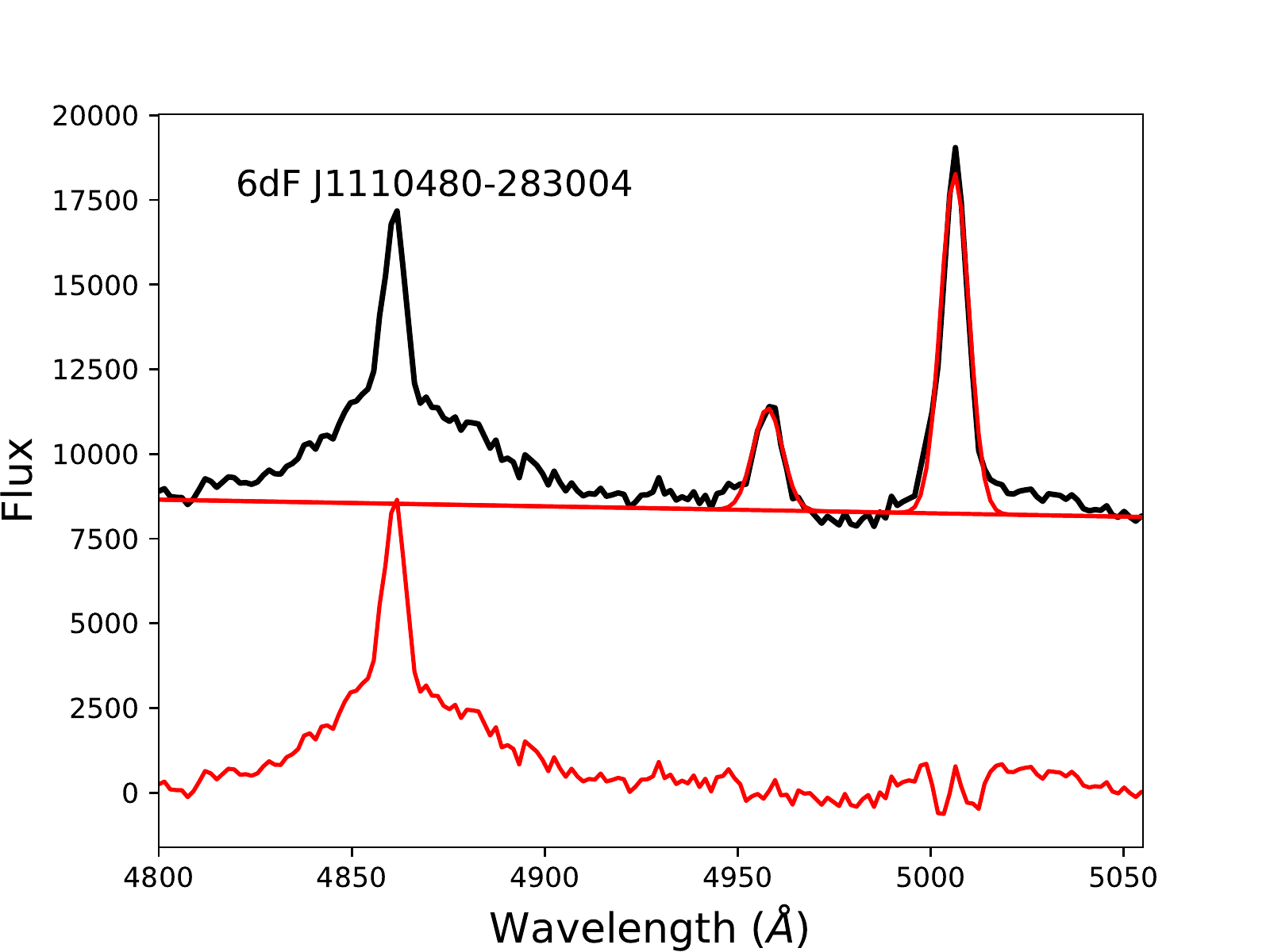}
    \includegraphics[width=0.49\textwidth,height=0.25\textwidth]{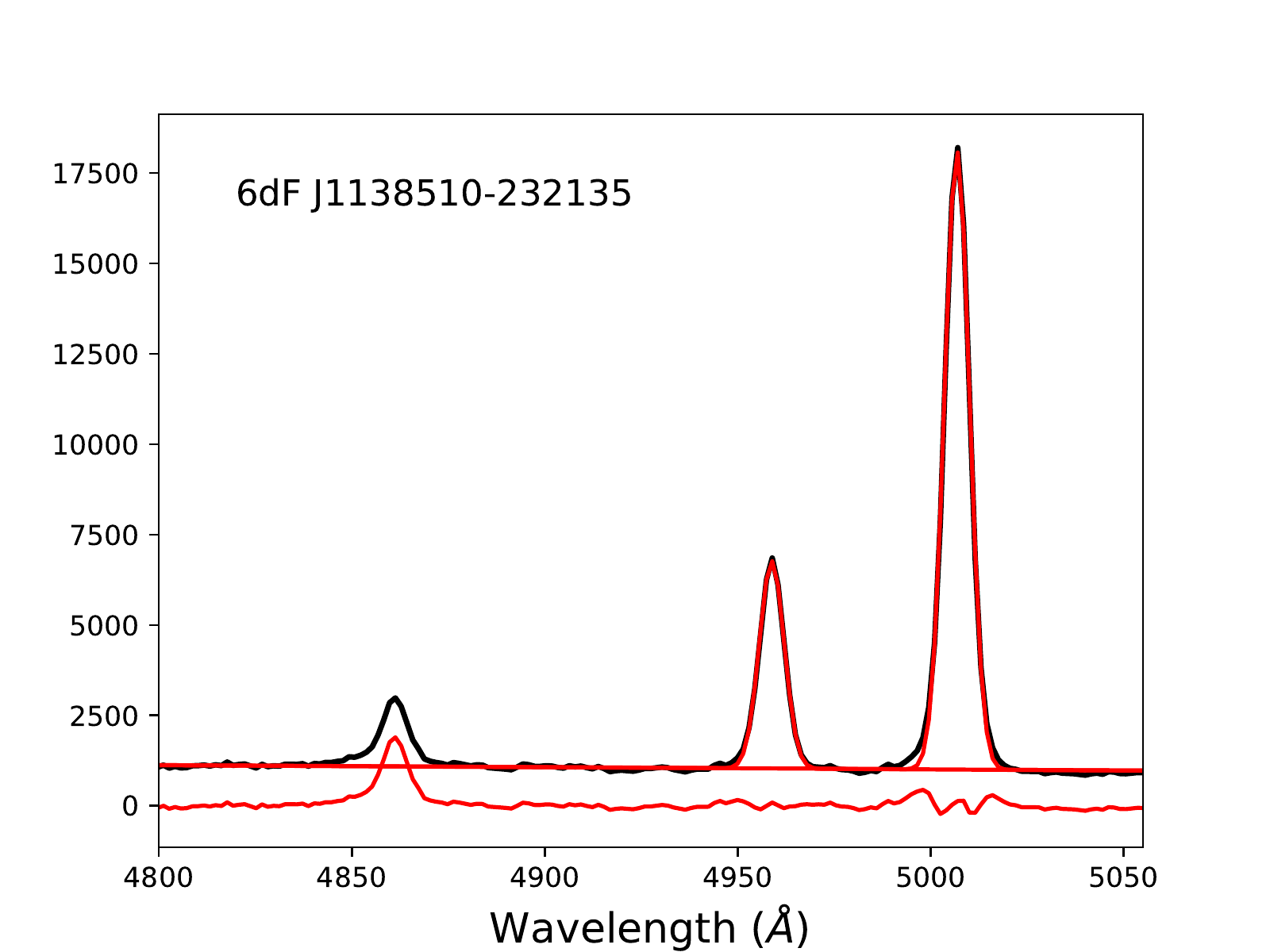}
\contcaption{}
\label{}
\end{figure*}

\begin{figure*}
    \includegraphics[width=0.49\textwidth,height=0.25\textwidth]{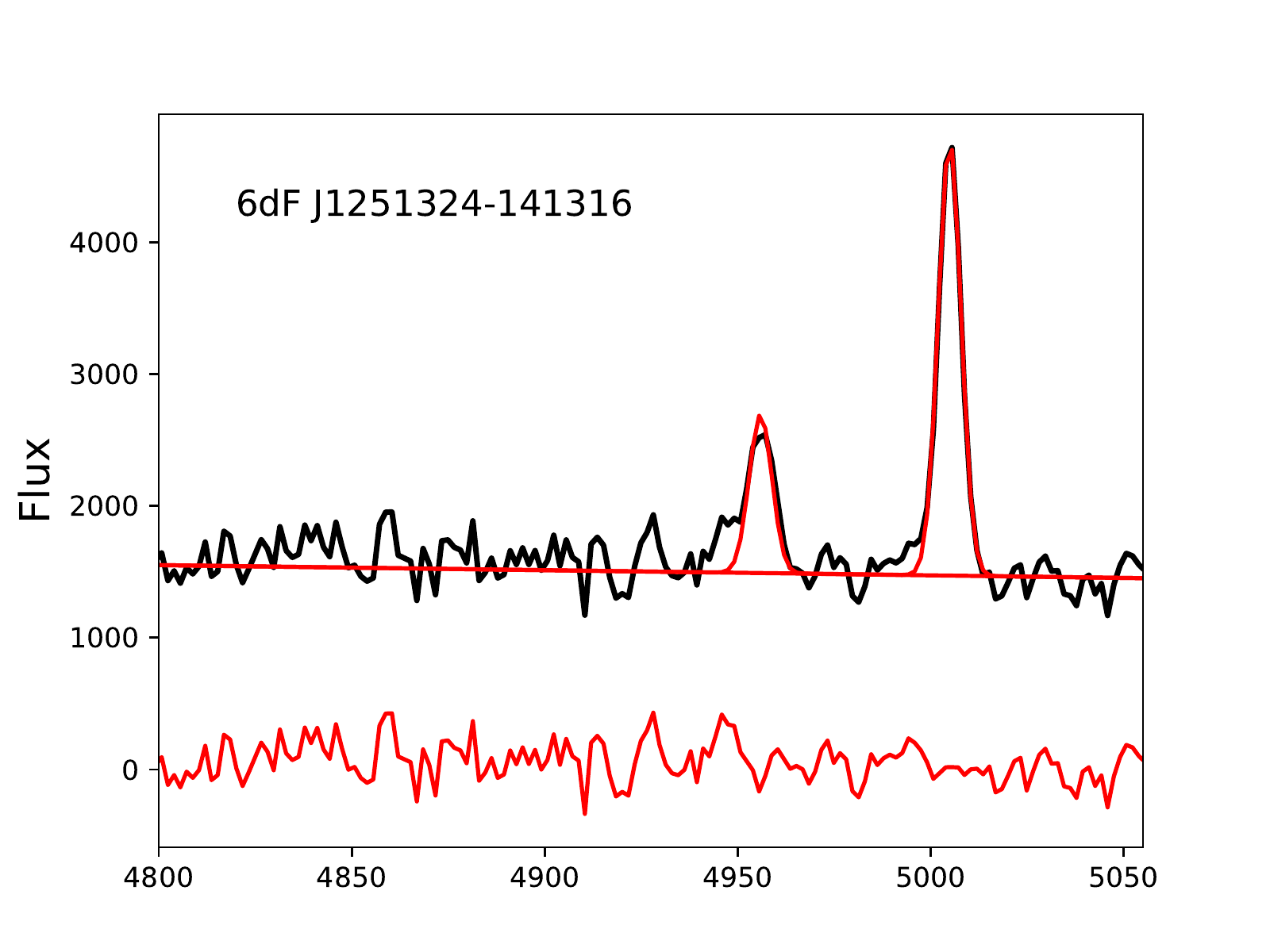} 
    \includegraphics[width=0.49\textwidth,height=0.25\textwidth]{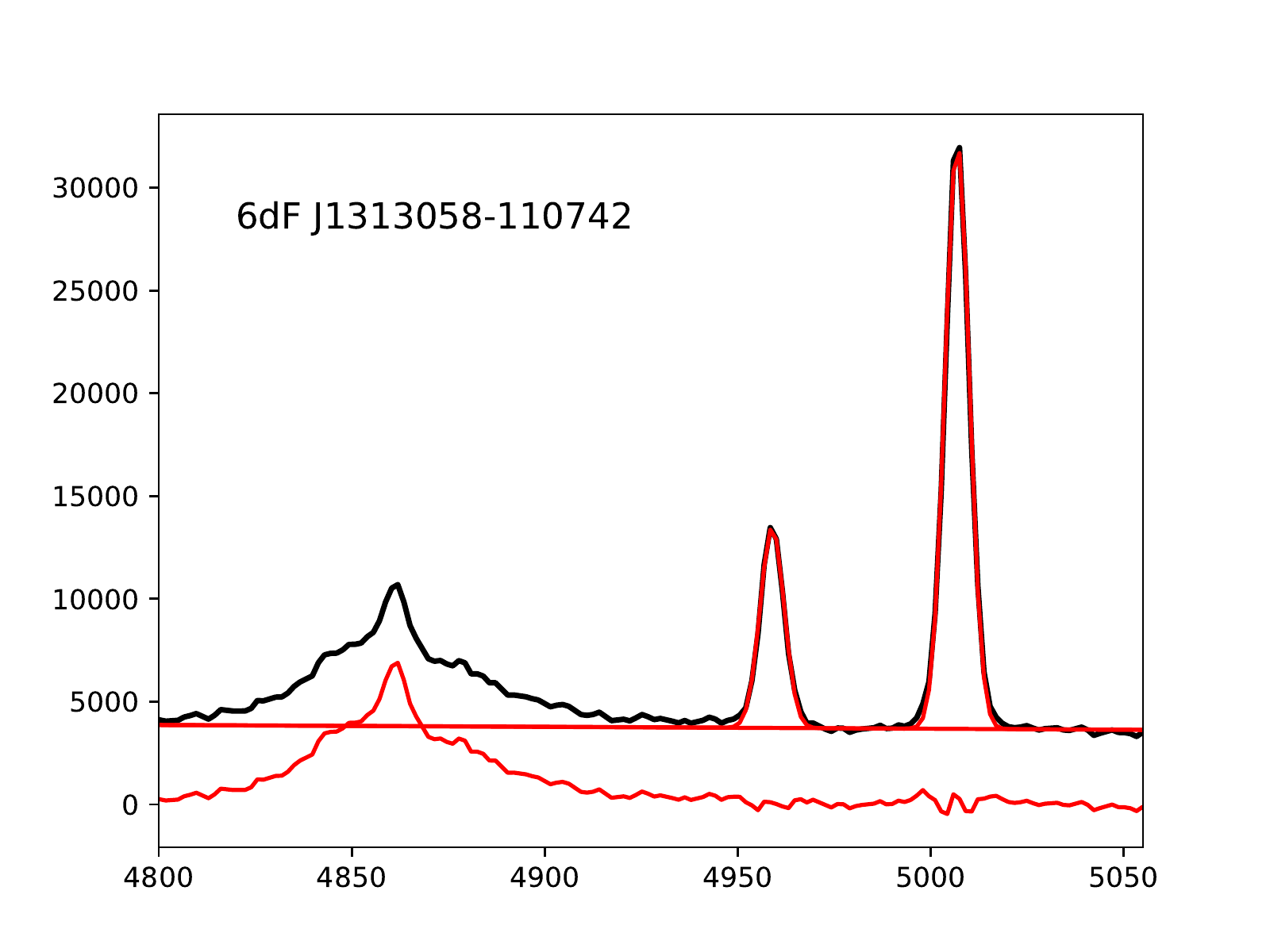} 
    \includegraphics[width=0.49\textwidth,height=0.25\textwidth]{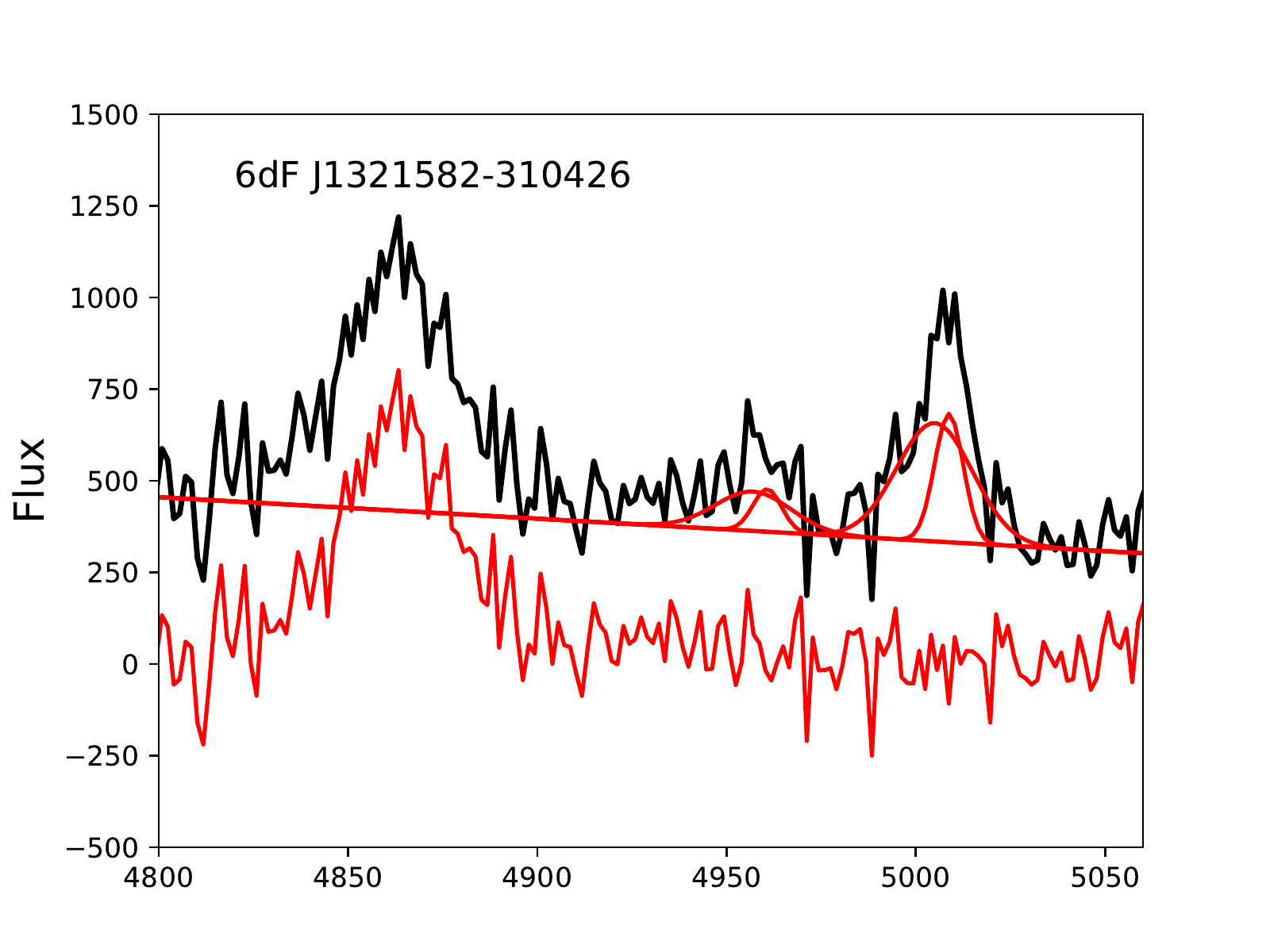}
    \includegraphics[width=0.49\textwidth,height=0.25\textwidth]{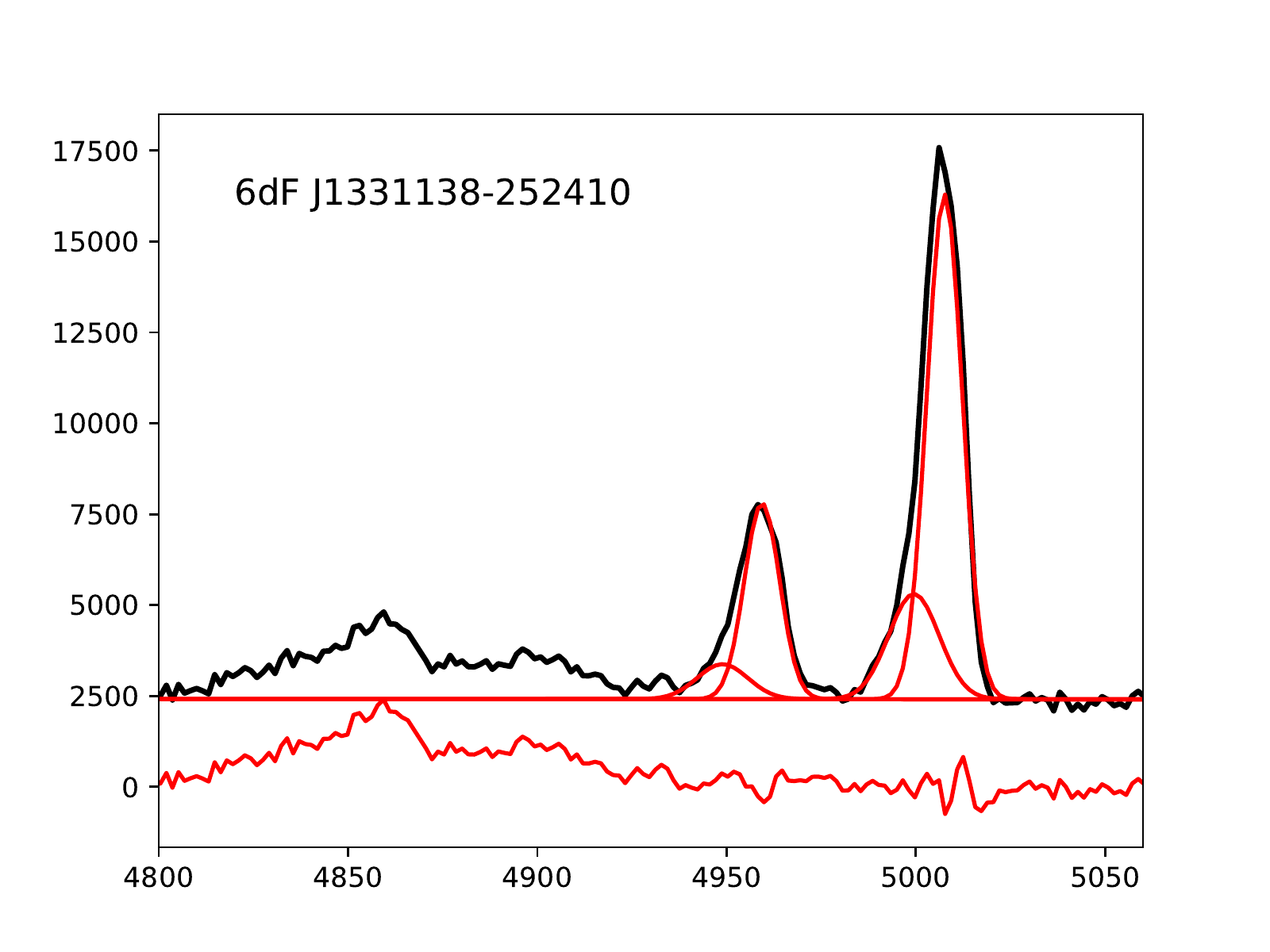}
    \includegraphics[width=0.49\textwidth,height=0.25\textwidth]{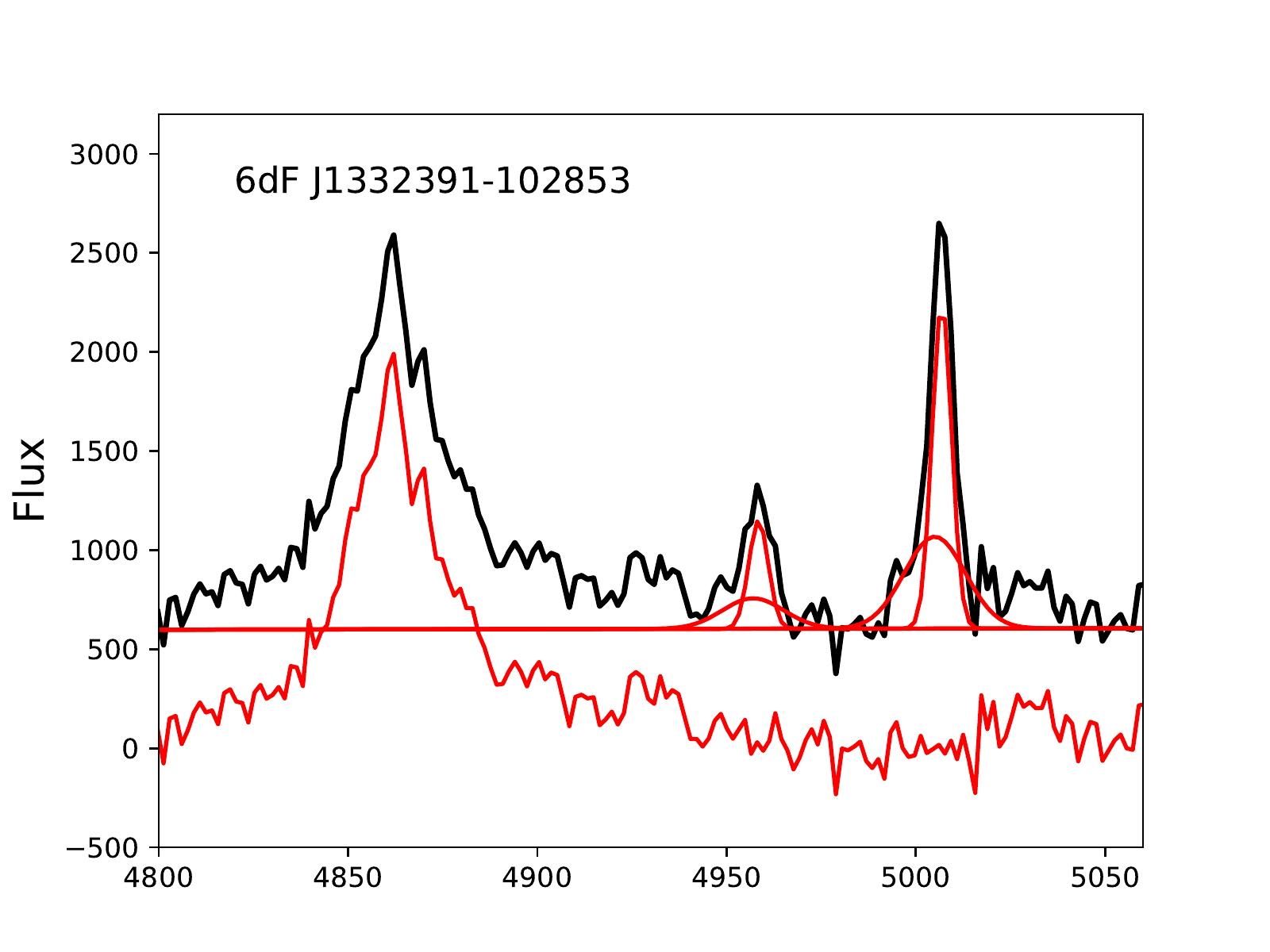}
    \includegraphics[width=0.49\textwidth,height=0.25\textwidth]{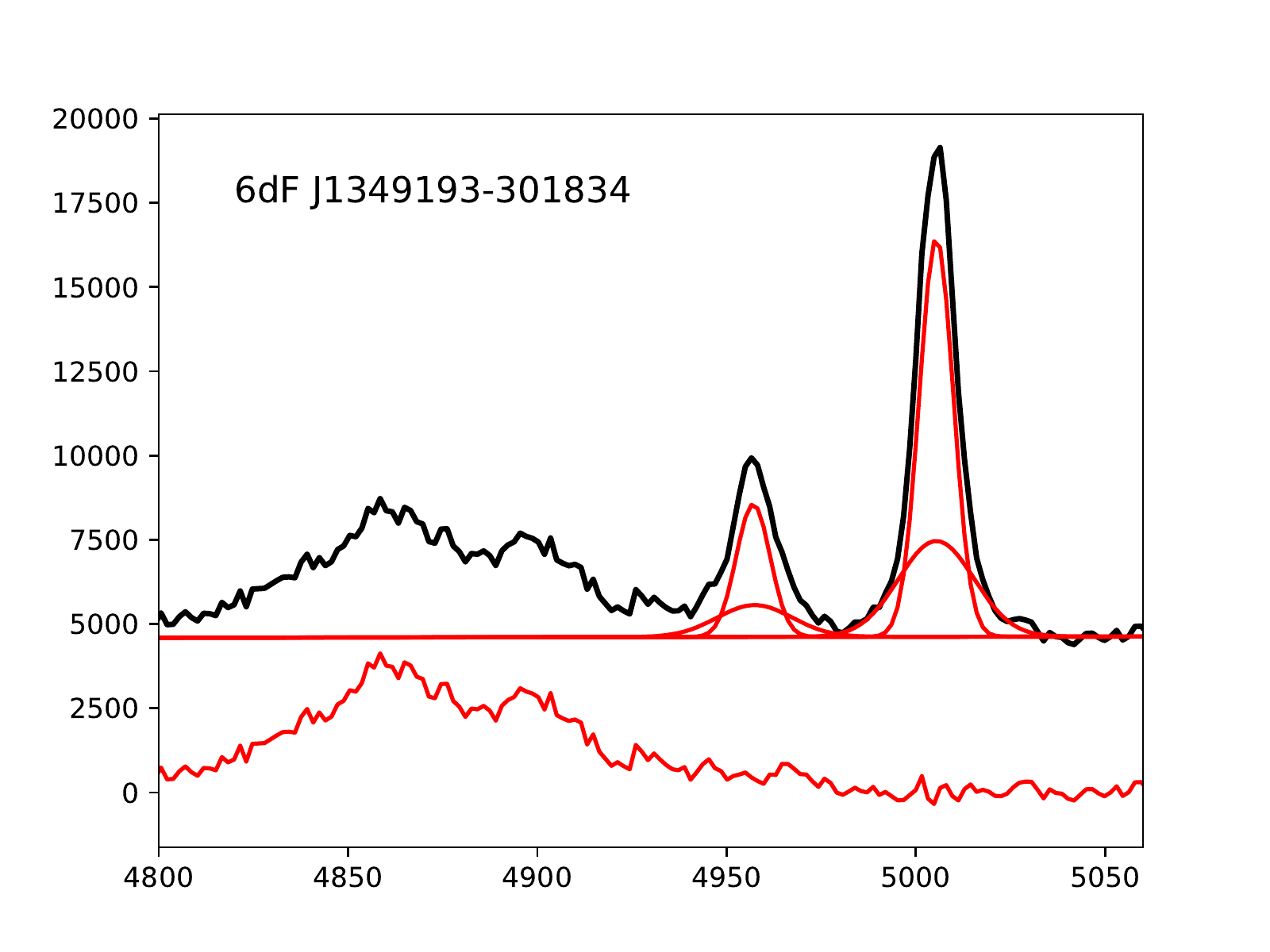}
    \includegraphics[width=0.49\textwidth,height=0.25\textwidth]{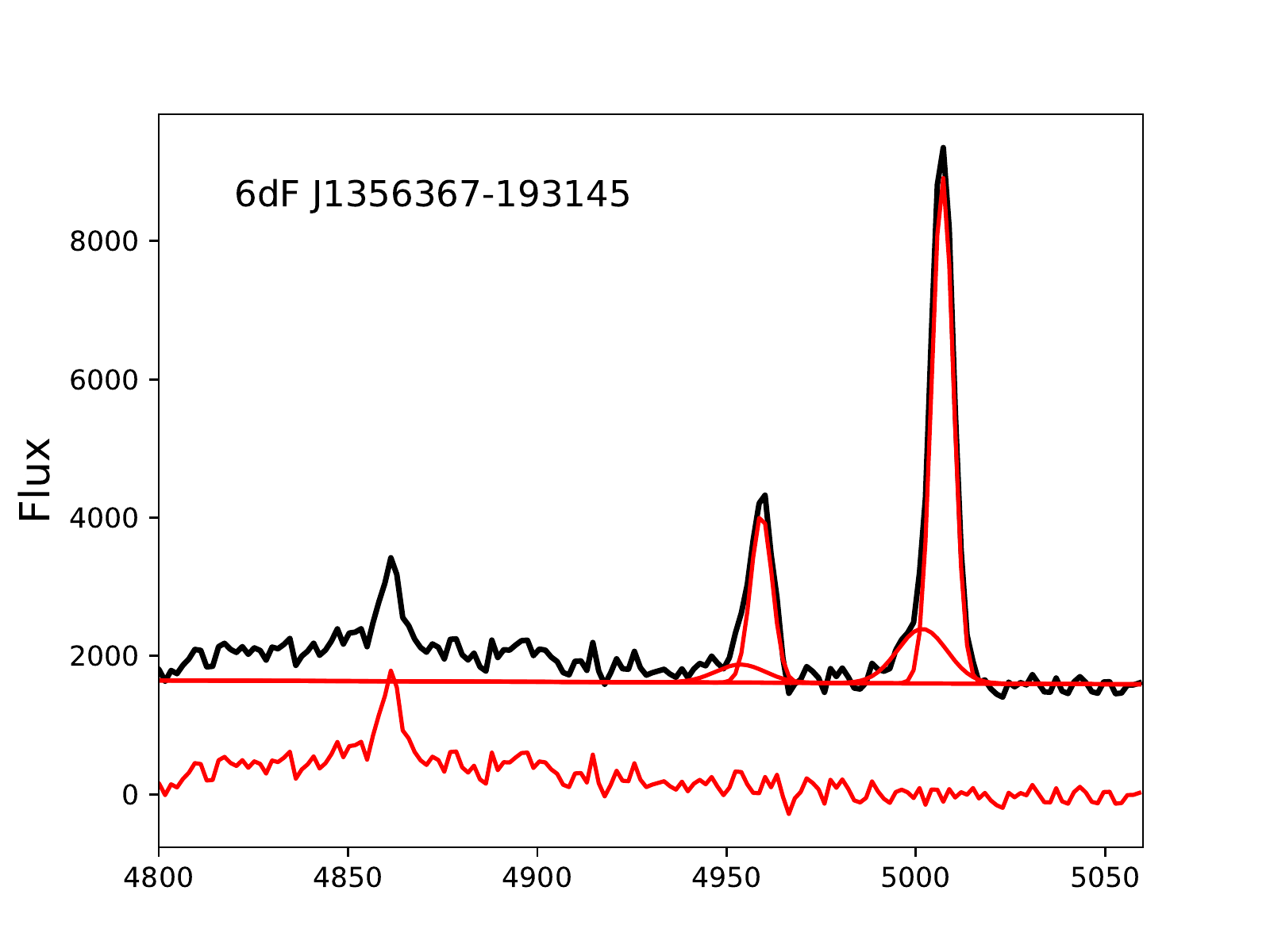}
    \includegraphics[width=0.49\textwidth,height=0.25\textwidth]{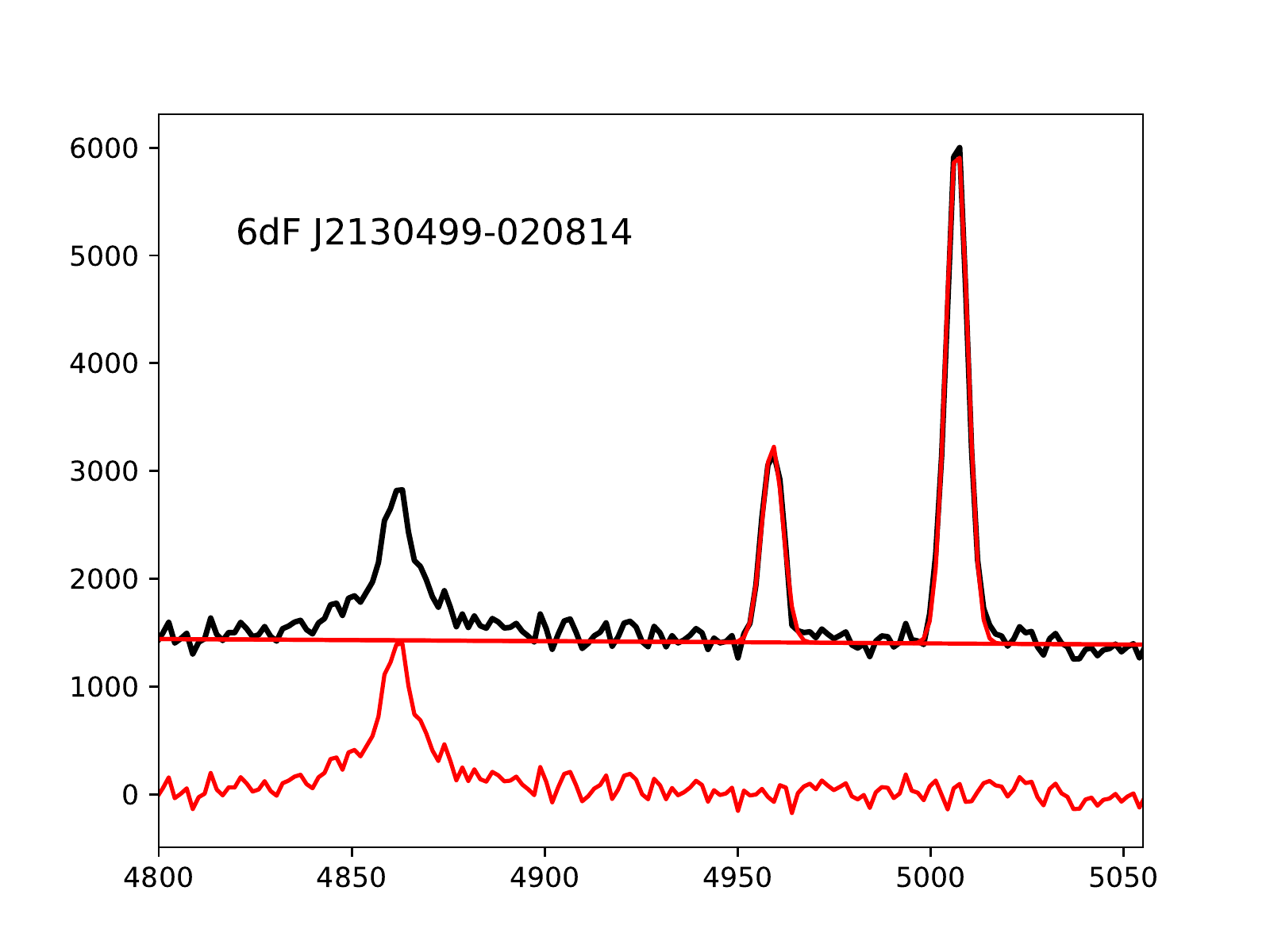}  
    \includegraphics[width=0.49\textwidth,height=0.25\textwidth]{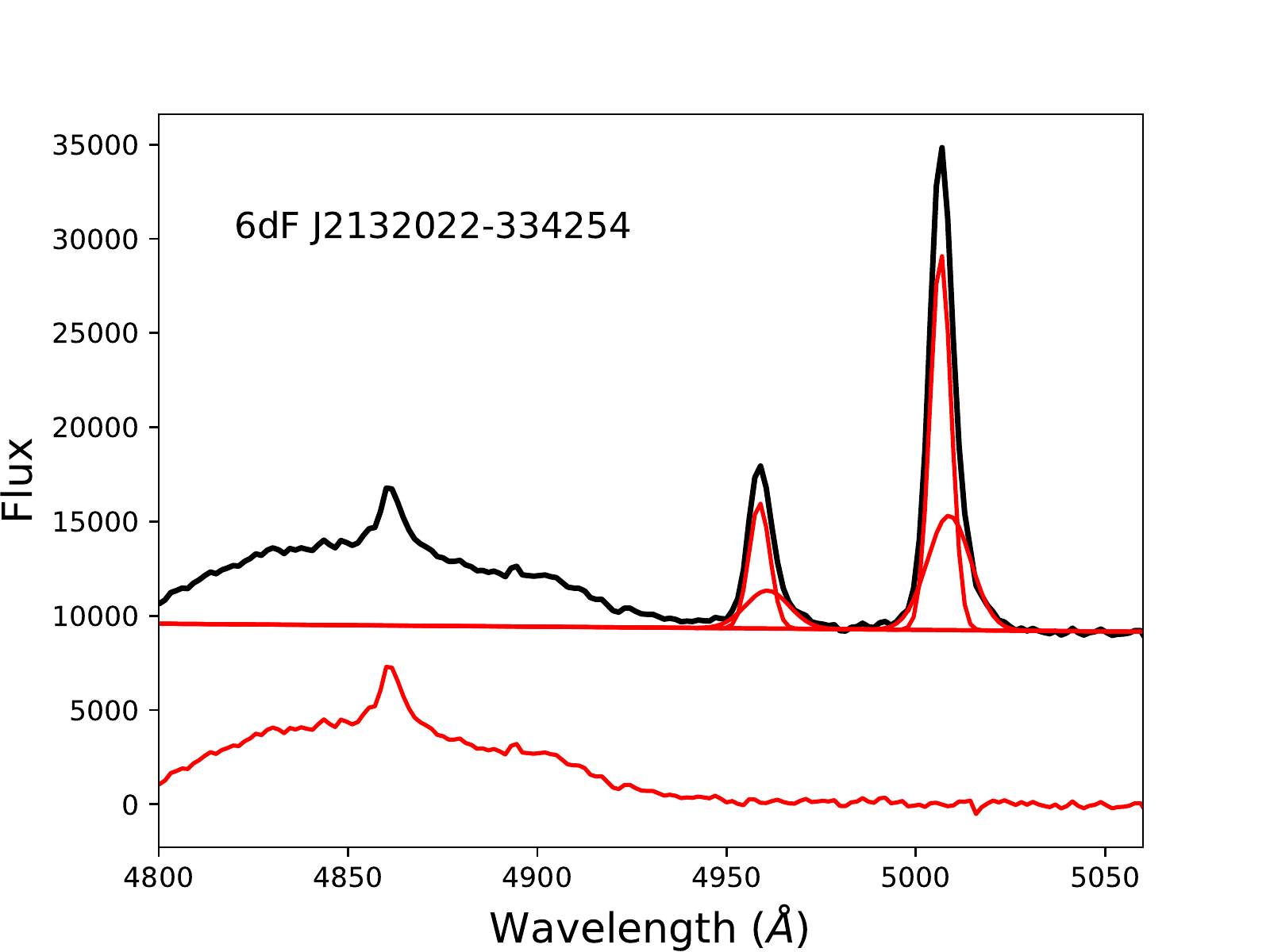} 
    \includegraphics[width=0.49\textwidth,height=0.25\textwidth]{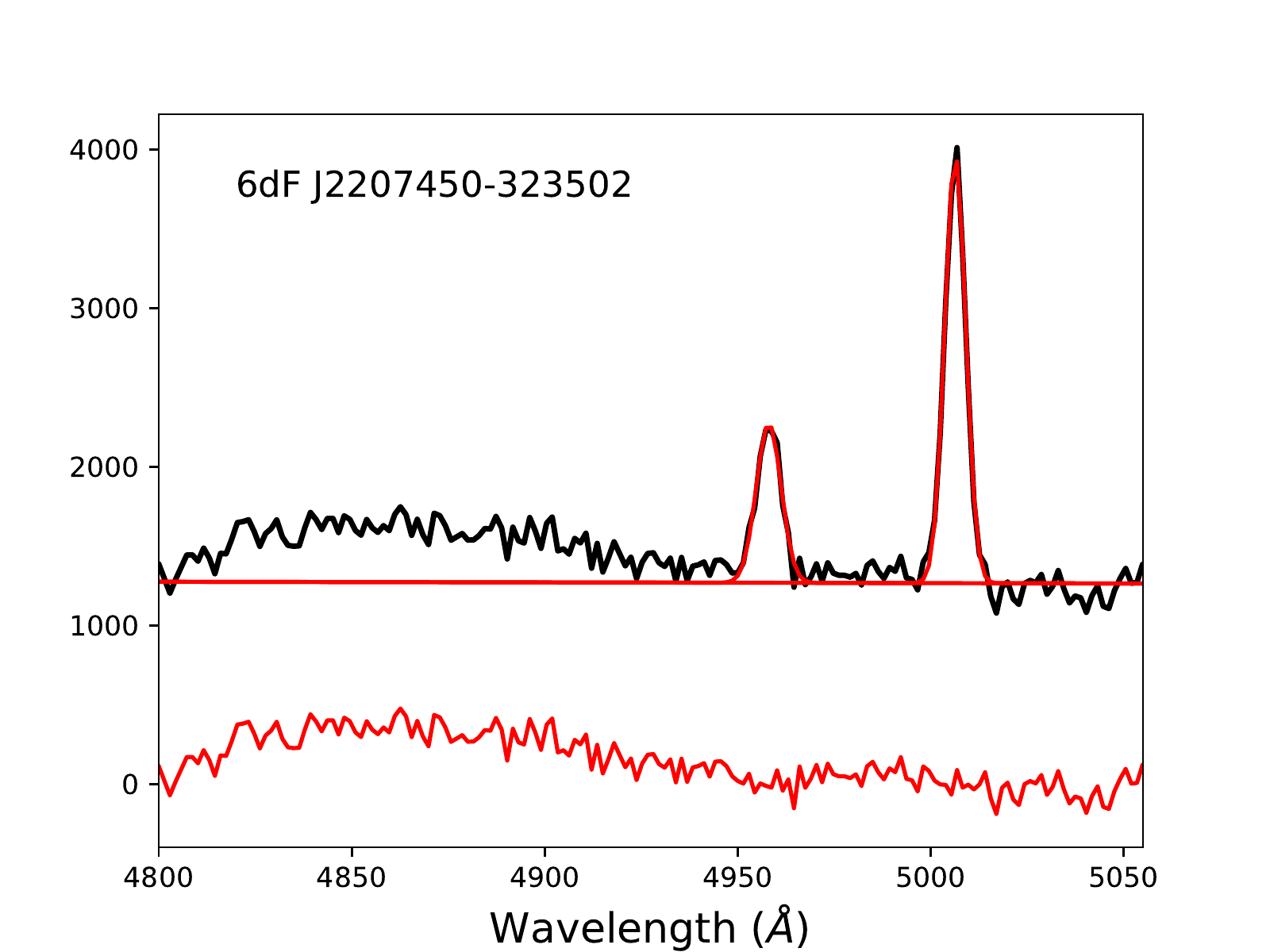}
\contcaption{}
\label{}
\end{figure*}

\begin{figure*}
    \includegraphics[width=0.49\textwidth,height=0.25\textwidth]{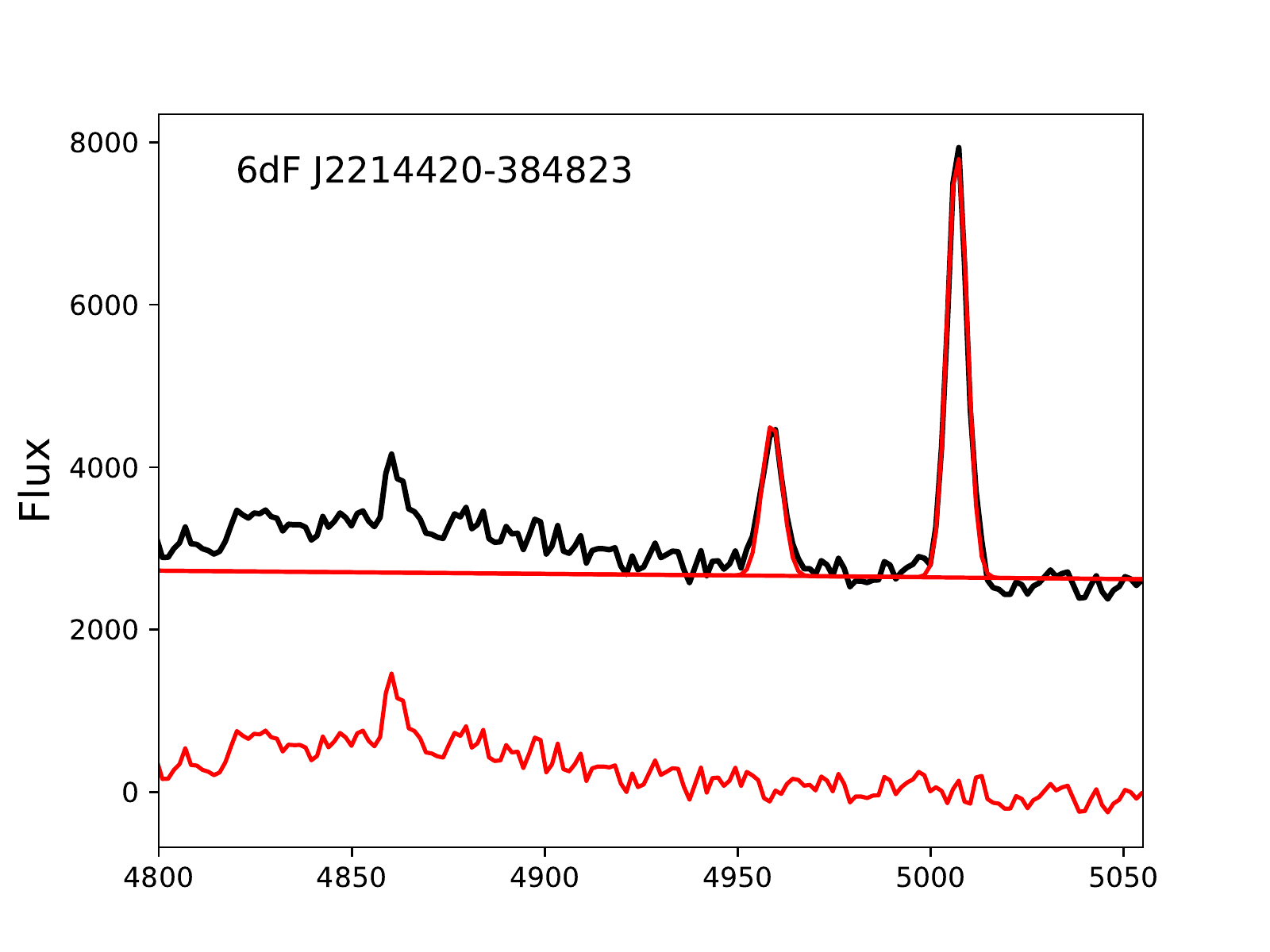} 
    \includegraphics[width=0.49\textwidth,height=0.25\textwidth]{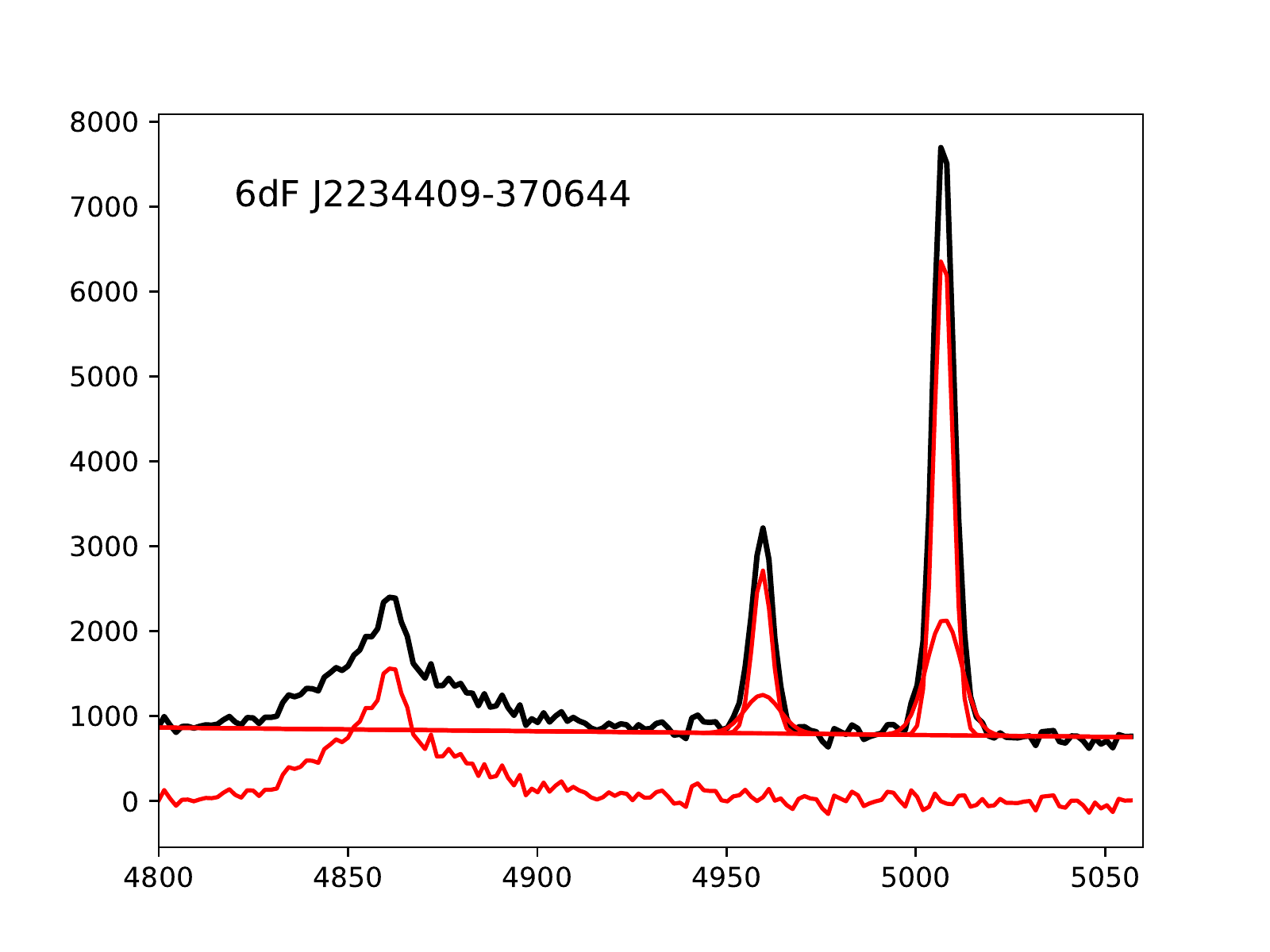}
    \includegraphics[width=0.49\textwidth,height=0.25\textwidth]{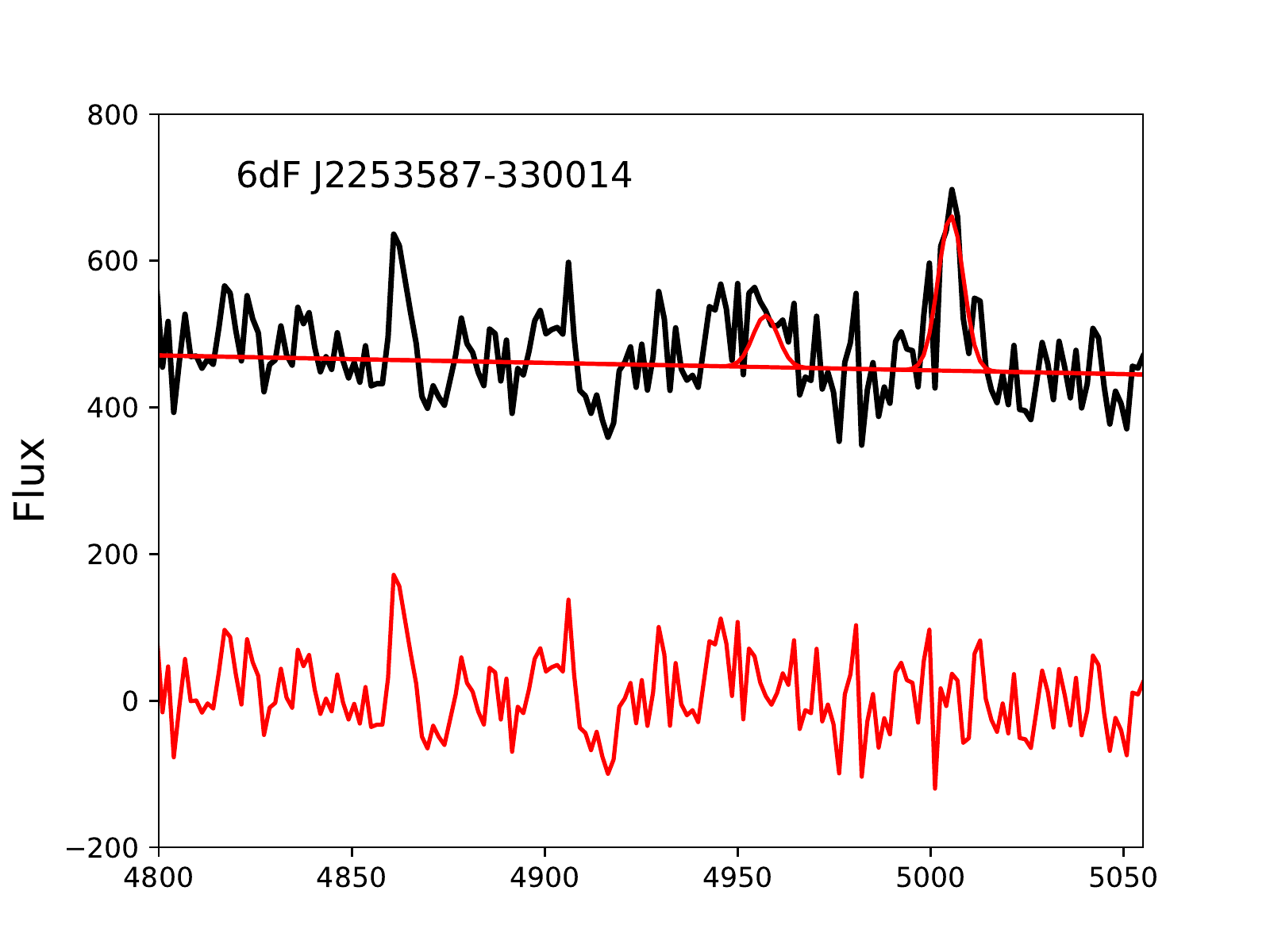}
    \includegraphics[width=0.49\textwidth,height=0.25\textwidth]{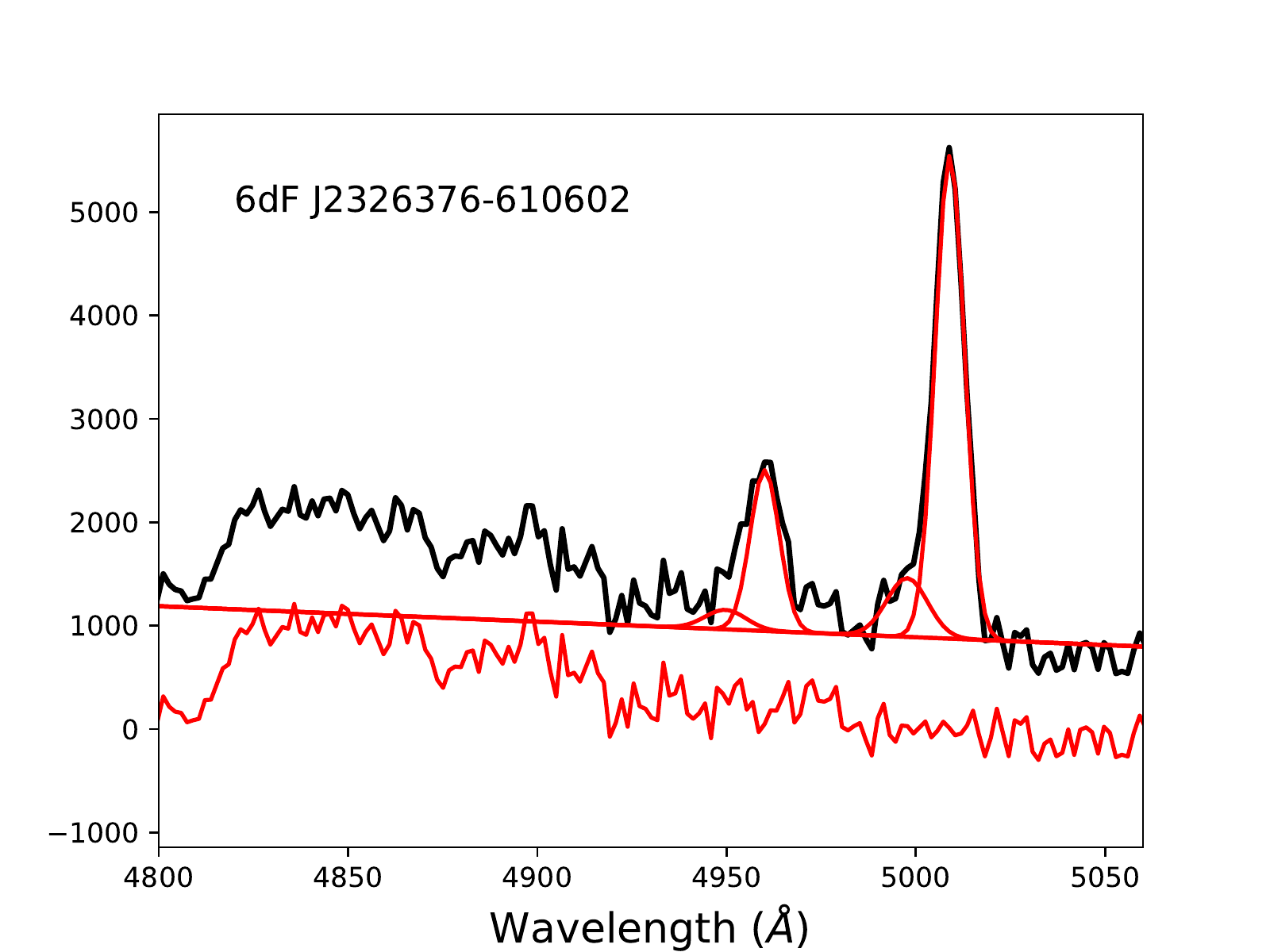}
    \includegraphics[width=0.49\textwidth,height=0.25\textwidth]{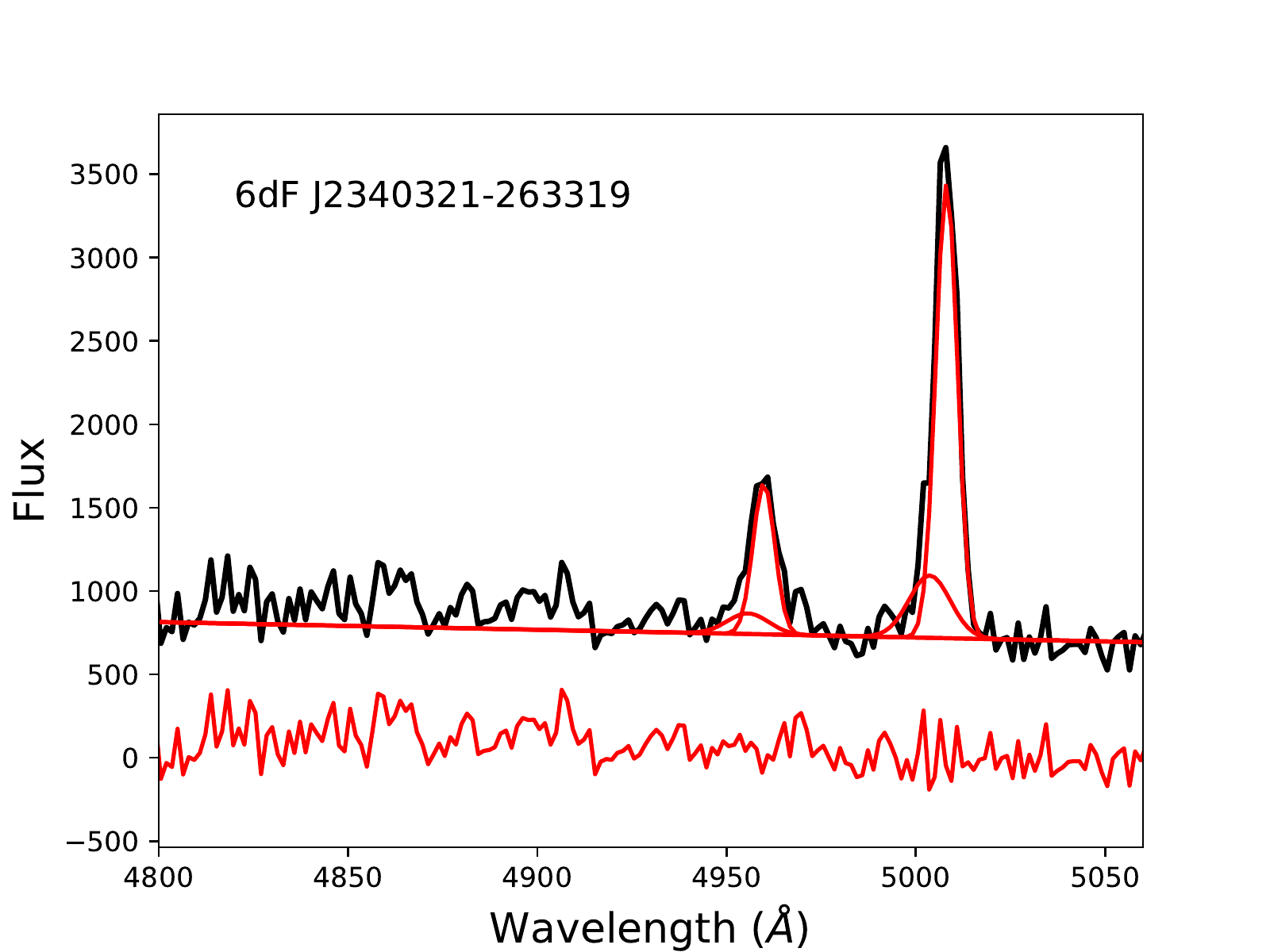}
\contcaption{}
\label{}
\end{figure*} 


\bsp	
\label{lastpage}
\end{document}